\documentclass[aps,twocolumn,floats,nofootinbib,prd,psfig]{revtex4}
\usepackage{graphicx, epsfig, natbib}

\newcommand{\beq}{\begin{equation}}
\newcommand{\eeq}{\end{equation}}
\newcommand{\beqa}{\begin{eqnarray}}
\newcommand{\eeqa}{\end{eqnarray}}
\newcommand {\ds}{\displaystyle}

\newcommand {\ode}{\Omega_{\rm DE}}

\newcommand {\odestart}{\Omega_{\rm DE}^{\rm start}}
\newcommand {\wstart}{w^{\rm start}}
\newcommand {\epsstart}{\epsilon^{\rm start}}
\newcommand {\etastart}{\eta^{\rm start}}
\newcommand {\xistart}{\xi^{\rm start}}

\newcommand {\odenow}{\Omega_{\rm DE}^0}
\newcommand {\wnow}{w^0}
\newcommand {\epsnow}{\epsilon^0}
\newcommand {\etanow}{\eta^0}
\newcommand {\xinow}{\xi^0}

\newcommand {\mpl}{m_{\rm pl}}
\newcommand {\LCDM}{$\Lambda$CDM\ }
\newcommand {\zstart}{z_{\rm start}}
\newcommand {\boldtheta}{\mbox{\boldmath $\theta$}}

\newcommand {\myrule}{\rule[-3mm]{0mm}{8mm}}

\begin{document}

\title{Dynamical behavior of generic quintessence potentials: constraints on
key dark energy observables} \author{Dragan Huterer and Hiranya
V. Peiris\footnote{Hubble Fellow}} \affiliation{Kavli Institute for
Cosmological Physics, Department of Astronomy and Astrophysics, and Enrico
Fermi Institute, University of Chicago, Chicago, IL 60637}

\begin{abstract}
We perform a comprehensive study of a class of dark energy models -- scalar
field models where the effective potential can be described by a polynomial
series -- exploring their dynamical behavior using the method of flow equations
that has previously been applied to inflationary models. Using supernova,
baryon oscillation, CMB and Hubble constant data, and an implicit theoretical
prior imposed by the scalar field dynamics, we find that the \LCDM model
provides an excellent fit to the data. Constraints on the generic scalar field
potential parameters are presented, along with the reconstructed $w(z)$
histories consistent with the data and the theoretical prior. We propose and
pursue computationally feasible algorithms to obtain estimates of the principal
components of the equation of state, as well as parameters $w_0$ and $w_a$. Further,
we use the Monte Carlo Markov Chain machinery to simulate future data
based on the Joint Dark Energy Mission, Planck and baryon acoustic oscillation
surveys and find that the inverse area figure of merit improves nearly by an
order of magnitude. Therefore, most scalar field models that are currently
consistent with data can be potentially ruled out by future experiments. We
also comment on the classification of dark energy models into ``thawing'" and
``freezing" in light of the more diverse evolution histories allowed by this
general class of potentials.
\end{abstract} 

\maketitle


\section{Introduction}

Is the universe accelerating because it is dominated by vacuum energy?  The
answer to this question has become a holy grail of cosmology; in addition to
the increasing improvements in existing constraints \cite{Perlmutter_1999,
Riess_1998, Knop,Riess_2004,Astier}, there are multiple ambitious experimental
efforts currently planned to help find the answer. Vacuum energy would be
spatially homogeneous and time-invariant, possessing a dark energy equation of
state, $w\equiv p_{\rm DE}/\rho_{\rm DE}$, equal to $-1$ identically and at all
times. Finding robust evidence for a deviation from this prediction would be
tremendously important and would suggest yet another cosmological mystery: it
would indicate that dark energy is more complicated than the simplest model,
Einstein's cosmological constant.

Mapping out the history of the equation of state of DE (or dark energy
density), and in particular any variation in redshift of $w(z)$, is therefore
one of the fundamental goals in cosmology. The simplest way to do this is to
measure a single parameter that describes the time-dependence of $w(z)$, such
as its derivative (e.g.\
\cite{Cooray_Huterer,Linder_wa,Corasaniti_Copeland,Hannestad_Mortsell}), while
the most general approach is to directly reconstruct $w(z)$ from the
distance-redshift or expansion rate-redshift data
\cite{reconstr,Nakamura_Chiba,Starobinsky}.  With any given test, a difficult
question arises: if we measure that the equation of state at some pivot
redshift is consistent with $-1$ with some error, is it worth spending a large
amount of time and resources in measuring the temporal variation of $w$ ---
which, one could naively guess, would then be small or zero in any realistic
model of dark energy?

In this paper we address the following question: given the current constraints
on the equation of state of dark energy that are increasingly tightening around
the \LCDM value of $-1$, and assuming a well-defined class of models that
allows diverse behavior in the dark energy sector, is it worth pursuing better
and more expensive cosmological probes in hope of detecting any deviation from
the \LCDM value?  Our analysis differs from a large body of previous work in
that we approach the question in a very general way. Instead of considering
very specific examples of DE models (say, specific power law scalar field
potentials), we encompass {\it all} models within a prescribed framework,
constrain this entire class with the currently available data, and address how
future improvements upon these constraints will affect our ability to test the
\LCDM paradigm for dark energy. To perform these tasks, we use the formalism of
dynamical trajectories in model parameter space that was previously applied to
slow-roll inflation models.

We make no attempt to study the individual or comparative merits of specific
cosmological probes; nor do we consider optimal survey design for any
particular probe. Such studies have been carried out by various authors in the
past (e.g.,
\cite{EHT,Huterer_Turner,Weller_Albrecht,Weller:2000pf,Maor,Kujat,Bassett_compression,
Bassett_optimize,Knox_Song_Zhan,Virey}), and many others have obtained
constraints on standard descriptions of the dark energy sector
\cite{Melchiorri_state,Corasaniti_foundations,Spergel_2003,Spergel_2006,
Tegmark_SDSS,Seljak_SDSS,Upadhye,Xia,Zhao,Nesseris,Jarvis,Doran} as well as
various parametric descriptions of the energy density
\cite{Saini_reconstr,Huterer_thesis,Takada_Jain,Seo_Eisenstein,Alam,
Jonsson,Feng,Jassal,Wang_Tegmark_2005,
Wang_Mukherjee,Shafieloo,Huterer_Cooray}.  Moreover, comparison of cosmological
probes strongly depends on the correct characterization of the systematic
errors. Instead, we concentrate on asking specific questions about a large,
precisely defined physical class of dark energy models using a compilation of
current and (simulated) future data. In the process we propose several recipes
for converting between different relevant parameter bases, and also for
describing the cosmological data. In this sense, we significantly extend the
work of Sahlen et al.\ \cite{Sahlen} who adopted a similar approach of Monte
Carlo reconstruction of the scalar field dark energy models.

The paper is organized as follows. In Sec.~\ref{sec:method} we describe the
motivation for this work and briefly describe the class of dark energy models
that we consider.  In Sec.~\ref{sec:methodology} we lay out the equations
necessary to compute the dark energy history of each model, and outline our
main assumptions that further define our framework. In Sec.~\ref{sec:MCMC} we
describe the Markov Chain formalism we adopt, specify the initial conditions,
and discuss extrapolation between the low redshift and the CMB era. 
In Sec.~\ref{sec:derived} we define some of the derived parameters that
describe the dark energy history, such as the principal components and $w_0$,
$w_a$ and $w_{\rm pivot}$. In Sec.~\ref{sec:results} we present the
cosmological constraints on the various parameters and functions, while in
Sec.~\ref{sec:implications} we discuss the cosmological implications and
figures of merit. We conclude in Sec.~\ref{sec:conclusions}. Appendices A and B
describe in detail, respectively, the current and future cosmological data we
have used, as well as the likelihoods assigned to them.

\section{Motivation and Framework}\label{sec:method}

We would like to ``scan'' through a variety of dark energy models with a
time-dependent equation of state in order to explore the generic allowed ranges
of dark energy sector parameters at redshifts where the observations are
powerful, and then test how well \LCDM can be distinguished from dark energy
models in terms of current and future constraints.  In most general sense,
scanning through dark energy models is impossible as a single physical
framework for dark energy is currently nonexistent. Therefore, we have to
specialize to one of the several classes of 
dark energy models, or else describe the background evolution of the universe
without recourse to a physical model of dark energy.  Here we adopt the former
approach and choose perhaps the most widely considered model: a rolling scalar
field, or quintessence \cite{Ratra_Peebles,Wetterich,Frieman_PNGB,
Coble,Ferreira_Joyce,Zlatev,Liddle_Scherrer}. To scan through the space of
scalar field models, we adapt the Monte Carlo reconstruction formalism that has
previously been applied to inflationary dynamics.  Note that the application of
the flow equations to the DE case requires some modifications compared to the
case of inflation.  Even though the acceleration of the universe implies that
we recently entered an inflationary period, the densities of other components
such as matter are not negligible, and this prevents us from relating the
Hubble parameter to the slow-roll parameters in an easy way as with
inflation. In particular, so-called Hubble slow roll parameters (e.g.,
\cite{LPB, Kinney_2003}) cannot be used since the Hubble parameter is {\it not}
changing very slowly because of the matter component.  In fact, there are no
small parameters in the DE flow equation formalism, as the potential slow-roll
parameters are not necessarily small. Even for a steep potential with large
``slow-roll" parameters much greater than unity, Hubble friction may still
enable slow roll with small kinetic energy of the field and $w(z)\simeq
-1$. Therefore, the allowed ranges for the analogous parameters in the dark
energy case are determined entirely by the data.

\section{Methodology}\label{sec:methodology}

\subsection{Scalar field equations}\label{sec:scalar_field_eqs}

We start with  the Klein-Gordon equation for the single scalar field
\begin{equation}
\ddot{\phi}+3H\dot{\phi} + {dV\over d\phi}=0,
\label{eq:KG}
\end{equation}
where the overdot represents a time-derivative. We impose two
requirements on the rolling field:
\begin{enumerate}
\item The field is not allowed to turn around during its roll on the potential
(i.e.\ models where $d\phi$ changes sign are rejected). If the turnaround
happens, the model is assigned zero likelihood.

\item At initial time, the field is only allowed to roll {\it down}
the potential slope.
\end{enumerate}

These requirements are in general not required; however they simplify the
initial assumptions while not weakening the generality of the framework. Fields
with rapidly changing sign of $d\phi$ --- violating (1) --- describe potentials
with the time-averaged equation of state near zero (e.g., an axionic field),
therefore not describing the DE.  Dark energy models therefore obey the first
requirement. The requirement (2) was imposed to speed up scanning of the
models. We have explicitly considered fields initially rolling up the
potential, and found that nearly all of them turn around, thereby violating
requirement (1).

We can rewrite Eq.~(\ref{eq:KG}) in terms of dark energy's equation of state
and energy density as a function of scale factor

\begin{equation}
{d^2 \tilde{\phi}\over d\ln a^2} + {3\over 2} 
\left [ 1-w\ode\right ] {d \tilde{\phi}\over d\ln a} + 
\frac{3}{16\pi}\,\left ({V'\over V}\right ) \ode (1-w) = 0
\label{eq:2nd_order}
\end{equation}

\noindent where $\tilde{\phi}\equiv \phi/\mpl$. Hereafter, we consider all
dimensionful quantities in units of $\mpl$ and drop the tilde for simplicity.

We integrate Eq.~(\ref{eq:2nd_order}) starting at some redshift $\zstart$. To
do so, we apply initial conditions as values of the equation of state and
energy density of the field at $\zstart$, $\wstart\equiv w(\zstart)$ and
$\odestart\equiv \ode(\zstart)$.  Using the Friedman equation and the definitions
of $w$ and $\ode$, we find the expressions for $w$ and $\ode$ at any given time
as

\begin{eqnarray}
   w &=& {2B(1+A) \over 2A+B} -1 \label{eq:w_ode_A}\\[0.1cm] 
\ode &=& {2A+B \over 2(1+A)}     \label{eq:w_ode_B}
\end{eqnarray}

\noindent where 

\begin{eqnarray}
   A &\equiv & {V(\phi)\over V_0}\, {\odestart (1-\wstart)\over 2(1-\odestart)} 
   \left ({a\over a_{\rm start}}\right )^3  \label{eq:A} \\[0.1cm] 
   B &\equiv & {8\pi\over 3} \left ({d\phi\over d\ln a}\right )^2.
\label{eq:B}
\end{eqnarray}

\noindent and where $a_{\rm start}$ is scale factor at the initial
time and $V_0=V(\phi=0)$ is the potential at initial time of
integration.  We need one further initial condition:
\begin{equation}
\left({d \phi \over d\ln a}\right)_{\rm start} = \left[{3\over{8\pi}}
\odestart(1+\wstart)\right]^{1/2}.
\label{eq:phidot_ini}
\end{equation}

Integrating Eq.~(\ref{eq:2nd_order}) with initial conditions $\wstart$,
$\odestart$ and Eq.~(\ref{eq:phidot_ini}), together with relations
(\ref{eq:w_ode_A}-\ref{eq:B}), gives the full dark energy
history out to arbitrary redshift.  In this work, we integrate the
equation starting at $\zstart=3$. While we can obtain constraints on
the parameters of interest at any given redshift, we choose to
consider them at $z=0$ which is close to the redshift where the
sensitivity of observations is greatest.  We have checked
that the constraints at $z=0$ do not change appreciably if we start
integrating either at $\zstart=2$ or $\zstart=5$.

Eq.~(\ref{eq:2nd_order}) can be integrated forward or backward in
time.  We have explicitly checked that the forward and backward
integration give precisely the same dark energy history (and values of
all parameters as a function of redshift) provided that the final
conditions of integration forward in time were used as initial
conditions of the integration backward in time, or vice versa.  While
is it not explicitly needed to solve the system of equations specified
above, we will also need the following quantity
\begin{eqnarray}
{dw\over d\ln a} 
&=& {1\over \ode}\left[{16\pi \over 3} \left({d\phi \over d\ln a}\right)
  \left({d^2\phi \over d\ln a^2}\right) \right. \nonumber \\[0.2cm]
& & \left . + 3 w (1+w)\ode(1-\ode) \right ].
\end{eqnarray}

\subsection{The flow equations as a potential generator}

The one extra ingredient we require in order to incorporate this formalism into
cosmological parameter estimation is a {\it potential generator} which
parameterizes the potential $V(\phi)$ in as general a manner as possible
\cite{Grivell_Liddle}. In order to accomplish this task we adapt the
inflationary flow equation formalism. Since it was first proposed a few years
ago \cite{Hoffman_Turner, Kinney_2003}, this formalism has been used to
generate a large number of inflationary models in a relatively
model-independent way for the purpose of Monte-Carlo reconstruction (e.g.\
\cite{Easther_Kinney, Peiris_2003}). Its principal advantage is that it does
not rely on any specific particle physics model, but rather starts with the
slow-roll formalism and provides an approximate expansion of the effective
potential in a hierarchy of slow-roll parameters. It has recently been
incorporated directly into cosmological parameter estimation in a way minimizes
the effects of the lack of knowledge of the measure of initial conditions
\cite{Peiris_Easther, PE2}, and we follow the spirit of this latter work.

The dark energy case differs from the inflationary case because it is not
described by slow roll parameters. However we can still specify an
infinite hierarchy of parameters in terms of the derivatives of the potential:
\begin{eqnarray}
\epsilon &=& {\mpl^2 \over 16\pi} \left( {V' \over V} \right)^2 \nonumber \\
^\ell \lambda &=& \left({\mpl^2 \over 8\pi}\right)^\ell {(V')^{\ell-1} \over V^\ell} 
	 {d^{\ell+1} V \over d\phi^{\ell+1}}; \ \ell \geq 1,
 \label{eq:hier}
\end{eqnarray}
where prime denotes derivatives with respect to $\phi$, and later we will refer
to $^1\lambda$ as $\eta$ and $^2\lambda$ as $\xi$. These parameters are related
to each other through an infinite system of coupled first order differential
equations. The trajectory specified by this infinite system is {\it exact};
however, in practice we must truncate it at some finite order, and obtain an
approximate potential. Truncating the hierarchy of flow parameters at the term
$^M\lambda$ means that $^{M+1}\lambda = 0$ at all times as well. From
Eq.~(\ref{eq:hier}), it also follows that $d^{(M+2)} V/d\phi^{(M+2)} = 0$ at
all times. This truncated hierarchy is therefore closed and has an analytic
solution \cite{Liddle_flow}; it simply describes a polynomial of order $M+1$ in
$V(\phi)$ with
\begin{equation}
V(\phi)=V_0\left [1+A_1\phi + A_2\phi^2+\ldots+A_{M+1} \phi^{M+1} \right ].
\label{eq:Vphi}
\end{equation}
The coefficients $A_i$, with $i > 1$, are written in terms of the starting values 
of the flow parameters as
\begin{eqnarray}
A_{\ell+1} &=& \frac{(8\pi)^\ell \ ^{\ell}\lambda_{\rm start}}{(\ell+1)! 
\ A_1^{\ell-1}} \, , \label{eq:coeffs}
\end{eqnarray}
where $A_1 = -\sqrt{16\pi\epsstart}$ specifies the direction the field is
rolling initially (an opposite choice of this direction and $A_1$ will lead to
identical results provided the initial velocity in Eq.~(\ref{eq:phidot_ini}) is
given a minus sign).  The expansion is in field space, not redshift space, and
without loss of generality we assign $\phi=0$ at the starting redshift,
$\zstart$. As you move away from $\zstart$, different $\phi$ values correspond
to different redshifts for different models; thus the expansion is not coeval
between models. Further, the specification of the parameters
$^{\ell}\lambda_{\rm start}$ at $\zstart$ specifies the potential over {\it
all} redshifts. Constraints obtained can be translated to any desired redshift,
and therefore the formalism does not contain an additional assumption of a
pivot-redshift.

In the inflationary context, the fact that the flow parameters are small
(because the inflaton has to roll slowly over $\sim 60$ e--folds of inflation)
means that one can make an argument for retaining terms to a high order in the
hierarchy (say $M+1=10$), assuming that successive higher order flow parameters
can be chosen from narrower and narrower priors. That is, a {\it slow roll
expansion} like Eq.~(\ref{eq:Vphi}) must describe {\it any} generic smooth,
flat potential quite accurately over a wide range of the field roll.

In the dark energy formalism, on the other hand, the flow parameters are not
necessarily small any more, and one can no longer make the assumption that
successive higher order flow parameters can be chosen from narrower and
narrower priors -- i.e. the potential is not necessarily described by a slow
roll expansion.  However the dark energy field does not need to roll for a 
large number of e--folds in order to explain the observed acceleration (unlike in inflation),
and therefore Eq.~(\ref{eq:Vphi}) only needs to describe the field over a narrow range in
its evolution. Also, there are no currently conceived cosmological observations
that can provide us with information to constrain more than a handful of potential
parameters. Therefore we don't necessarily need to make a slow roll expansion
valid over a wide range of field evolution.

Instead, Eq.~(\ref{eq:Vphi}), truncated at a low order, now serves as a
generator of potentials, where the values of the flow parameters at $\zstart$
are generated in a Monte-Carlo fashion to simulate large numbers of potentials
valid over the redshift range $\left[0, \zstart \right]$. We explicitly check
below, using wide ranges for the priors on the initial values of the potential
parameters (i.e. making no assumptions of smallness), that the $w(z)$ histories
generated by truncating the hierarchy at $M+1=2$, $M+1=3$ and $M+1=5$ are
generically qualitatively similar, and therefore our parameter estimation is
not very sensitive to the order of this truncation. We posit that this
similarity is due to two reasons: firstly because the $w(z)$ history only
captures a very limited amount of information about the detailed shape of the
potential, and secondly because such dysmorphic potentials have a greater
chance of the scalar field changing direction, thus causing those models to be
excluded due to our theoretical prior.

In the parameter estimation we carry out below, we choose to close the
hierarchy at $M+1=2$ and $M+1=3$ and consider the models defined in
Table~\ref{tab:summary}. This is equivalent to setting $V^{(3)}(\phi)$ and all
higher derivatives to zero in the first case, and setting $V^{(4)}(\phi)$ and
all higher derivatives to zero in the second. Finally, note that proxies for
the initial potential energy of the field (equivalently, parameter $V_0$) and
initial field velocity $\dot{\phi}_0$ are the parameters $\odestart$ and
$\wstart$, respectively.

\begin{table}[!t]
\begin{large}
  \caption{The classes of dark energy models are shown in the first column.
The second and third columns list the numbers and names of the parameters
describing the effective potential. Note that proxies for the initial field
value and initial potential energy of the field are  parameters not listed in this table
--- $\odestart$ and $\wstart$, respectively. }
  \label{tab:summary}
  \begin{tabular}[t]{|c|c|c|c|}
    \hline
\myrule     Model  &     M+1           &      Potential Parameters       \\\hline\hline
\myrule 	2 $V(\phi)$ parameters   & 	 2	&  $\epsstart$, $\etastart$ \\\hline
\myrule  	3 $V(\phi)$ parameters   & 	 3 	&  $\epsstart$, $\etastart$, $\xistart$ \\\hline
 \end{tabular}
\end{large}
\end{table}

\section{Markov Chain Monte Carlo Analysis}\label{sec:MCMC}

\begin{figure*}[]
\includegraphics[scale=0.5]{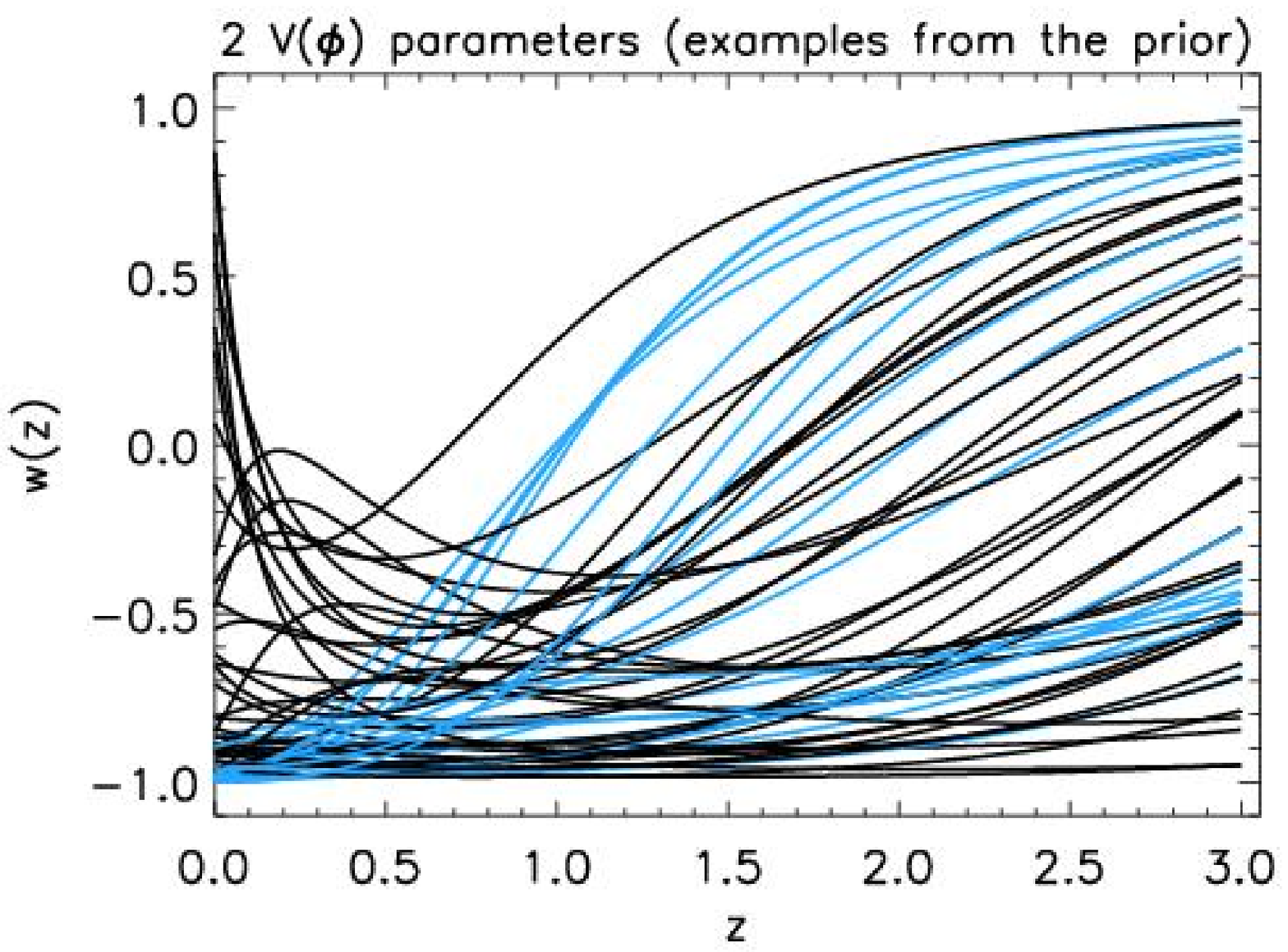} \hfill
\includegraphics[scale=0.5]{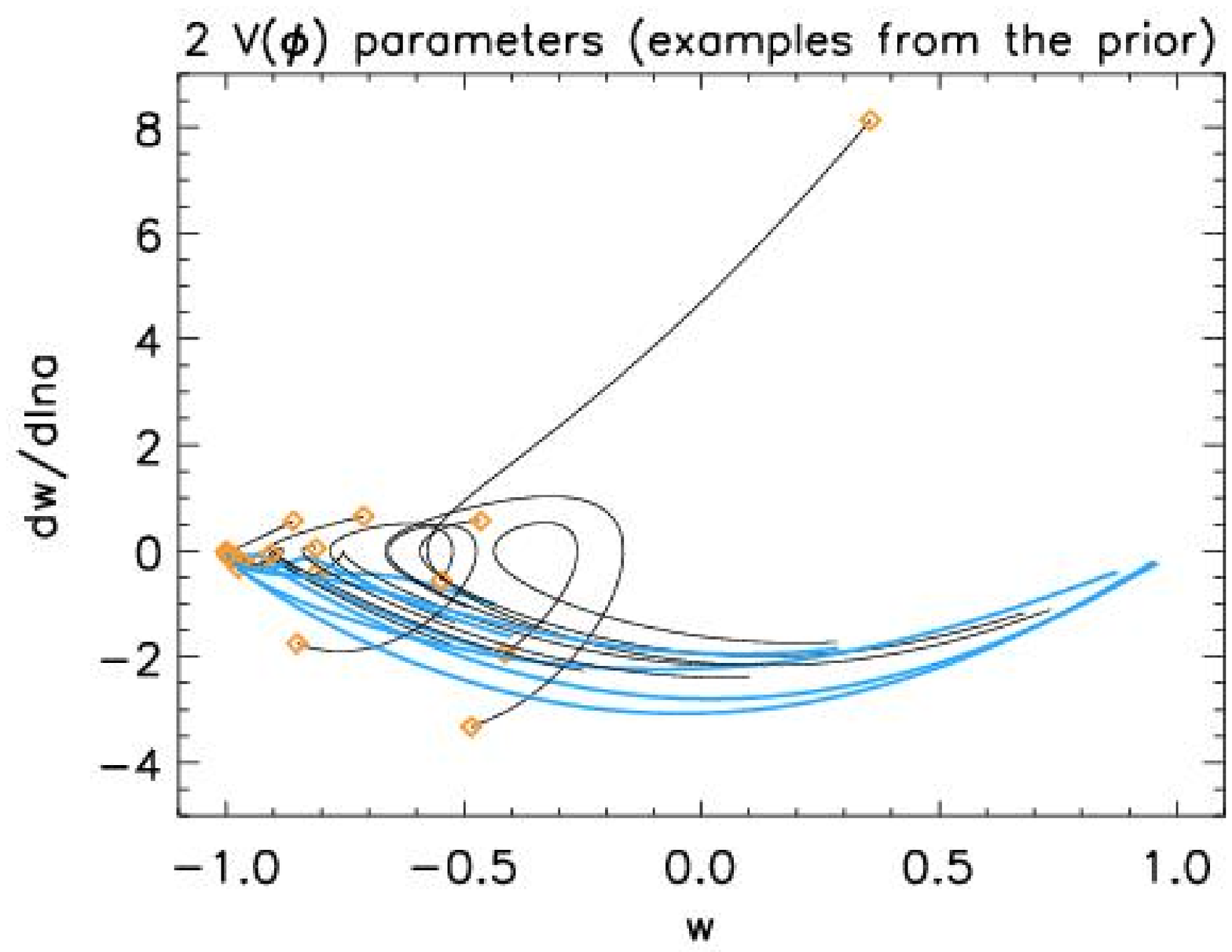}\\[0.1cm]
\includegraphics[scale=0.5]{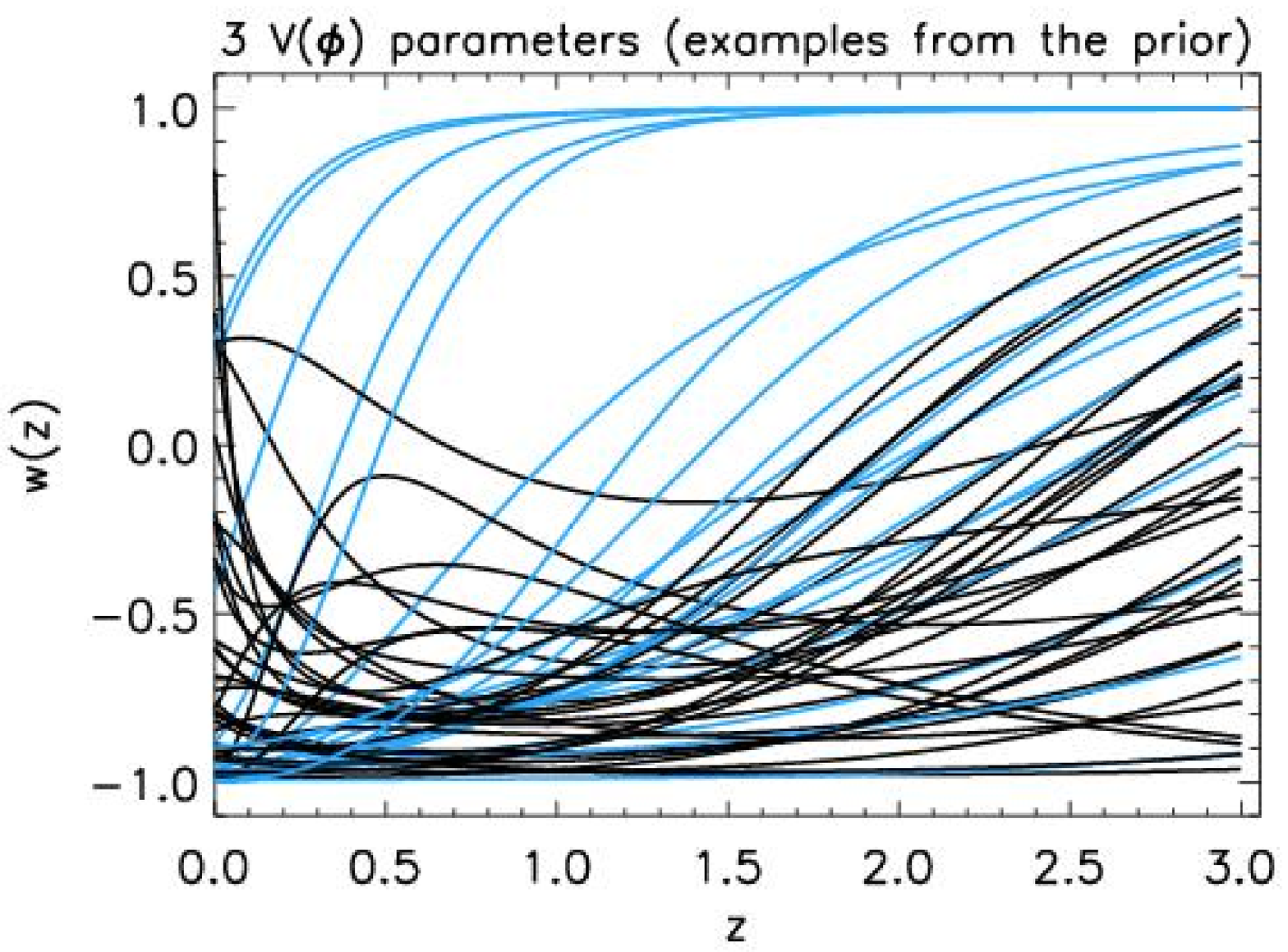}\hfill
\includegraphics[scale=0.5]{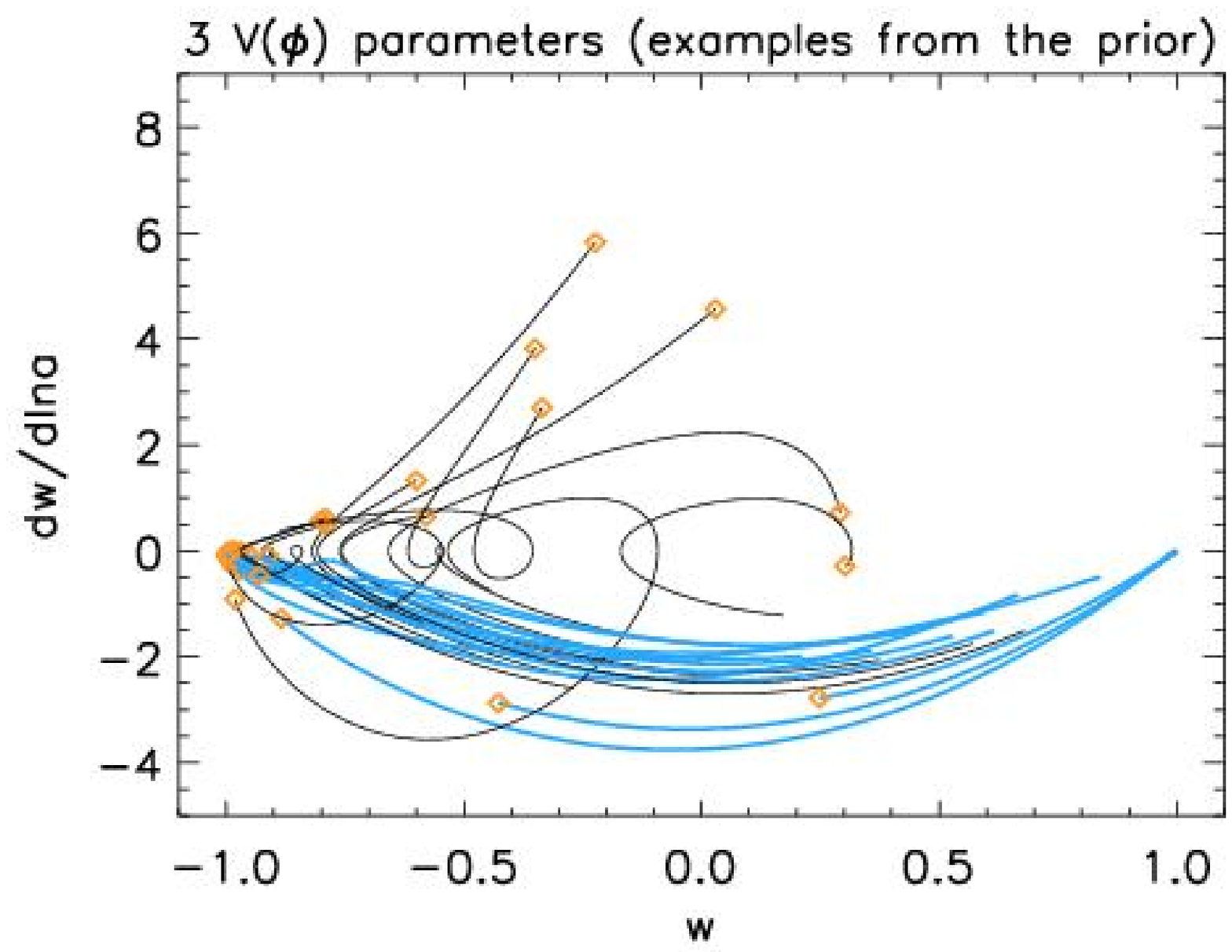}\\[0.1cm]
\includegraphics[scale=0.5]{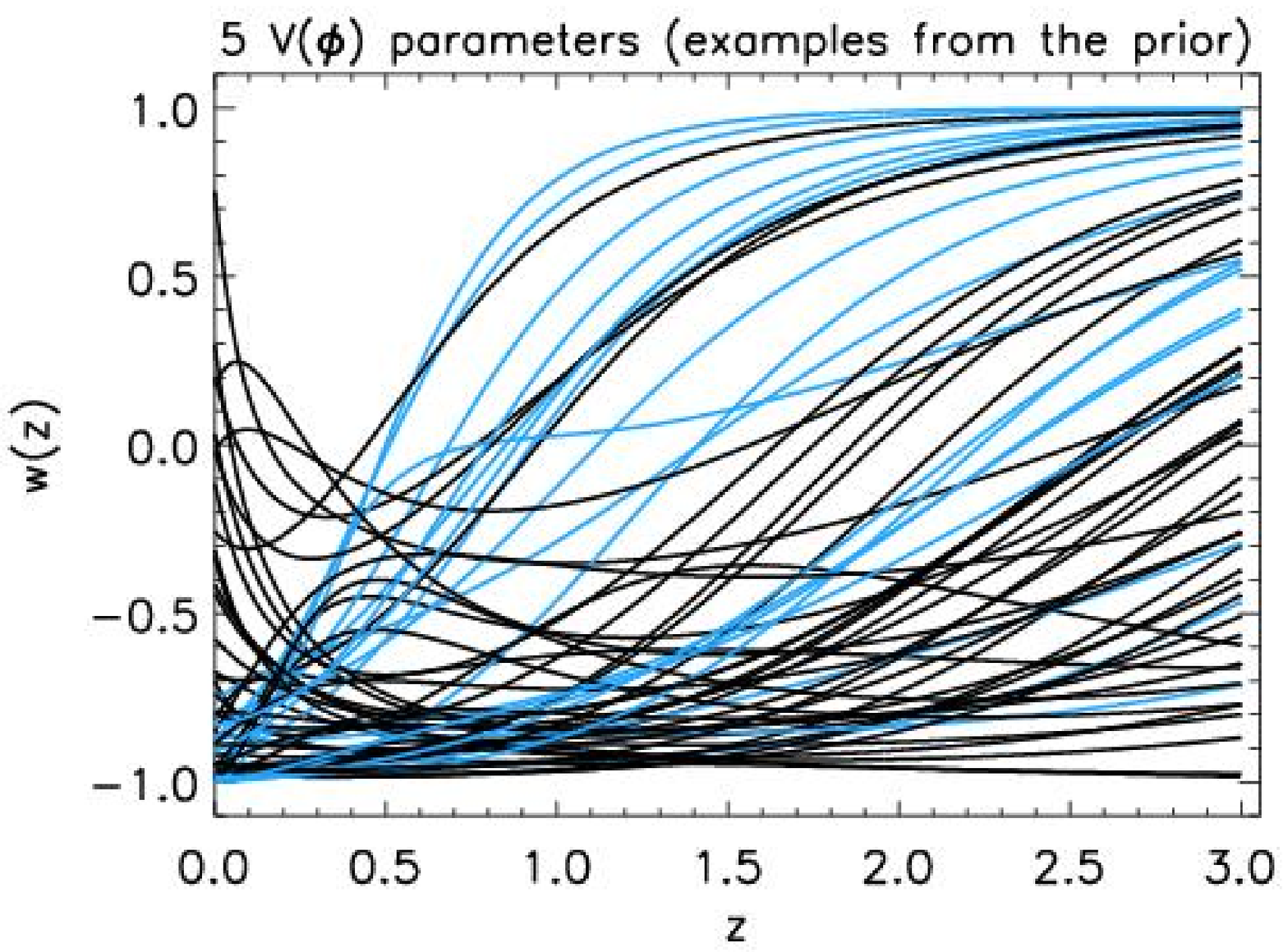} \hfill
\includegraphics[scale=0.5]{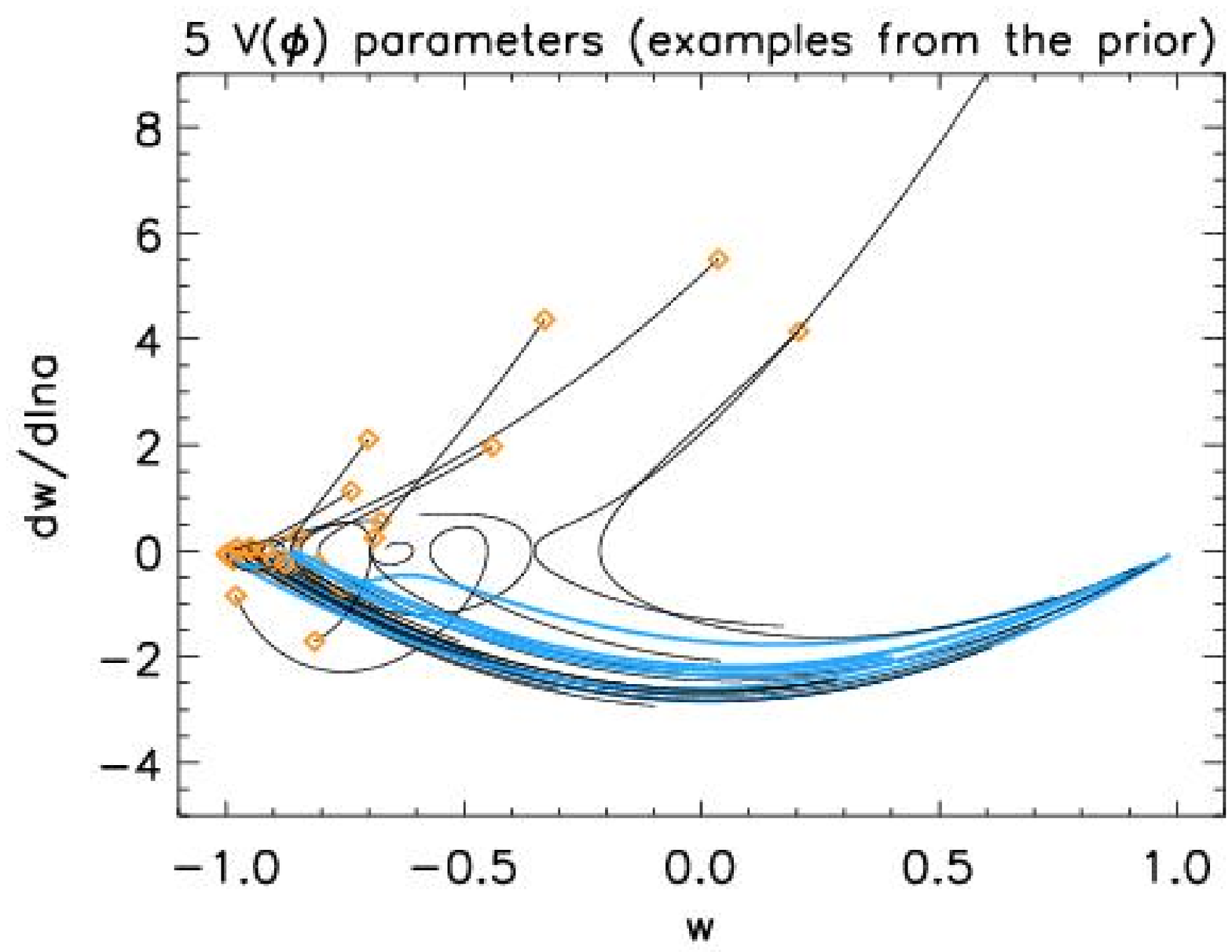}
\caption{Left panels: representative histories $w(z)$ for a number of randomly
chosen sets of initial conditions, shown for illustration without applying
cosmological data constraints. Models which are freezing (see
Sec.~\ref{sec:thawfreeze}) are color-coded in blue, while the models color-coded
in black are neither thawing nor freezing. We find that thawing models are very
rare. Right panels: the corresponding phase plots in the $w-dw/d\ln a$ plane,
showing the evolution of these parameters as a function of redshift. The
evolution histories originate at $\zstart=3$. The orange diamonds represent
$z=0$. The parameterization assumes 2-parameter (top row), 3-parameter (middle
row) and 5-parameter (bottom row) descriptions of $V(\phi)$.}
\label{fig:wz_prior}
\end{figure*}

We use a Markov Chain Monte Carlo (MCMC) technique \cite{Christensen:2000ji,
Christensen:2001gj,Knox:2001fz,Lewis:2002ah,Kosowsky:2002zt,Verde:2003ey} to
evaluate the likelihood function of model parameters. The MCMC is used to
simulate observations from the posterior distribution ${\cal P}({\bf
\alpha}|x)$, of a set of parameters $\boldtheta$ given event $x$, obtained
via Bayes' Theorem,
\begin{equation}
{\cal P}(\mbox{\boldmath $\boldtheta$}|x)=
\frac{{\cal P}(x|\boldtheta){\cal P}(\boldtheta)}{\int
{\cal P}(x|\boldtheta){\cal P}(\boldtheta)d\boldtheta},
\label{eq:bayes}
\end{equation}
\noindent where ${\cal P}(x|\boldtheta)$ is the likelihood of
event $x$ given the model parameters $\boldtheta$ and ${\cal
P}(\boldtheta)$ is the prior probability density. The MCMC
generates random draws (i.e. simulations) from the posterior distribution that
are a ``fair'' sample of the likelihood surface. From this sample, we can
estimate all of the quantities of interest about the posterior distribution
(mean, variance, confidence levels). A properly derived and implemented MCMC
draws from the joint posterior density ${\cal P}(\boldtheta|x)$
once it has converged to the stationary distribution.

We use four chains per model and a conservative Gelman-Rubin convergence
criterion \cite{gelman/rubin} to determine when the chains have converged to
the stationary distribution. For our application, $\boldtheta$ denotes a set of
cosmological parameters: the starting value of the energy density and equation
of state of dark energy, $\odestart$ and $\wstart$, and the starting values of
the potential parameters as specified in Table~\ref{tab:summary}.  This set of
parameters fully determines the dark energy history from $z=\zstart$ until
today. In addition, we marginalize over the angular size of the acoustic horizon
$\theta_A$ (a proxy for the Hubble constant $H_0$) and the fractional physical energy
density in baryons, $\Omega_b h^2$ as nuisance parameters. The universe is
assumed to be to be flat; thus the fractional matter density is given by
$\Omega_m = 1-\Omega_\Lambda$.

The full details of the cosmological data used, along with their likelihood
functions, are given in Appendix A.  Essentially, we use a combination of
supernova data (Supernova Legacy Survey \cite{Astier}), baryon oscillation
results from the Sloan Digital Sky Survey \cite{Eisenstein}, cosmic microwave
background constraints from the WMAP experiment \cite{Spergel_2006}, and the
Hubble Key Project \cite{HKP} measurement of the Hubble constant.  Appendix B
details the projected {\it future} cosmological data, where we assume the same
cosmological probes but with the expected, smaller statistical errors (as well
as estimates of the systematic errors).

Note that we do not know {\it a priori} the ranges of the initial conditions, since
experimental measurements probe a specific temporal average of $\ode$ and
$w$. Therefore, we try to keep the initial conditions as general as possible,
and fully marginalize over the maximally uninformative flat priors, 

\begin{eqnarray}
\odestart            &\in & [0, 1]  \nonumber \\
\wstart              &\in & [-1, 1] \nonumber \\
\epsstart            &\in & [0, \infty] \nonumber \\
\etastart            &\in & [-\infty, \infty] \nonumber \\
\xistart             &\in & [-\infty, 30]  \label{eq:prior}.
\end{eqnarray}
Note that we had to impose an upper limit on $\xistart$ in order to get the
chains to converge: the upper limit in $\xistart$ is essentially unconstrained,
both for current and future data, and our imposed limit assures convergence
without affecting the results. We then apply the following Monte-Carlo
algorithm:

\begin{enumerate}

\item Pick random initial values from the priors in Eq.~(\ref{eq:prior}),
together with values of $\theta_A$ and $\Omega_b h^2$. 

\item Integrate the Klein-Gordon equation; Eq.~(\ref{eq:2nd_order}) to get
$\ode$ at the present epoch and $w(z)$. Compute the derived parameters
$\Omega_m h^2$ and $H_0$, and also the principal components and $w_0$ and $w_a$
(see next Section).

\item Compute the likelihood of this model (using the likelihood functions
defined in the Appendices).

\item Move to the new model using the Metropolis-Hastings sampler.

\item Repeat from step 2. 

\end{enumerate}

Figure \ref{fig:wz_prior} shows randomly chosen DE histories $w(z)$ assuming 2,
3 and 5-parameter descriptions of $V(\phi)$ (left panels; top, middle, and
bottom rows respectively).  For illustration, we show here the results of Monte
Carlo draws from the prior, before applying cosmological constraints. To a
first approximation, the three cases look qualitatively very similar. In
particular, both show examples of freezing models --- where $w(z)$ is
asymptotically approaching $-1$ (blue curves), using the language of Caldwell
and Linder \cite{Caldwell_Linder}, while we see very few thawing models ---
where $w(z)$ is asymptotically receding from $-1$.  We also see examples of
models that are neither purely freezing nor thawing (black curves); we comment
on this in more detail in Sec.~\ref{sec:thawfreeze}.  The right panels of
Figure \ref{fig:wz_prior} show the corresponding phase plots (i.e. the
evolution of each of the models above on the $w-dw/d\ln a$ plane). These plots
therefore show typical histories of $w(z)$ which are generated by the prior on
the parameters which are going to be subsequently confronted with the data
using the MCMC method.

Note that we have cases where $w(\zstart)>0$. While these models may in
principle not be ruled out by the low-z data (SNe, BAO), they may well not be
viable if the energy density in DE is high enough at $z>\zstart$ to spoil the
successful formation of structure. A strong test of such models is imposed by
the CMB measurement of the distance to recombination, encapsulated by the
angular size of the acoustic horizon $\theta_A$. To apply the CMB
acoustic scale test, it is clear that we need to know $w(z)$ at $z>\zstart$;
however this is difficult as it does not seem reasonable to extrapolate our
local expansion of $V(\phi)$ to the early universe and integrate the equation
for $\phi$ backward to $z\approx 1000$. Instead, we choose a simpler
extrapolation in the equation of state, and assume that the equation of state
is constant beyond our starting redshift, $w(z>\zstart) =w(\zstart)$. While
somewhat ad hoc, this choice seems reasonable, as the only way that most models
with $w(z>\zstart)>0$ could survive the data cut is if their potential had a
feature, so that $w(z)$ at high $z$ becomes again sufficiently negative. We
implicitly ignore this small class of potentials that is unconstrained by the
data. Further, we have repeated our analysis without the $\theta_A$ constraint
(that is, without the constraint from distance to the last scattering surface) and
found small differences at $z=\zstart$ --- dark energy models with
$w(\zstart)>0$ are allowed without the $\theta_A$ essentially because the early
history of dark energy is completely unconstrained in that case. However, the
low-redshift constraints are nearly unchanged with or without $\theta_A$
imposed. Models that would be ruled out with the $\theta_A$ (that is, those
with the ``incorrect'' distance to the last scattering surface) are often ruled
out by the SNe+BAO+CMB+$H_0$ combination alone because they do not revert to
being close to \LCDM at low redshifts where the constraining power of the data
is the greatest. We conclude that our ansatz $w(z>\zstart)=w(\zstart)$ has no
discernible effects on the final low-redshift constraints.

The following sections detail the dark energy observables on which we obtain constraints.

\section{Derived parameters: definitions}\label{sec:derived}

In what follows we adopt the 2-parameter description of $V(\phi)$ (see
Eq.~(\ref{eq:Vphi})) as our fiducial case, and also consider the 3-parameter
description as a test of the sensitivity of our results to the truncation of
$V(\phi)$.  Therefore, we have a total of at least four parameters that
describe the background cosmology: at least two describing the potential
($\epsstart$ and $\etastart$), and then $\odestart$ and $\wstart$. Note that we
will be mainly interested in constraints on the {\it present-day} values of the
first four parameters, which we denote with the superscript ``0''. Finally, we
will {\it not} be very interested in the constraints on the remaining
parameters, $\theta_A$, $H_0$, $\Omega_m h^2$ and $\Omega_b h^2$, because we
found that, as expected, the constraints on these parameters largely reproduce
the prior given to them by the observational measurements (the details of which
are in the Appendices).

In addition to presenting constraints on our fiducial parameters and
looking at dark energy histories via $w(z)$, we would like to report errors on
the commonly used 2-parameter description of the equation of state $(w_0, w_a)$
where \cite{Chevallier_Polarski,Linder_wa}

\begin{equation}
w(z)=w_0+ w_a {z\over 1+z }.
\label{eq:w0wa_def}
\end{equation}

However, as seen in Figure \ref{fig:wz_prior}, the scalar field models
evidently do not follow the relation (\ref{eq:w0wa_def}) over the redshift
range considered, though they (presumably) do over the range of redshifts
strongly probed by the data.  Therefore, we need an algorithm to assign the
best-fitting $w_0$ and $w_a$ to a given dark energy history. We do that in two
steps: we first determine the best-fitting principal components computed from
an idealized Fisher matrix analysis, then convert the first two of them into
the familiar equation of state parameters. We now describe this procedure in
detail.

\subsection{Principal components}\label{sec:PC_define}

A convenient way to describe a given dark energy model is to compute the
best-measured principal components (PCs) of $w(z)$ for each dark energy model history
\cite{Huterer_Starkman}. This is a form of data compression, where many
parameters needed to describe the dark energy sector are replaced by a few that
describe the quantities that are best measured by a given set of cosmological
probes. The main advantage of the PCs is that they are model independent --- in
other words, they are independent of parameterizations of dark energy density or
equation of state.  The weight of the best-measured principal component is a
model-independent predictor of what redshift range is best probed by a given
cosmological probe --- the first PC peaks at $z\sim 0.2$ for future SNe
measurements \cite{Huterer_Starkman, Crittenden_Pogosian}, $z\sim 0.5$ for weak
lensing \cite{Linder_Huterer_howmany,Simpson_Bridle_PC} and $z\sim 0.7$ for baryon
oscillations. The shapes of principal components also depend, albeit more
weakly, on the survey specifications and the fiducial cosmological model.

\begin{figure}[!t]
\psfig{file=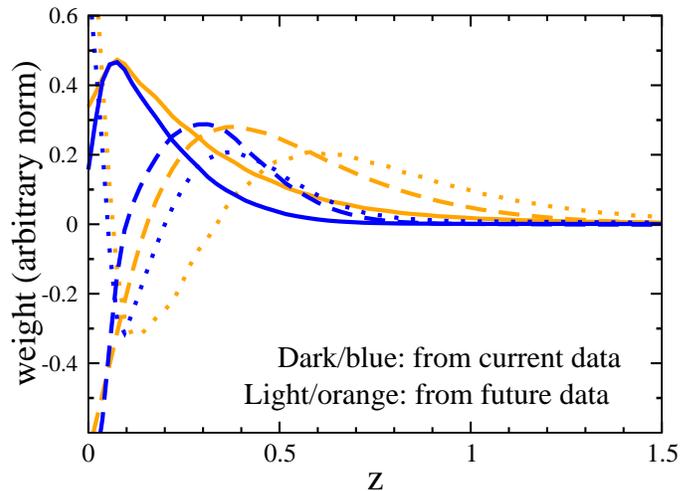,width=3.5in}
\caption{The first three principal components that are best measured by the
current data (solid, dashed and dotted dark/blue line) and the future data
(solid, dashed and dotted dark/blue line light/orange line).  The components
were computed starting with 40 piecewise constant values of the equation of
state linearly distributed in the redshift range $0\leq z\leq 2$ and using the
Fisher matrix formalism, as described in \cite{Huterer_Starkman}.  Note that
the future PCs have peak weight at somewhat higher redshifts, which directly
reflects the extended redshift reach of future SNe and BAO observations.  }
\label{fig:PC_splined}
\end{figure}

In order to compute the principal components from a given survey, one needs to
compute joint constraints on a large number of parameters that determine the
function $w(z)$, then diagonalize their covariance matrix. While relatively
straightforward to do in the Fisher matrix formalism and with simulated data
\cite{Huterer_Starkman,
Hu_PC,Linder_Huterer_howmany,Crittenden_Pogosian,Stephan-Otto,Dick}, this task
is very challenging with actual data as accurate computation of the covariance
matrix of a large number ($\sim 10$-$50$) of highly correlated parameters is
required, although the computational requirements are significantly less severe
if one only wants a rough resolution of the principal components in redshift as
done in Refs.~\cite{Huterer_Cooray} and \cite{Shapiro_Turner}.

Note that we could in principle compute the exact principal
components of our scalar field model equations of state, since we have the full
$w(z)$ of each model in our Markov chains; it is just a matter of
outputting these quantities at a number of redshifts, computing their covariance
matrix from the information in the chains, and diagonalizing it.  However, the
equation of state values at different redshifts {\it in scalar field models} are highly
correlated since the scalar field paradigm together with the few-parameter
description of the effective potential models only allows specific dark energy
histories. These correlations render the principal components of the scalar
field equation of state very different from the usual PCs of a general
$w(z)$. While it may be of interest to work with the scalar field PCs, this is
largely {\it terra incognita}, and in this work we instead concentrate on the
usual PCs for a completely general $w(z)$ computed for a combination of surveys
that we consider. We leave the study of scalar-field-specific PCs  for future work.

Therefore, we adopt a hybrid approach here: we compute the principal components
assuming a Fisher matrix as an approximation to the inverse of the true
covariance matrix, and also assuming theoretically unconstrained dark energy
histories. We use the actual cosmological data, current or future (from
Appendices A and B), approximating the CMB information by the distance to the
last scattering surface which has been marginalized over the sound horizon
parameters; this analysis closely parallels that in \cite{Huterer_Starkman,
Linder_Huterer_howmany}.  We parameterize the equation of state in 40 piecewise
constant values uniformly distribution in $0<z<z_{\rm max}$. We evaluate the
original Fisher matrix, invert it, eliminate the parameters not
corresponding to $w(z)$ (such as $H_0$ and $\Omega_m$), and diagonalize the
resulting matrix in the equation of state parameters to obtain the principal
components.

Figure \ref{fig:PC_splined} shows the shapes $e_i(z)$ of the first three
principal components of the current data (solid, dashed and dotted dark/blue
line) and the future data (solid, dashed and dotted light/orange line). The
curves have been smoothed in redshift with a cubic spline, and are
normalized arbitrarily in this plot.  Note that the weights of the future PCs
peak at somewhat higher redshifts, which directly reflects the
extended redshift reach of the future SNe and BAO observations.

Given a dark energy history $w(z)$, the amplitude of $i$th principal component,
$\alpha_i$, is given by \cite{Huterer_Starkman}
 
\begin{equation}
\alpha_i = \int_0^{\infty} w(z)\, e_i(z)\, dz
\label{eq:alpha_i}
\end{equation}

\noindent where the principal components are normalized so that
$\int_0^{\infty} e_i(z)\, dz=1$, and the error in $\alpha_i$ is determined by
the cosmological datasets used. For each model from the Markov chains, we
compute the principal component coefficients according to
Eq.~(\ref{eq:alpha_i}).  Given that we are now using the pre-computed principal
components that are obtained for a general $w(z)$, the parameters $\alpha_i$ will
{\it not} be uncorrelated in our scalar field dark energy analysis.

\subsection{The equation of state parameters}\label{sec:w0wa_define}

We would also like to show our constraints in terms of more familiar
2-parameter description of dark energy $w_0$ and $w_a$. Of course, our goal is
not to simply fit $(w_0, w_a)$ to the cosmological data as often done in the
literature --- instead, we seek to describe each scalar field model that we
generate with some effective $(w_0, w_a)$.  Mapping the true redshift-dependent
equation of state of a model into a finite number of parameters is not trivial.
The most obvious way to proceed would be to fit $(w_0, w_a)$, together with
other cosmological parameters, to each dark energy model in our
chains. Unfortunately, this procedure would be too time-consuming, as we would
need to perform a multi-parameter fit to each one of millions of dark energy
models that we generate in our MCMC procedure. Clearly, a faster way to construct
$(w_0, w_a)$ is needed. 

Here we propose and adopt an alternative, extremely simple algorithm: we simply
convert the first two principal components $(\alpha_1, \alpha_2)$ into $(w_0,
w_a)$.  This approach is justified because the first two PCs carry essentially
all the necessary information about the effects of dark energy dynamics on the
expansion of the universe over the observable scales. From
Eqs.~(\ref{eq:w0wa_def}) and (\ref{eq:alpha_i}) it follows that we can define

\begin{figure*}[t]
\psfig{file=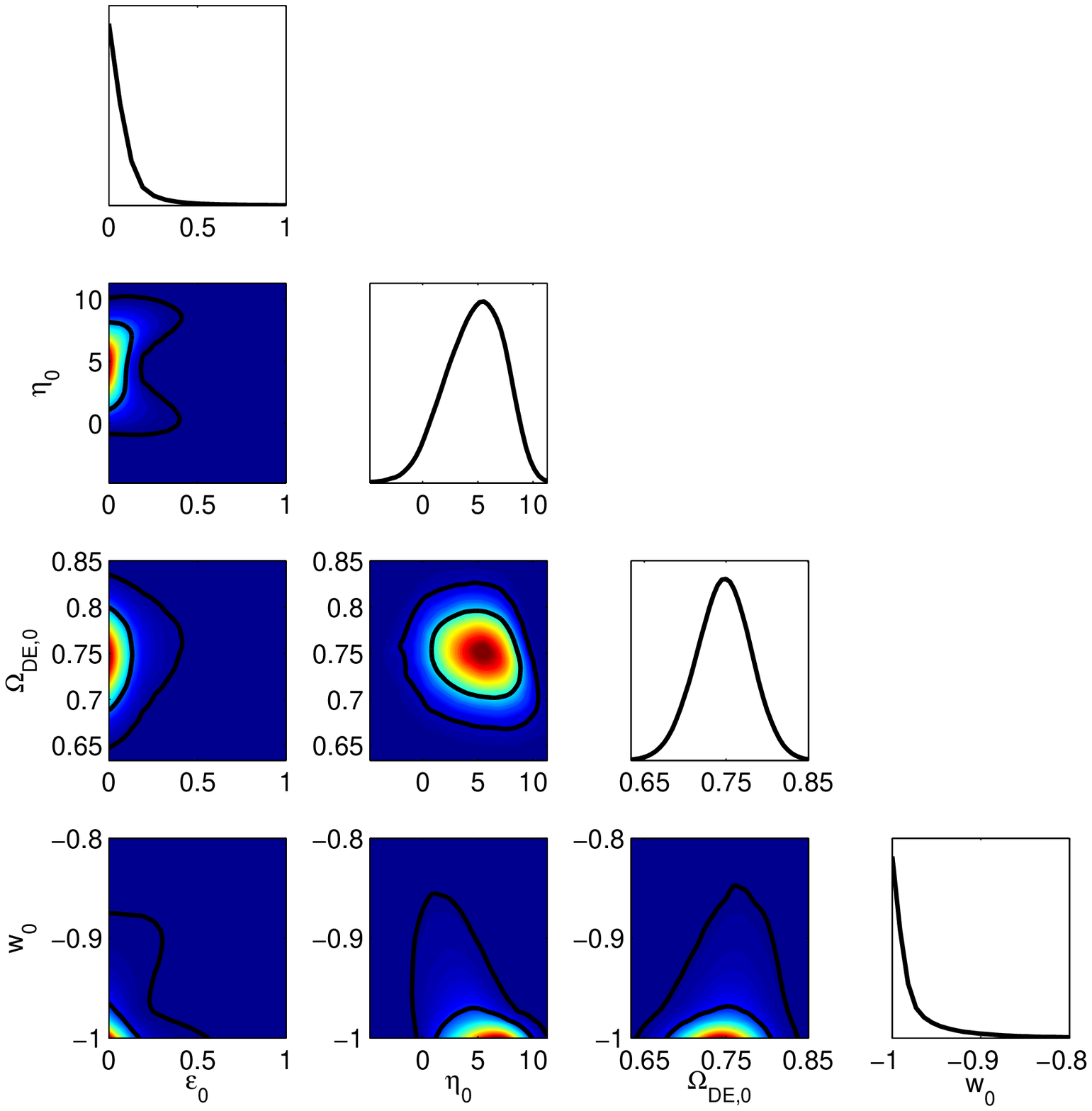,width=3.5in}\hfill
\psfig{file=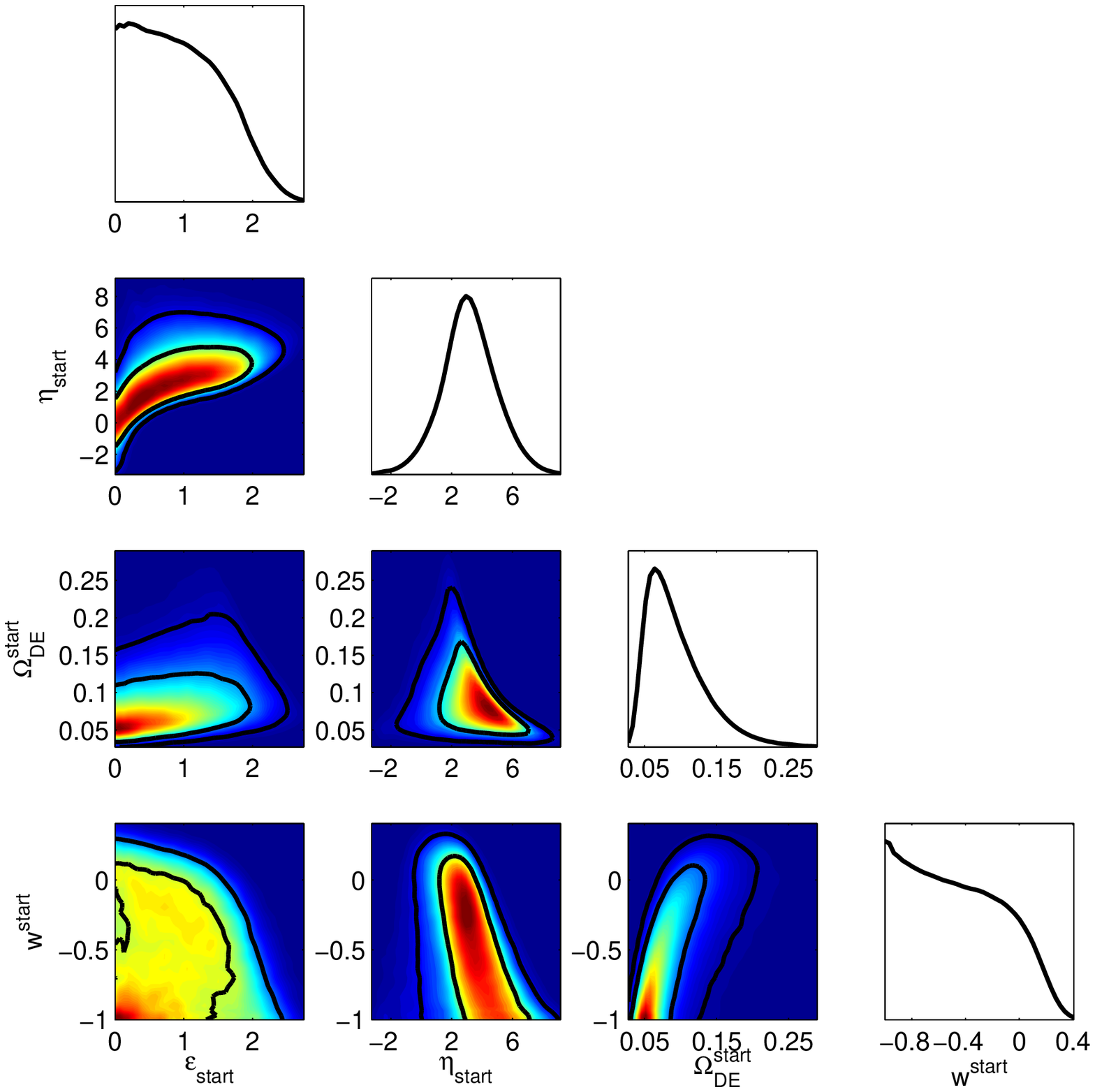,width=3.5in}
\caption{Solid lines show the marginalized 2D-joint 68\% and 95\% probability
contours (off-diagonal panels) and 1D marginalized probability distribution
(diagonal panels) for the two potential parameters $\epsilon$ and $\eta$, as
well as $\ode$, and $w$. The color coding in the off-diagonal panels shows the
marginalized probability density in these 2D parameter spaces, ranging from red
for the highest density to blue for the lowest. The left and right sets of
panels show constraints at $z=0$ and $\zstart=3$ respectively, using the
current cosmological data as described in Appendix \ref{app:current}. We do not
show the parameters $H_0$, $\Omega_m h^2$ and $\Omega_b h^2$ which largely reproduce the
applied measurement priors. Note that the constraints at $\zstart=3$ are
considerably weaker than at $z=0$, since the constraining power of the data is
concentrated at lower redshift.}
\label{fig:like_4x4_2params}
\end{figure*}

\begin{figure*}[t]
\psfig{file=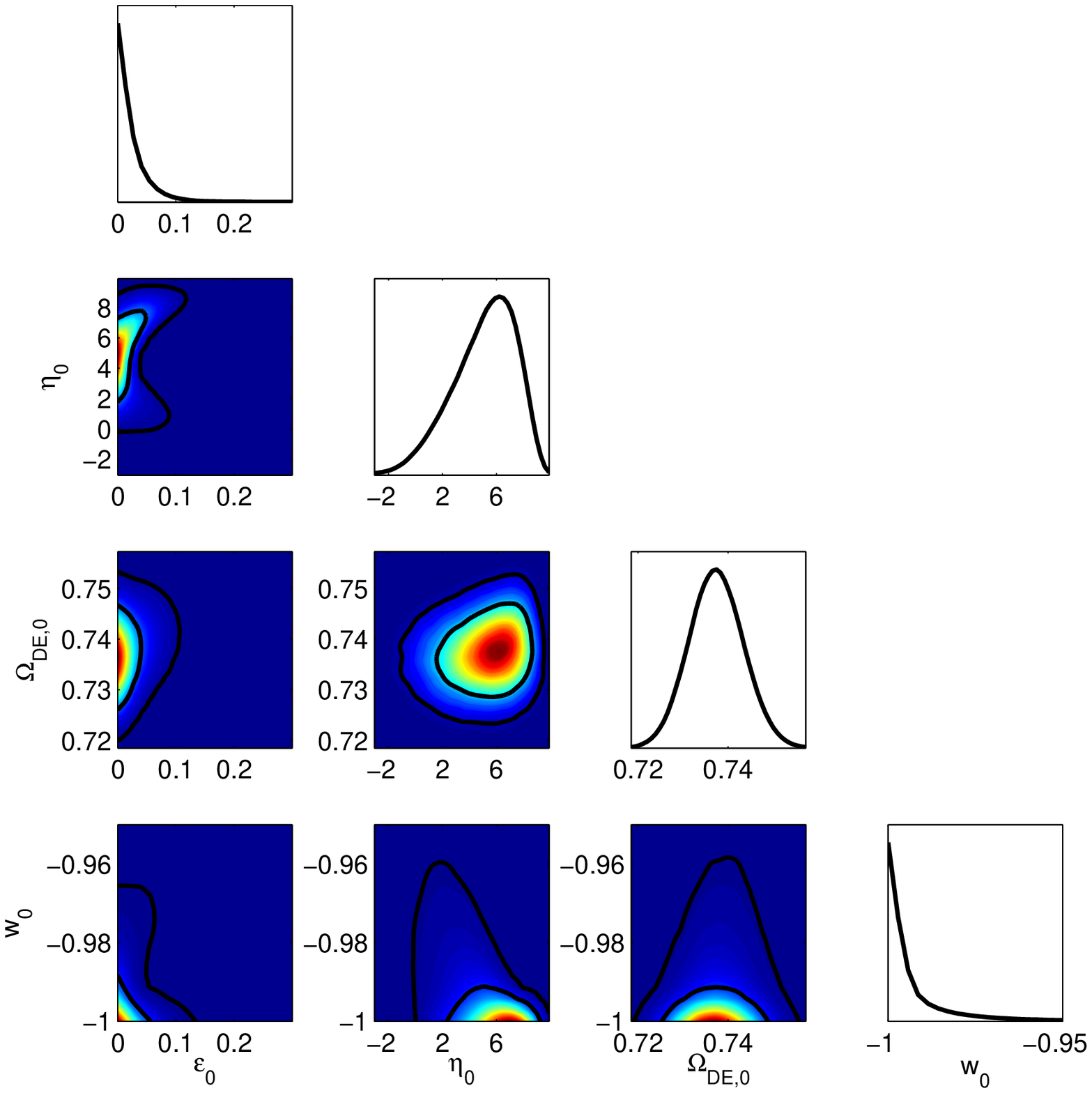,width=3.5in}\hfill
\psfig{file=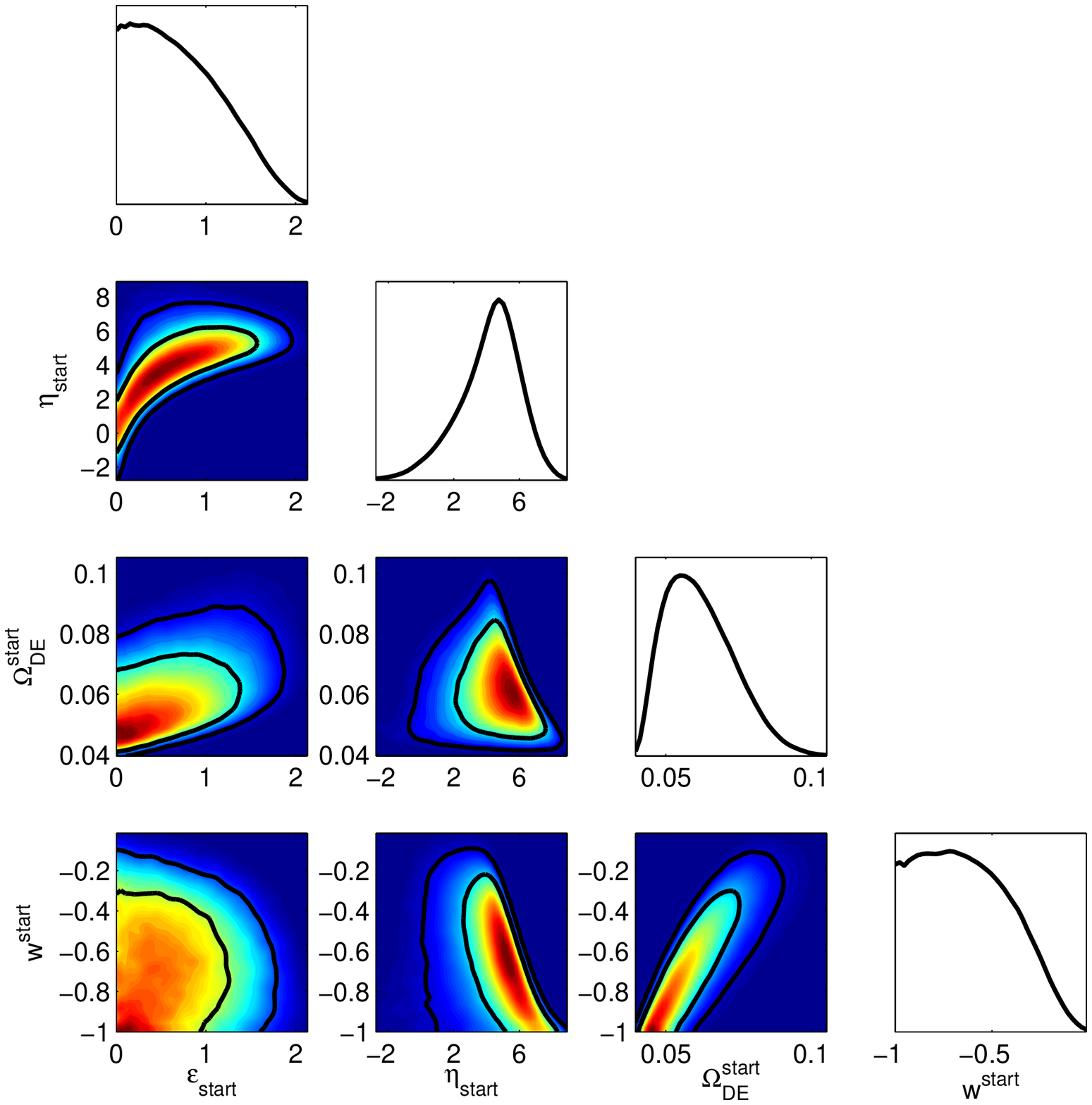,width=3.5in}
\caption{Same as Fig.~\ref{fig:like_4x4_2params}, but parameter constraints are
for future data as described in Appendix~\ref{app:future}, defined at $z=0$
(left) and $\zstart=3$ (right). It is interesting to note that constraints on
the potential parameters $\epsilon$ and $\eta$ improve to a significantly
lesser degree than those on $\ode$ and $w$ when compared with the constraints
from the current data.}
\label{fig:like_4x4_2params_future}
\end{figure*}

\begin{eqnarray}
w_0&\equiv &{\alpha_2\beta_1-\alpha_1\beta_2 \over \beta_1-\beta_2}
\label{eq:w0_from_alpha}
\\[0.1cm]
w_a&\equiv &{\alpha_1-\alpha_2               \over \beta_1-\beta_2}
\label{eq:wa_from_alpha}
\end{eqnarray}

\noindent where

\begin{equation}
\beta_i\equiv \int e_i(z)\, {z\over 1+z}\, dz. 
\label{eq:beta_i}
\end{equation}

\noindent Equations (\ref{eq:w0_from_alpha}) and (\ref{eq:wa_from_alpha}) are
now our definitions of the parameters $w_0$ and $w_a$, given a dark energy
history $w(z)$ which determines the $\alpha_i$.  We have explicitly checked on
individual examples that the two-parameter equation of state closely follows
the true $w(z)$ over the redshift range most effectively probed by the data.
Note too that the constraint $w_0\geq -1$, which follows from $w(z)\geq -1$, is
not strictly obeyed by $w_0$ obtained in this way since $w_0$ and $w_a$ are now
essentially a {\it fit} to the dark energy equation of state history. We will
return to the efficacy of this approximation later in Sec.~\ref{sec:wz_reconstr}.

\subsection{Equation of state pivot}\label{sec:wp_define}

The pivot value of the equation of state parameter is the value of $w(z)$ at
the specific redshift where the equation of state is best constrained and has
a ``waist'' \cite{Huterer_Turner}. In other words, we 
require that $w_{\rm pivot}=w_0+Aw_a$ and $w_a$ be uncorrelated, where $A$ is a
coefficient to be determined.  In practice, having obtained the $(w_0, w_a)$
corresponding to a model for each step in the Markov chains, we can compute the
covariance matrix of these two parameters and diagonalize it to extract
$(w_{\rm pivot}, w_a)$. It is easy to show that $A=- {\rm Cov(w_0, w_a)}/{\rm
Cov(w_a, w_a)}$ and

\begin{equation}
w_{\rm pivot} = w_0 - {{\rm Cov(w_0, w_a)}\over {\rm Cov(w_a, w_a)}}\, w_a
\label{eq:wp_from_w0wa}
\end{equation}

\noindent where ${\rm Cov}(x, y)$ stands for elements of the covariance matrix element. 
Although we make no specific use of the pivot redshift, for completeness, it is given by

\begin{equation}
z_{\rm pivot} = - {{\rm Cov(w_0, w_a)}\over {\rm Cov(w_0, w_a)}+{\rm Cov(w_a, w_a)}}.
\label{eq:zp}
\end{equation}

Now we will present our constraints on these dark energy observables.

\section{Basic results}\label{sec:results}

\subsection{Fundamental parameters}\label{sec:fund_constrain}

In Fig.~\ref{fig:like_4x4_2params} we show the constraints from the current
data on $\epsilon$, $\eta$, $\ode$, and $w$ at redshift zero (left panel;
superscripts ``0'') and redshift $\zstart=3$ (right panel; superscripts
``3''). In Fig.~\ref{fig:like_4x4_2params_future} we show the equivalent constraints
for future data, assuming a \LCDM fiducial model.

The $z=3$ constraints (right panel of Fig.~\ref{fig:like_4x4_2params}) are not
particularly tight, as the only direct probe of this redshift comes from the
CMB.  Nevertheless, we see that the equation of state is limited to be
$\wstart\lesssim 0$; this constraint is directly due to the CMB location of the
first peak $\theta_A$, as models with $w(z)>0$ at high redshift are
increasingly dark energy dominated and produce an incorrect distance to the
last scattering surface.  Note that the best-fit \LCDM cosmology has
$\odestart\approx 0.05$.

In contrast, constraints at $z=0$ (left panel of
Fig.~\ref{fig:like_4x4_2params}) are excellent.  In particular the equation of
state $\wnow$ and the first slow-roll parameter $\epsnow$ are determined with
great accuracy. Our choice to show constraints at $z=0$ is somewhat arbitrary,
as we would expect similarly good constraints at $z\lesssim 0.5$ in
general. Rather than these quantities at particular time slices, it is of more
interest to show the reconstructed histories of dark energy, as well as
quantities that reflect the time-averaged behavior of dark energy (e.g.\ $w_0$,
$w_a$, principal components) as we do below.  Nevertheless, we pause to comment
that these constraints are complementary to the usual approach where the dark
energy sector is parameterized via a single, constant equation of state
parameter $w$. While neither one of our fundamental parameters in
Fig.~\ref{fig:like_4x4_2params} is constant in time, the constraints on them
are tight because of the combination of the theoretical prior and cosmological
data.  Constraints from these two different approaches, model-independent
$w={\rm const}$ and scalar field time-dependent $w(z)$, do not necessarily need
to be in agreement --- nevertheless, we do expect many salient features to be
the same, such as the preference for $\odenow\approx 0.75$ and $\wnow\approx
-1$, and this is reflected in the left panel of
Fig.~\ref{fig:like_4x4_2params}.

One interesting point to note is that the expected future constraints on the
potential parameters $\epsilon$ and $\eta$ improve to a significantly lesser
degree than those on $\ode$ and $w$ when compared with the constraints from the
current data. This suggests that our knowledge of the local shape of the
effective potential will not improve very much in the future compared to that
of the velocity and the potential energy of the field. 

In our formalism we can directly compute the field excursion from $z=3$ to
$z=0$ for each model in our chains by integrating Eq. (\ref{eq:2nd_order}), and
hence obtain constraints on it: we find $\Delta \phi / \mpl = 0.09\pm
0.03$. Thus the field evolution in the range probed best by the data is
small. Therefore it is not surprising to find that the constraints on the shape
of the potential are not strong, since such a narrow range of the potential is
probed. In addition, unlike in the case of inflation, there is no theoretical
slow-roll prior to help constrain the shape of the potential over a larger
$\phi$ range than that probed by observations.

\subsection{Principal components}\label{sec:PC_constrain}

\begin{figure*}[!th]   
\psfig{file=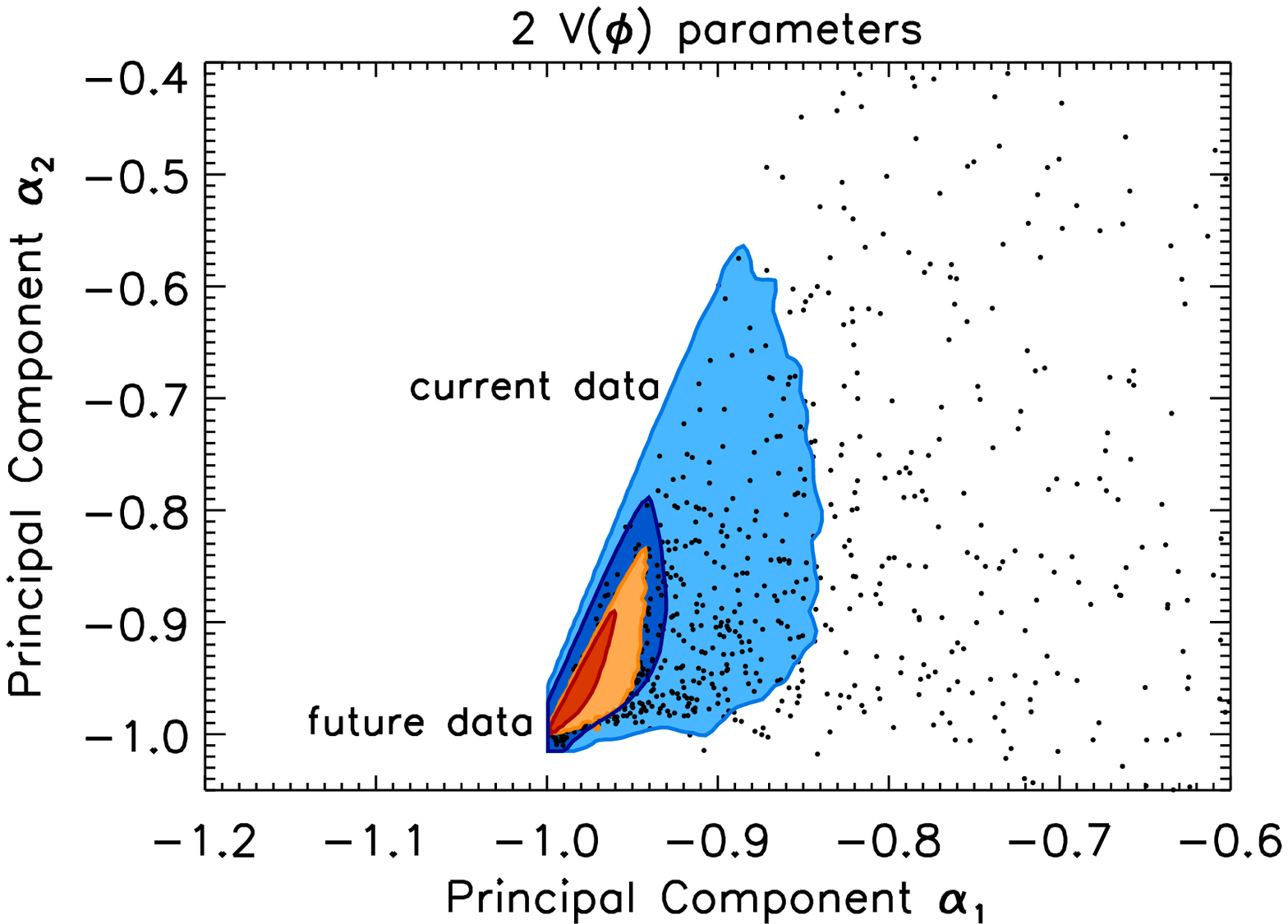,width=3.5in}  \hspace{-0.2cm}
\psfig{file=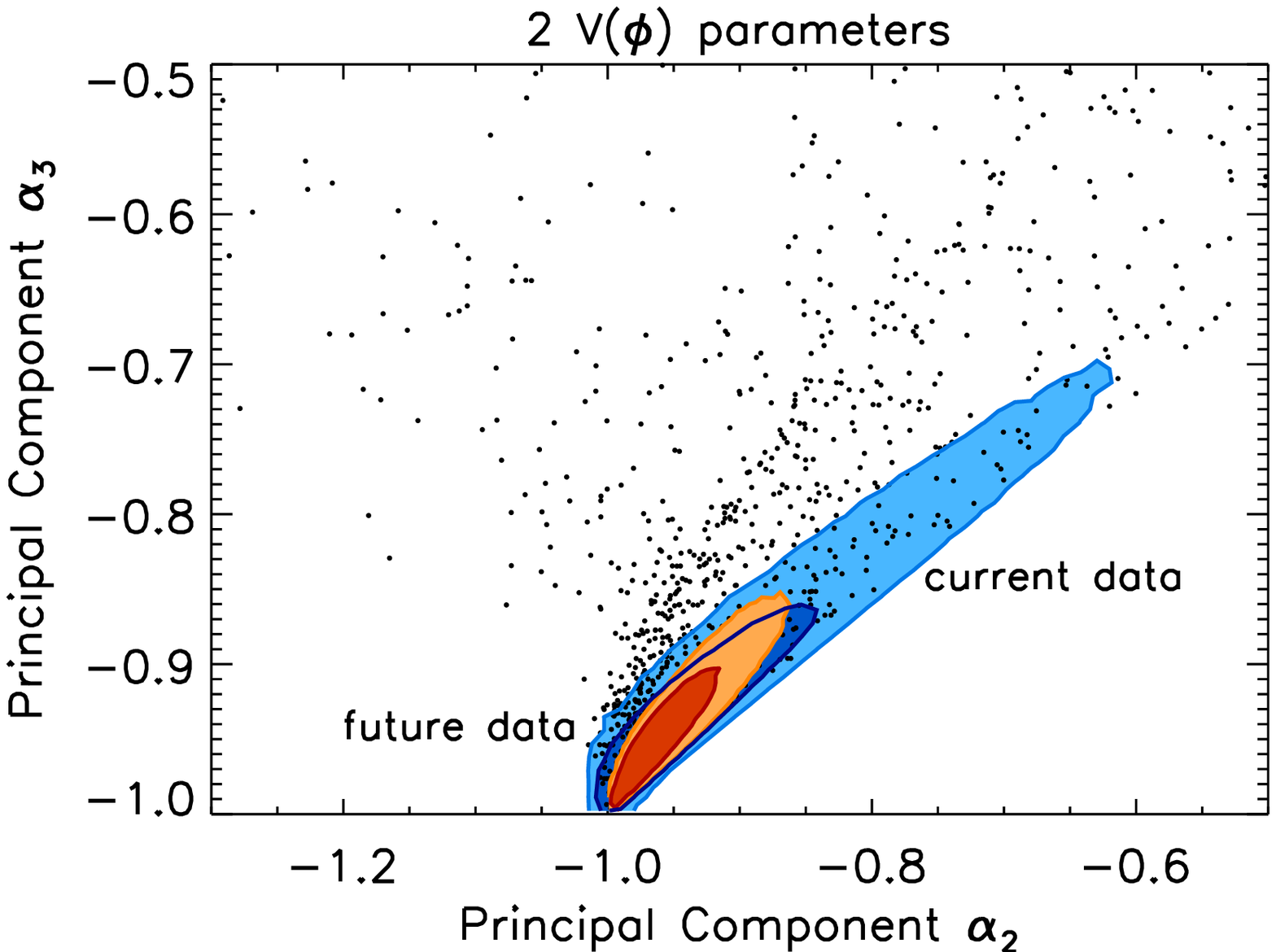,width=3.5in}  
\caption{Left panel: 68\% and 95\% CL constraints from current data
(larger/blue contours) and future data (smaller/orange contours) on the
principal components $\alpha_1$ and $\alpha_2$.  Right panel: same constraints,
but on the principal components $\alpha_2$ and $\alpha_3$.  The future data
have been centered on the \LCDM model ($\alpha_i=-1$ for all $i$), while the
current data reflect the actual constraints.  The points show $\sim 1000$ DE
models simulated randomly from the priors, that passed our basic requirements
set forth at the beginning of Sec.~\ref{sec:methodology} but have not been
subjected to the data. Note that the points lying within the region allowed by
the future data constraint have been overplotted by the constraint for
clarity --- models from the prior go all the way down to the \LCDM solution.
}
\label{fig:PC_constraints}
\end{figure*}

Figure \ref{fig:PC_constraints} shows the 2D-joint 68\% and 95\% CL constraints
in the $(\alpha_1, \alpha_2)$ (left panel) and $(\alpha_2, \alpha_3)$ (right
panel) planes. Constraints using the current data, described in Appendix
\ref{app:current}, are shown in blue/dark regions, while constraints from the
future data, described in Appendix \ref{app:future}, are in orange/light
regions.  As mentioned earlier, the future data are centered on the \LCDM model
($\alpha_i=-1$ for all $i$) by fiat. We overlay the actual constraints with
$\sim 1000$ models randomly selected from our prior without being subject to
the data constraints. Note that the future constraints are much better than the
current ones (this is further discussed in Sec.~\ref{sec:current_vs_future}).
While the density of the models itself depends on the prior, it is clear that
both the current and the future data exclude a significant fraction of
them. Finally, the marginalized constraints for the individual principal
component parameters are shown in Table \ref{tab:constraints}.

The constraints on the principal components exhibit features that are
interesting to comment on.  Principal components of general dark energy models
are by definition uncorrelated, and this is clearly not the case from
Fig.~\ref{fig:PC_constraints}; the reason is, as explained in
Sec.~\ref{sec:PC_define}, that we have used the general PCs and constrained
them assuming scalar field models. The principal components are therefore
correlated, but this does not change any of the scientific results in the
paper. 

Moreover, one can note four well-defined edges in the left panel of Fig.
 \ref{fig:PC_constraints} for either dataset. The condition $\alpha_1\geq -1$ is always
strictly obeyed since the first principal component is a non-negative
weight over the equation of state and $w(z)\geq -1$ for scalar fields; see
Fig.~\ref{fig:PC_splined}.  For the same reason the condition $\alpha_2\geq -1$
is obeyed, though not strictly --- the second PC has a both positive and
negative weight and, for a small fraction of our $w(z)$ models, this leads to
$\alpha_2<-1$ (a few of the dotted examples in the left panel of
Fig.~\ref{fig:PC_constraints}, around the value $\alpha_1\approx 0.7$, exhibit
this behavior). The long diagonal edge seen in this panel is also due to the
requirement $w(z)\geq -1$ and its precise slope and intercept are determined by
the shapes of the first two principal components. Finally, the fuzzy edge with
roughly $\alpha_1\lesssim -0.85$ (for the current data) is due to the cosmological
data that effectively restrict the weighted average of the equation of state to be
less than this value.  This last limit is more stringent for the future data, and
dotted models from the prior, which have not been subjected to the data, are
obviously not restricted in this direction.

\subsection{Parameters $w_0$, $w_a$ and $w_{\rm pivot}$}\label{sec:w0wa_constrain}  

\begin{figure*}[!th]
\psfig{file=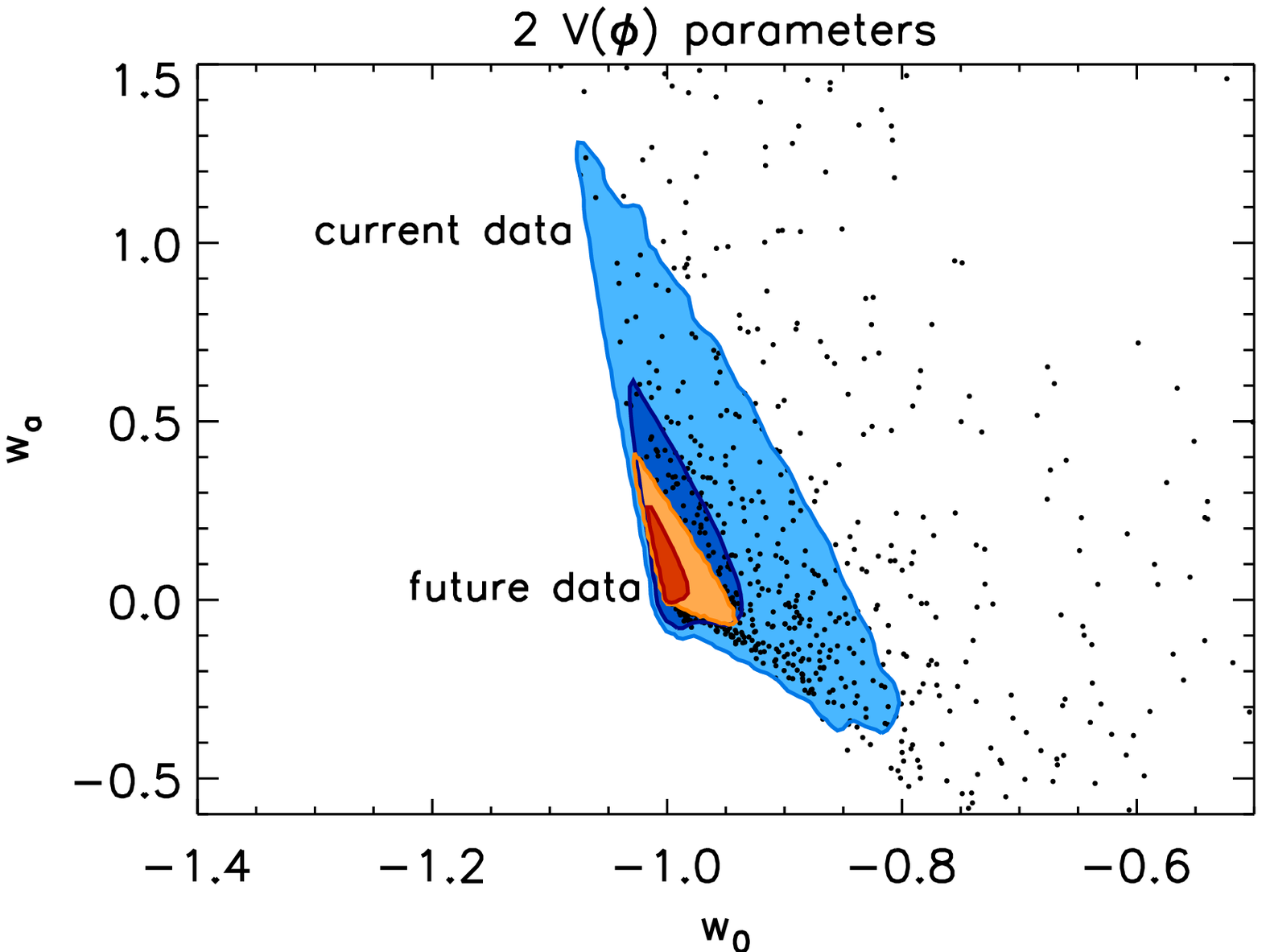,width=3.5in} \hspace{-0.2cm}
\psfig{file=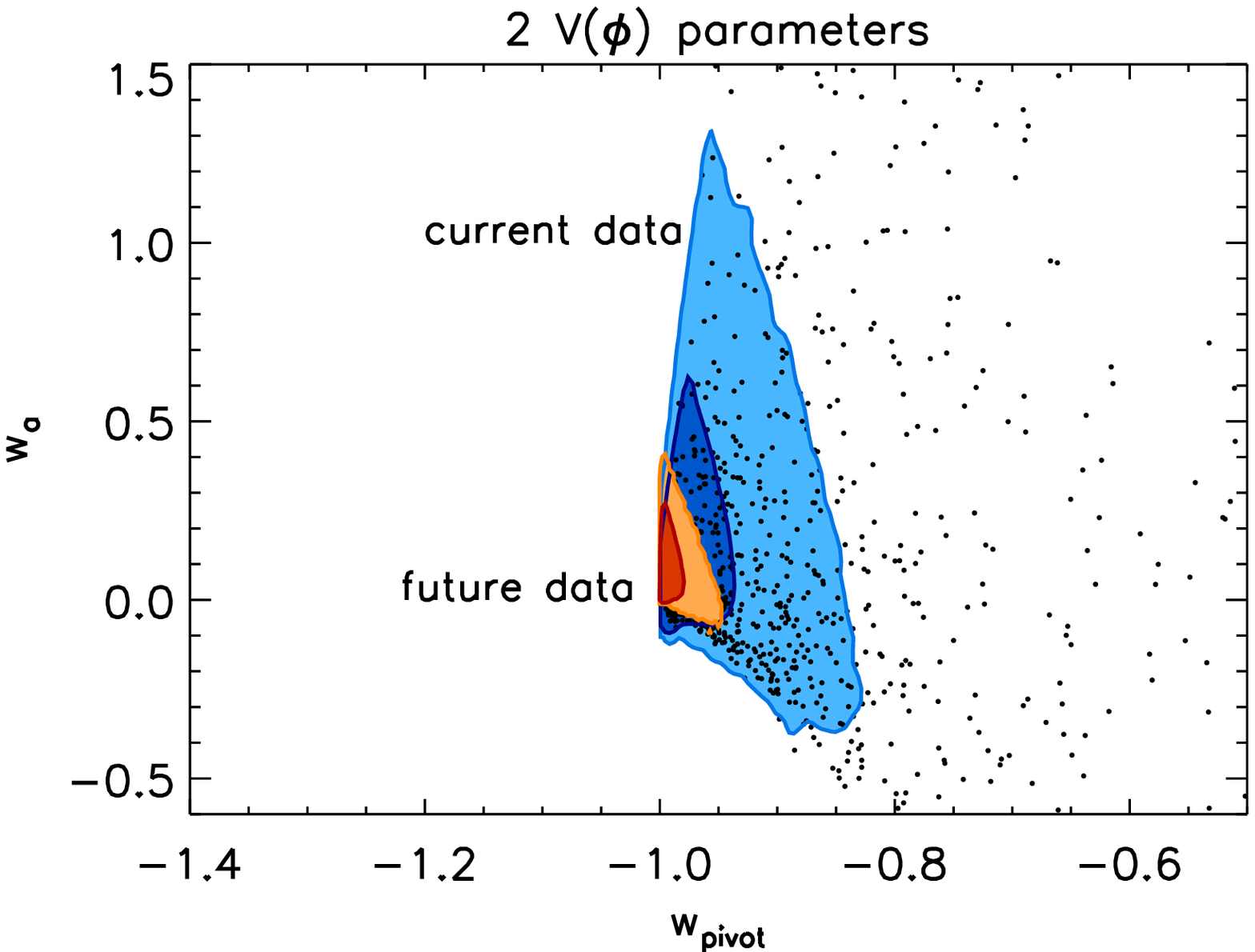,width=3.5in}
\caption{Left panel: 68\% and 95\% CL constraints from current data
(larger/blue contours) and future data (smaller/orange contours) on the equation
of state parameters $w_0$ and $w_a$. Right panel: same, but the parameters are
now $w_{\rm pivot}$ and $w_a$, where the pivot $w_{\rm pivot}$ is the
value of $w(z)$ at the specific redshift where $w_0$ and $w_a$ are decorrelated.  As in
the previous two figures, the future data have been centered on the \LCDM model
(coordinates $(-1, 0)$ in both panels); the current data reflect the actual
constraints, and the points show $\sim 1000$ DE models simulated randomly from
the priors.  }
\label{fig:w0wa_constraints}
\end{figure*}

Constraints in the $w_0$-$w_a$ plane are shown in the left panel of
Fig.~\ref{fig:w0wa_constraints}. As in Fig.~\ref{fig:PC_constraints}, we have
shown the current and future constraints (the latter assuming \LCDM) as well as
$\sim 1000$ individual models from the prior.  Moreover, there are two sharp
edges that all models obey due to the requirement $w(z)\geq -1$; this is
similar as with the principal components shown in
Fig.~\ref{fig:PC_constraints}.  Likewise, the third, long side of the triangle
contours are due to the cosmological data which, to a first approximation,
impose an upper limit to an average of the equation of state. For the current
data, this constraint is roughly $6w_0+w_a\lesssim -5$.

Following the simple procedure outlined in Sec.~\ref{sec:wp_define}, we can
obtain constraints on the pivot value of the equation of state.  Constraints in
the $(w_{\rm pivot}, w_a)$ plane are shown in the right panel of
Fig.~\ref{fig:w0wa_constraints}. These constraints are similar to those on
$(w_0, w_a)$ except that $w_{\rm pivot}$ and $w_a$ are uncorrelated by the
definition of $w_{\rm pivot}$.
As with the principal component $\alpha_1$, the relatively sharp edge at
$w_{\rm pivot}=-1$ is due to the fact that $w(z)\geq -1$ for scalar field
models.  Future constraints are about an order of magnitude better than the
current ones, as noted above.

Finally, Table \ref{tab:constraints} shows the constraints on $w_0$, $w_a$, and
$w_{\rm pivot}$. Note that $w_0$ is a two-tailed distribution while $w_{\rm
pivot}$ is a significantly skewed one-tail distribution with a hard prior
$w_{\rm pivot} \ge-1$, where the probability density abruptly falls at the
boundary. This is because while $w_0$ is allowed to go below $-1$ due to our
method for assigning ($w_0$, $w_a$) for a given model $w(z)$, decorrelating
these parameters recovers the prior on the original evolution histories,
$w(z)\geq-1$. Further, contrary to the expectation from the usual Fisher matrix
analyses, the strong non-Gaussianity of the probability distribution function
means that, while the $w_{\rm pivot}$ and $w_a$ are decorrelated, {\it it is
not true any more that we have the best constraint on the equation of state at
$z_{\rm pivot}$}. However, it is still useful to go to the uncorrelated
parameter basis $(w_{\rm pivot}, w_a)$ even without this latter benefit, as
$w_{\rm pivot}$ is a better approximation than $w_0$ to the real $w(z)$ over
the range probed by the data, in particular recovering the theoretical prior
$w_{\rm pivot}\geq -1$.

\subsection{Equation of state reconstruction}\label{sec:wz_reconstr}

\begin{figure*}[!th]
\psfig{file=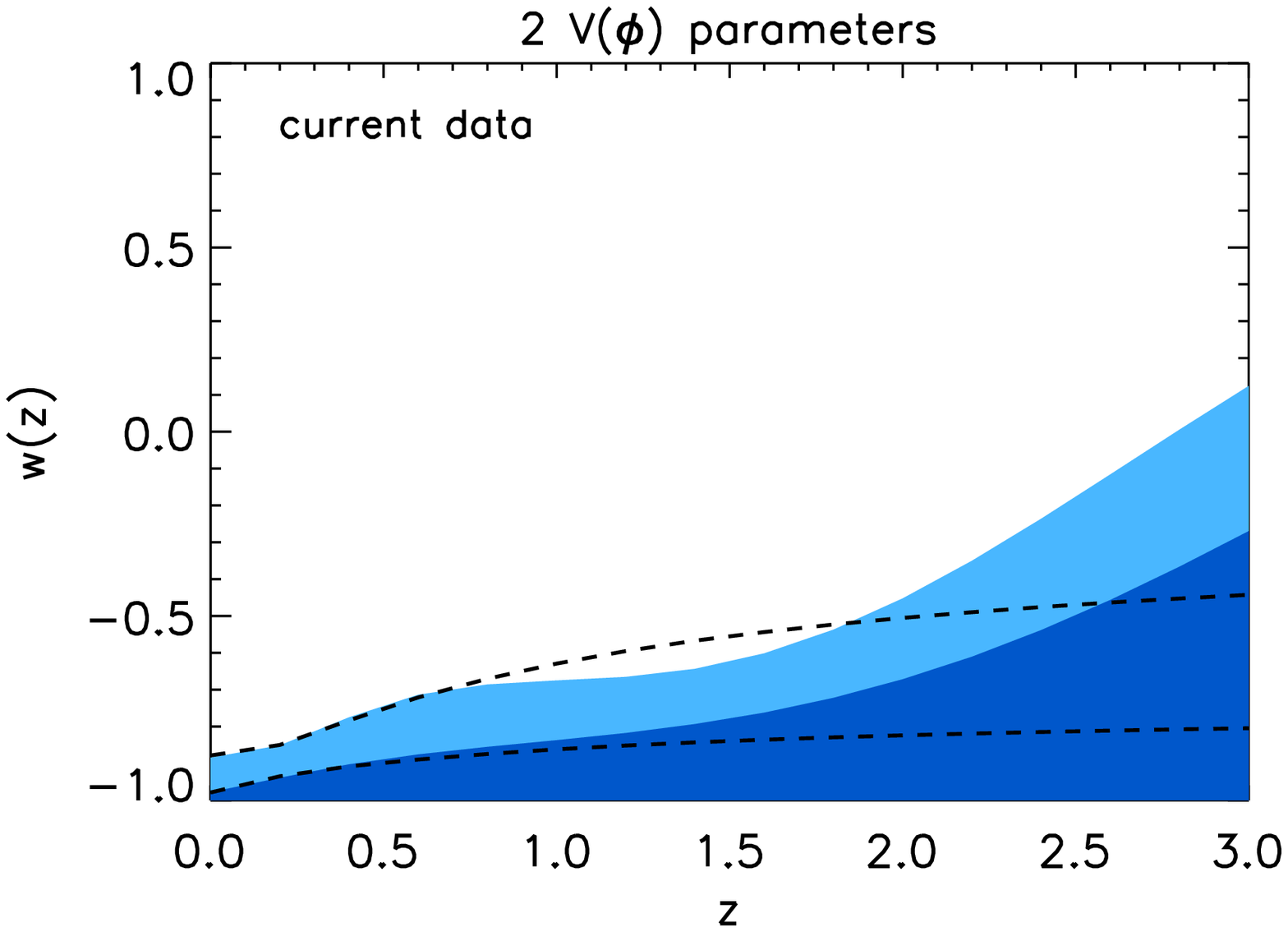,width=3.5in}  \hspace{-0.2cm}
\psfig{file=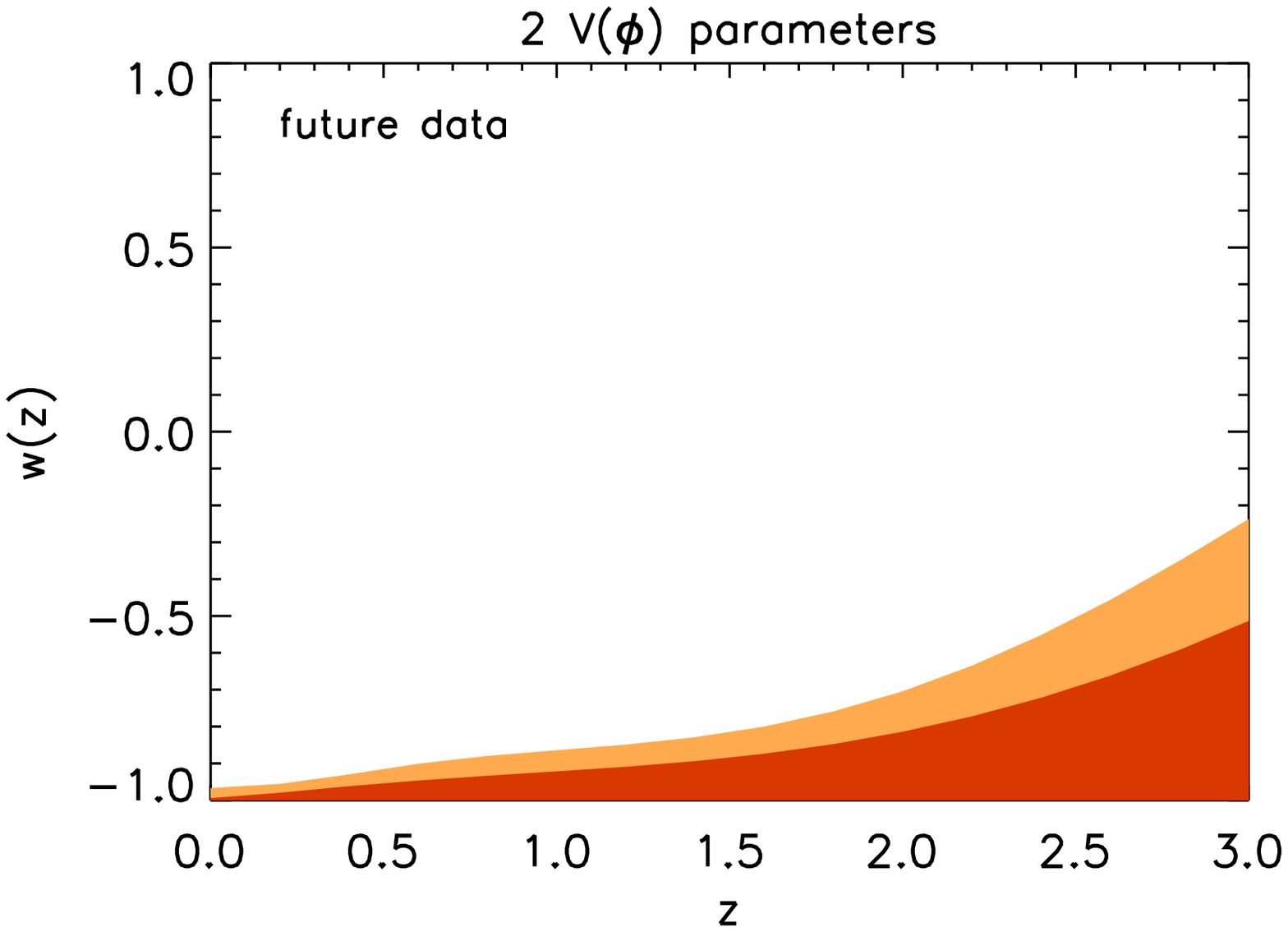,width=3.5in}
\caption{Shaded regions show the 68\% and 95\% CL constraints on the equation
of state $w(z)$ from the current data (left panel) and future data (right
panel).  The future data assume that the fiducial cosmology is precisely \LCDM
$w(z)=-1$.  In the left panel, the dashed lines show the 68\% and 95\%
reconstruction limits using the $(w_0, w_a)$ parametrization. Note that the
approximate $(w_0, w_a)$ reconstruction is in excellent agreement with the
actual constraints at $z\lesssim 0.7$. }
\label{fig:wz_reconstr}
\end{figure*}

Finally, in Fig.~\ref{fig:wz_reconstr} we show the 68\% and 95\% CL regions on
the reconstructed equation of state $w(z)$. The reconstruction is
straightforward, as the values of $w(z)$ at a number of redshifts are written
out for each of our models as derived parameters in the MCMC. The left panel
shows the constraints from the current data, while the right panel shows
reconstruction from the future data.  In the left panel, the dashed lines show
the 68\% and 95\% reconstruction limits using the $(w_0, w_a)$ parametrization.

The constraints are very good even for the current data set. They impose a
sharp upper limit on the equation of state at low redshift, and progressively
get weaker at higher z. Moreover, $w(z)\lesssim 0$ is required at any redshift
since otherwise, dark energy would be increasing in importance relative to
matter at early times and would spoil the distance to the last scattering
surface (as well as structure formation). Finally, note that the constraints
using the $(w_0, w_a)$ parametrization are in excellent agreement with the true
constraints out to $z\approx 0.7$ --- therefore, our definition of $w_0$ and
$w_a$, Eqs.~(\ref{eq:w0_from_alpha}) and (\ref{eq:wa_from_alpha}), does an
excellent job in reconstructing the dark energy history out to this
redshift. In fact, we have checked that the rms difference between the true and
$(w_0, w_a)$-reconstructed distance out to $z=3$ is only about 0.8\% (for
models from the posterior distribution for the current data), though it
degrades to 3\% by $z=1089$.  As discussed in Sec.~\ref{sec:distances}, the
fitting accuracy to $z=3$ is significantly better (by about a factor of 4) than
the accuracy to which the $z=3$ distance is determined from the current data,
and therefore there is no bias in using the $(w_0, w_a)$ fit in this redshift
range; however it is dangerous to extrapolate the fit out to higher redshift.

\begin{figure*}[!th]
\psfig{file=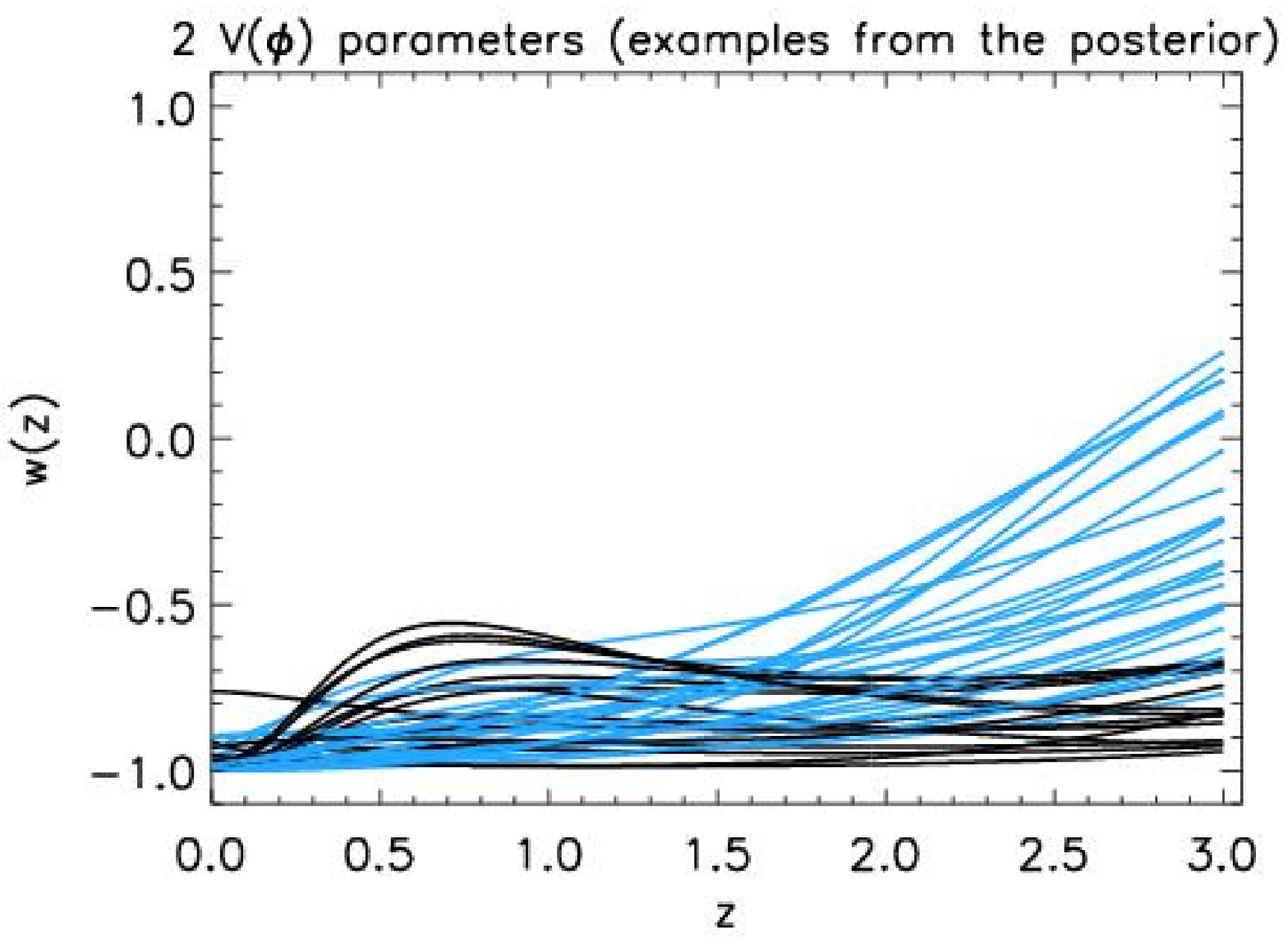,width=3.5in}\hspace{-0.2cm}
\psfig{file=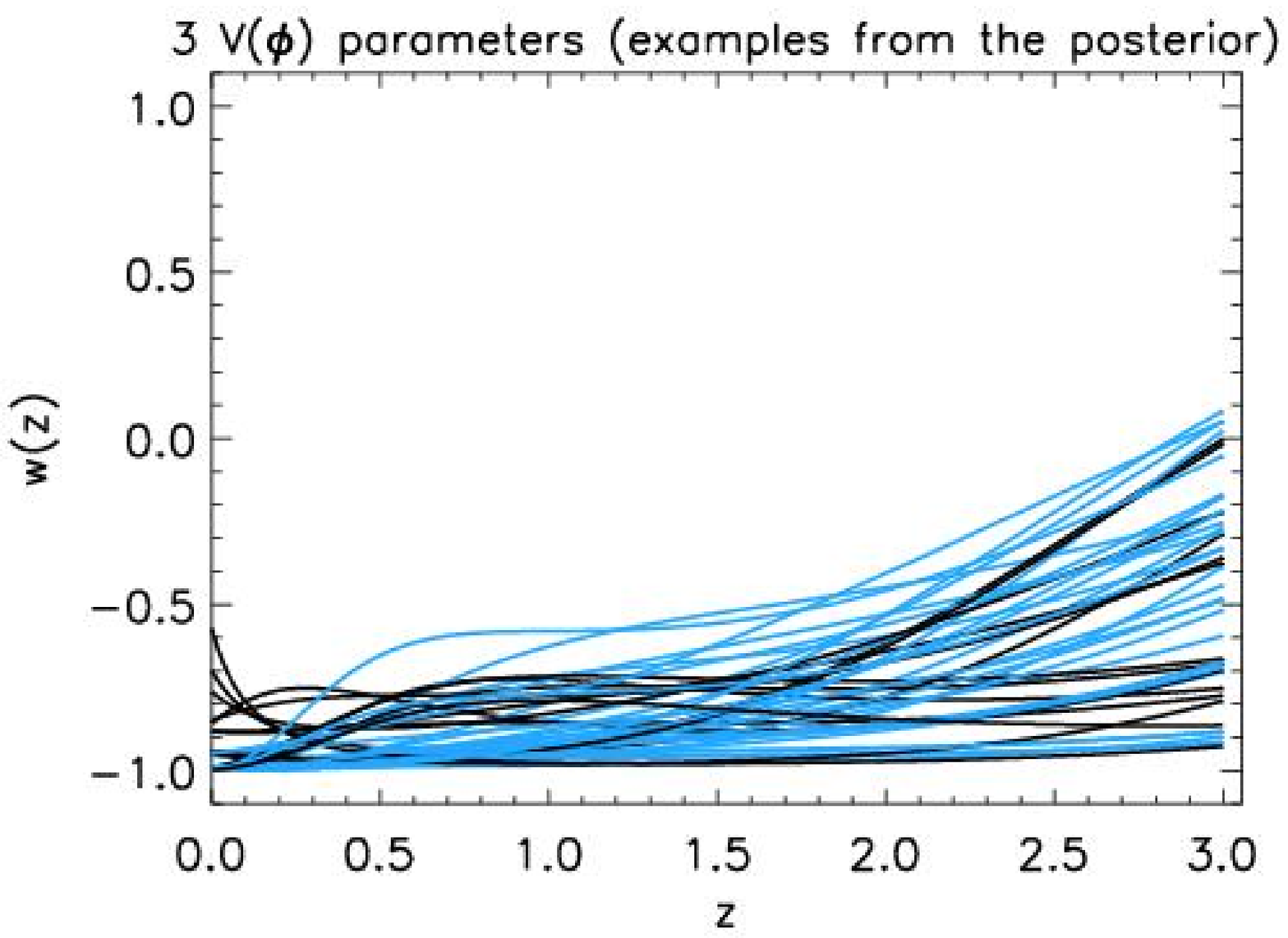,width=3.5in}
\caption{Follows the format of Fig.~\ref{fig:wz_prior}, except these figures
show the evolution histories of a number of models which were accepted steps in
the MCMC. Therefore, whereas Fig.~\ref{fig:wz_prior} shows examples of models
generated by the {\it prior}, this figure shows examples of models in the {\it
posterior}. }
\label{fig:wz_posterior}
\end{figure*}

\begin{table}[!b]
\begin{large}
  \caption{68\% C.L. constraints on the first three principal components, and
on parameters $w_0$, $w_a$, and $w_{\rm pivot}$. The second column shows the
constraints from the current data, while the third column shows errors from the
future data, assuming the underlying \LCDM cosmology (for which all parameters
are $-1$ except $w_a$ which is zero). All rows report the usual two-sided error
bars, except $w_{\rm pivot}$ where we report one-sided error bars because of
the mathematical constraint $w_{\rm pivot}\geq -1$ where the probability
density abruptly falls at the boundary. Recall that the constraints are
significantly better than in the general non-parametric case because of the 
implicit theoretical prior of using scalar fields. }
  \label{tab:constraints}
  \begin{tabular}[t]{|c|c|c|}
    \hline
\myrule      Parameter  &     Current data            &      Future data      \\\hline\hline
\myrule $\alpha_1$      & $-0.949^{+0.042}_{-0.050}$  & fiducial$^{+0.015}_{-0.014}$\\\hline
\myrule $\alpha_2$      & $-0.899^{+0.084}_{-0.081}$  & fiducial$^{+0.034}_{-0.033}$\\\hline
\myrule $\alpha_3$      & $-0.901^{+0.066}_{-0.065}$  & fiducial$^{+0.035}_{-0.031}$\\\hline\hline
\myrule $w_0$           & $-0.978^{+0.032}_{-0.031}$  & fiducial$^{+0.011}_{-0.012}$\\\hline
\myrule $w_a$           & $ 0.197^{+0.229}_{-0.199}$  & fiducial$^{+0.089}_{-0.081}$\\\hline
\myrule $w_{\rm pivot}$ & $<-0.957$ & $<-0.987$ \\\hline
  \end{tabular}
\end{large}
\end{table}

Note that this Monte Carlo reconstruction is very different in spirit from the
direct non-parametric reconstruction of the equation of state from distance
measurements
\cite{reconstr,Nakamura_Chiba,Saini_reconstr,Gerke_Efstathiou,Daly_Djorgovski,Simon,Sahni_review}
--- the former is done via models, scalar field in this case, while the latter
is completely independent of any models (although the non-parametric
reconstruction still requires parametric fits to smooth the noisy
distance-redshift curve observed by, say SNe Ia). Heuristically, the Monte Carlo
reconstruction follows the dark energy histories generated by the assumed
models, and the models correlate the equation of state values at different
redshifts. This is the reason that we obtain significantly better and more
reliable constraints than in the general non-parametric case. Conversely, the
reconstruction presented here is clearly model-dependent, and necessarily less
general than in the non-parametric case. Fig.~\ref{fig:wz_posterior} shows
explicit examples of $w(z)$ histories taken from the posterior.

\subsection{Three-parameter $V(\phi)$ constraints}\label{sec:3param}

There is no reason to limit the effective potential to be of second order: our
parametrization naturally allows potentials of arbitrary order, and we are only
limited by how many parameters can be constrained with current or future
data. We illustrate the constraints obtained with the polynomial of 3rd order
in Fig.~\ref{fig:3param_constraints}. The top left panel shows the constraints
on $\epsnow$, $\etanow$, $\odenow$, $\wnow$ and $\xinow$ (i.e., at $z=0$), while
the top right panel shows the same constraints at $\zstart=3$. The middle row shows
the constraints in the $(w_0, w_a)$ plane (left panel) and the $(w_{\rm pivot}, w_a)$
plane (right panel), while the bottom row show the reconstructed $w(z)$ for the
current data (left panel) and future data (right panel).

The three-parameter $V(\phi)$ constraints on the five fundamental parameters
are somewhat weaker than those for the two-parameter case (shown in
Fig.~\ref{fig:like_4x4_2params}). In particular, the three-parameter
constraints seem to get progressively worse as we go from $\epsnow$ to
$\etanow$ to $\xinow$. Nevertheless, the limits are still interesting,
especially those on $\odenow$ and $\wnow$.  However, constraints on the
parameters $w_0$, $w_a$, $w_{\rm pivot}$ and the principal components $a_i$ are
essentially unchanged from the two-parameter case. This means that, despite the
more complicated potential that is now harder to constrain, the range of dark
energy histories constrained by the data is largely unchanged. This could also
be seen by comparing the range of models present in the posterior distribution;
see the left three panels in Fig.~\ref{fig:wz_prior}. Thus, we conclude that
our constraints are only weakly sensitive to the truncation order of the
polynomial series.

\subsection{(Lack of) sensitivity to initial conditions}\label{sec:lack_of_sens}

\begin{figure}[!th]
\psfig{file=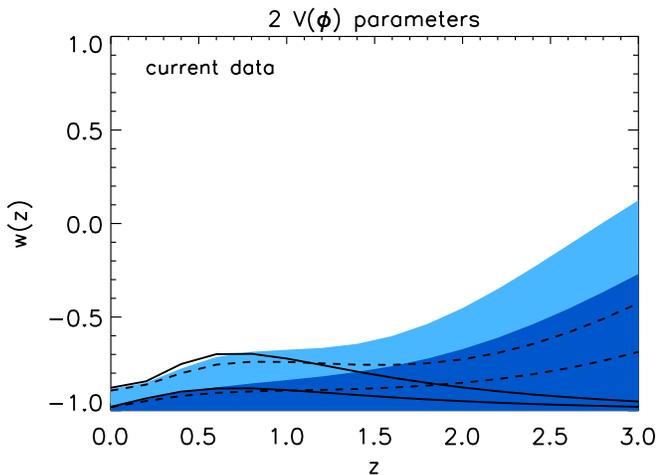,width=3.5in}  
\caption{Shaded regions show the 68\% and 95\% CL constraints on the equation
of state $w(z)$ from the current data, and are identical to those in the left
panel of Fig.~\ref{fig:wz_reconstr}. Dashed lines show the reconstruction
assuming the starting redshift $\zstart=5$ rather than $\zstart=3$ (the epochs
$3\leq z\leq 5$ are not shown for clarity). Solid lines show the reconstruction
also with $\zstart=5$, but now forcing the scalar field to start at rest (so
that $\wstart=-1$).  Clearly, the constraints at $z\lesssim 0.7$ are
insensitive to changing the initial conditions.  }
\label{fig:wz_rec_sensitivity}
\end{figure}

\begin{figure*}[]
\psfig{file=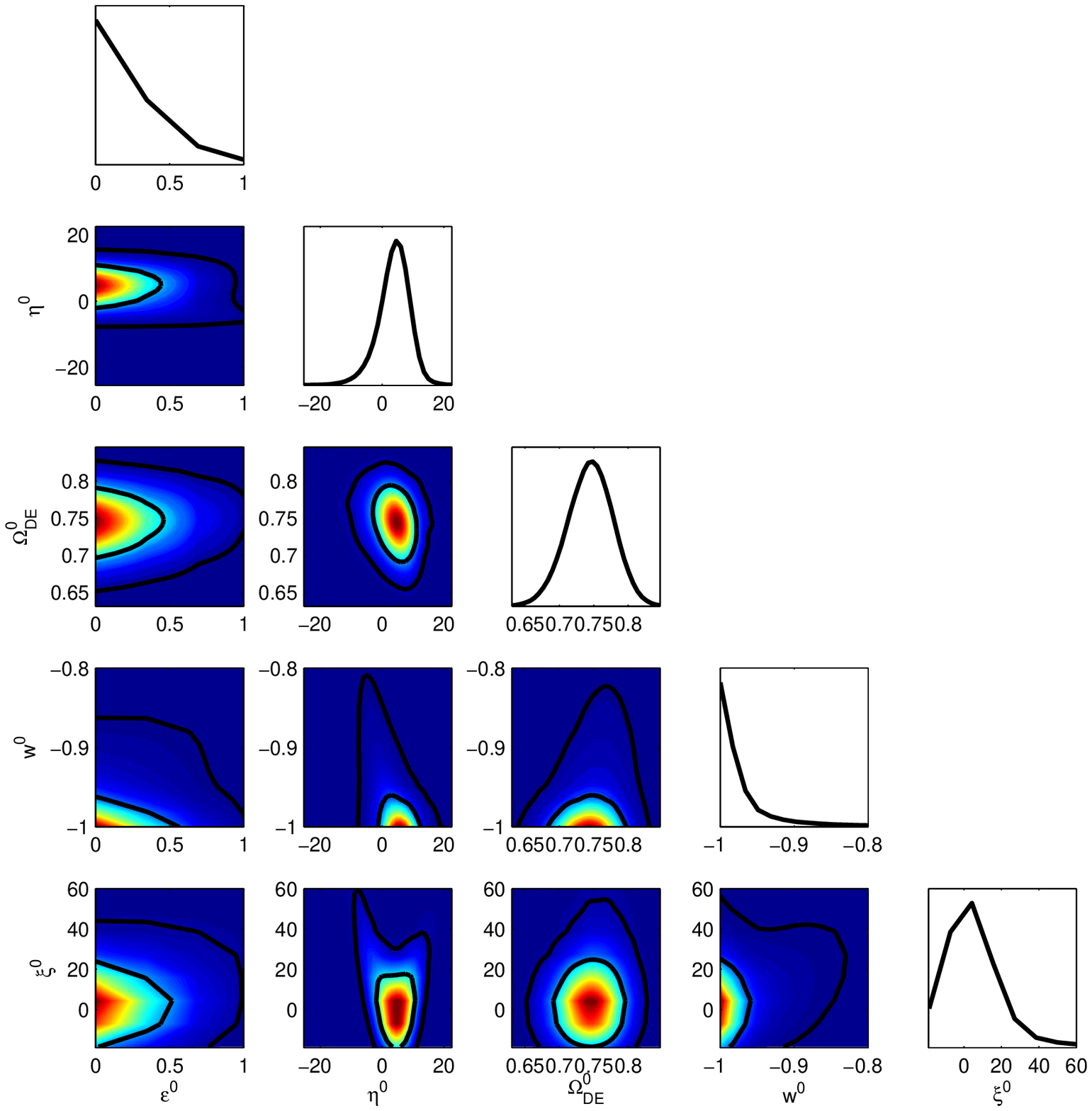,width=3.5in}\hfill
\psfig{file=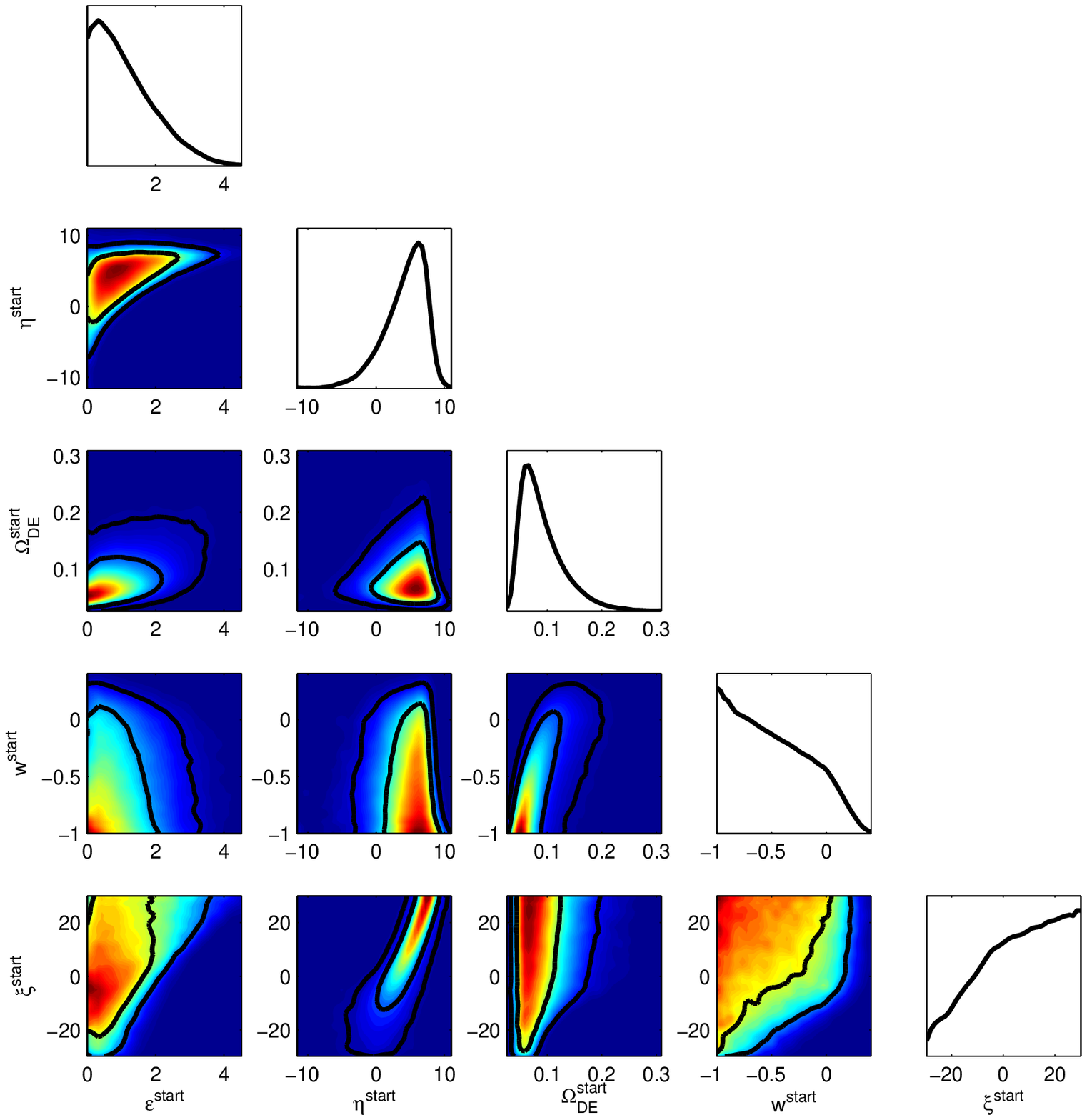,width=3.5in}\\[0.6cm]
\psfig{file=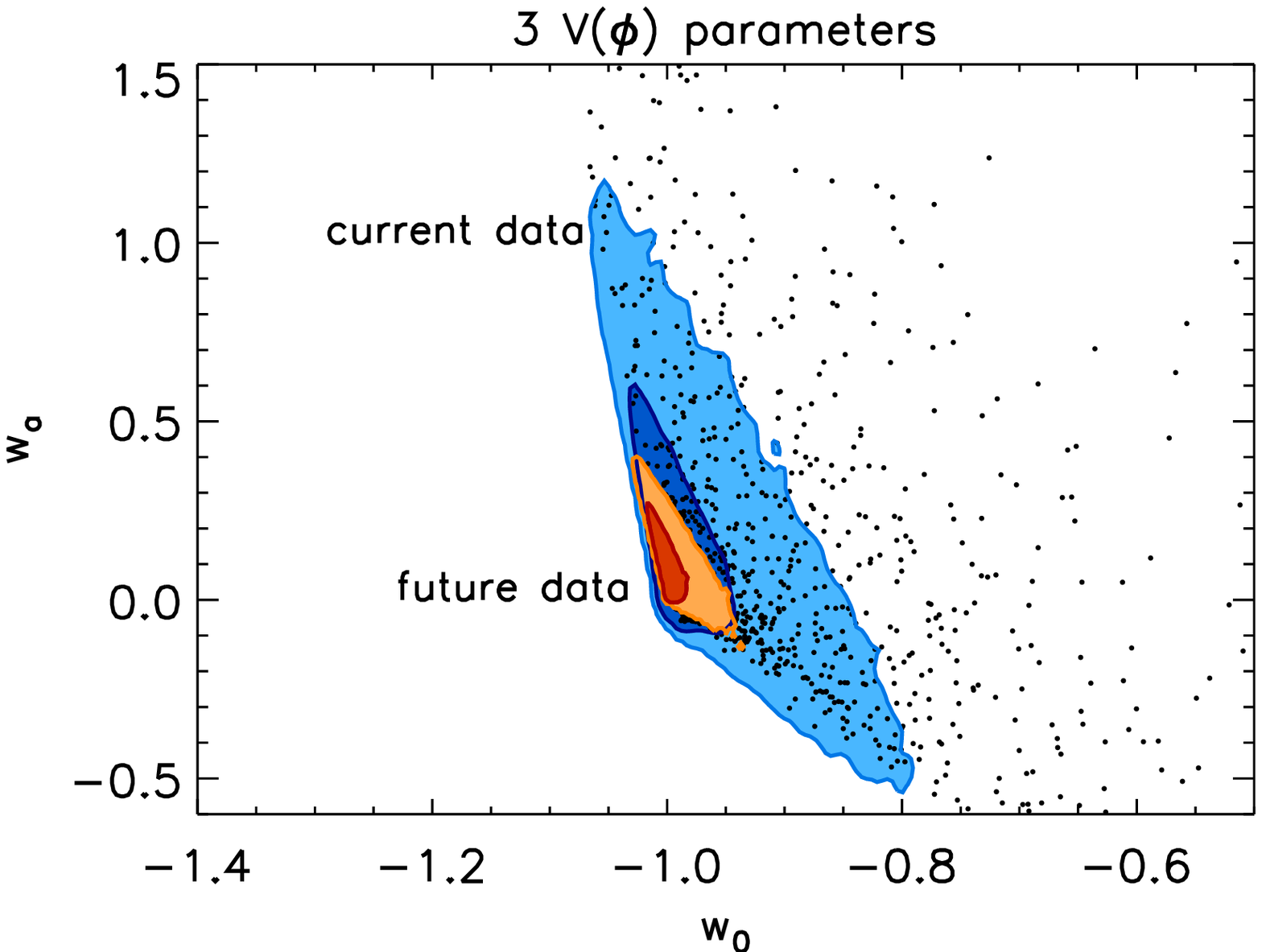,width=3.5in} \hspace{-0.2cm}
\psfig{file=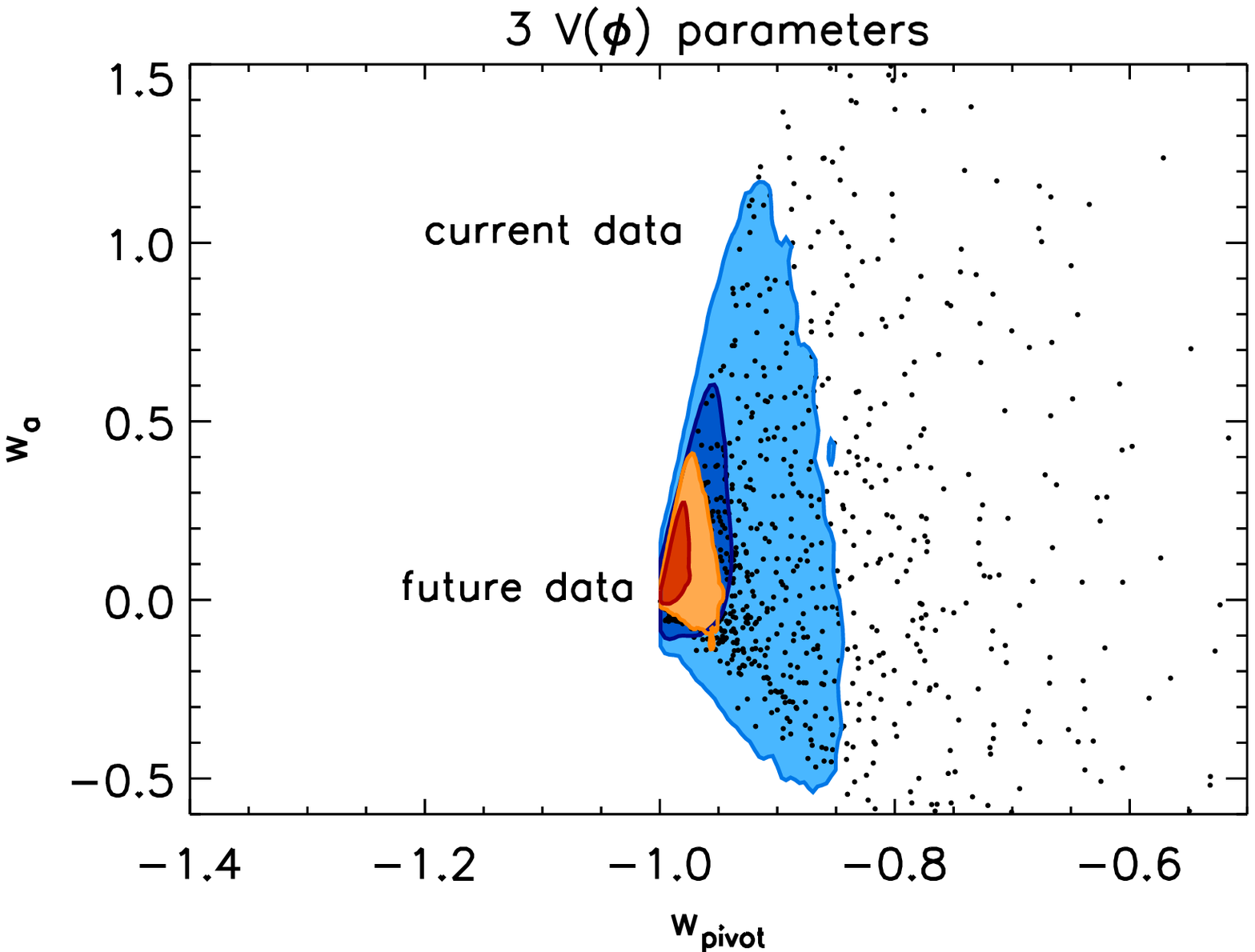,width=3.5in}\\[0.2cm]
\psfig{file=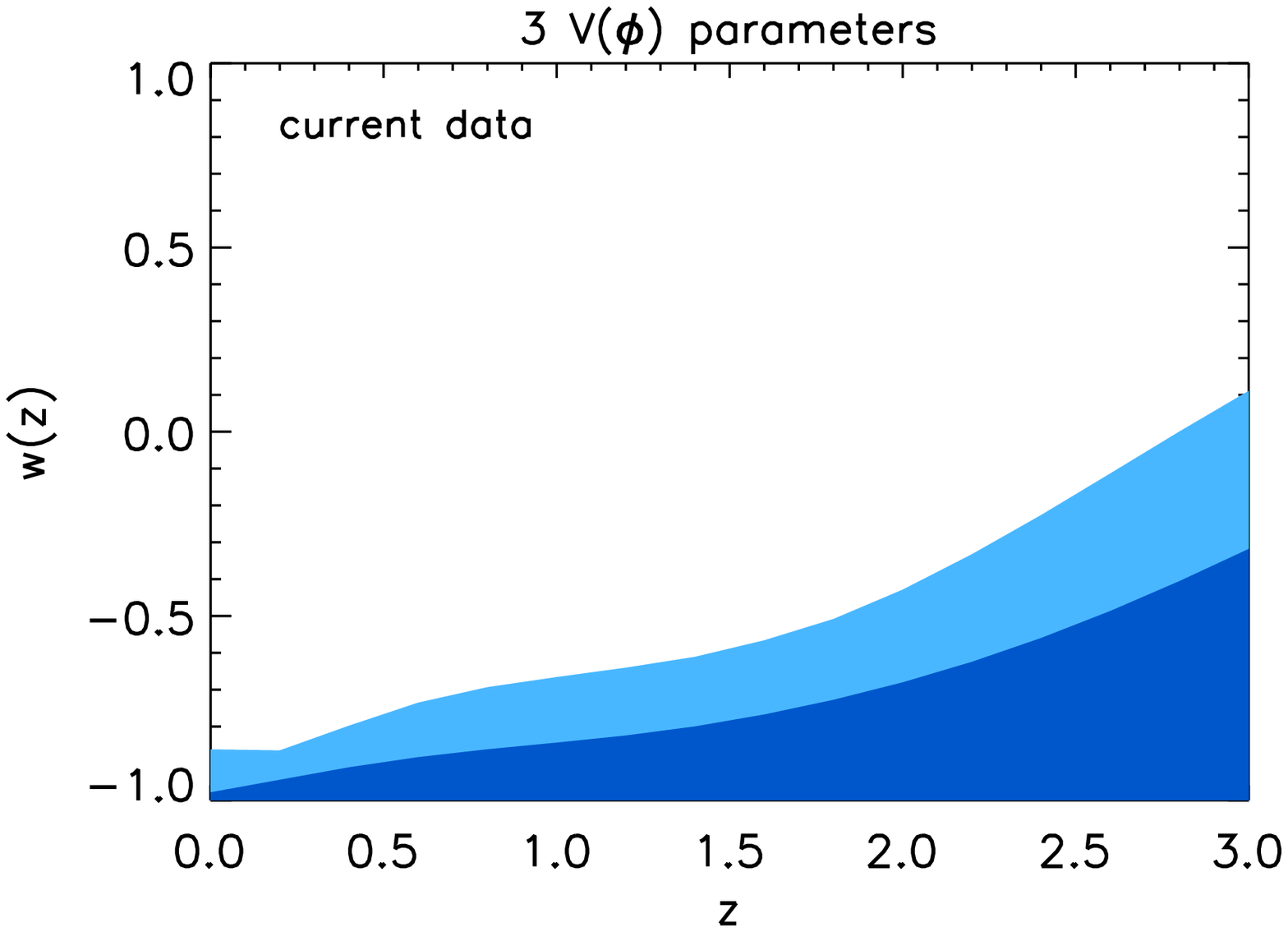,width=3.5in}  \hspace{-0.2cm}
\psfig{file=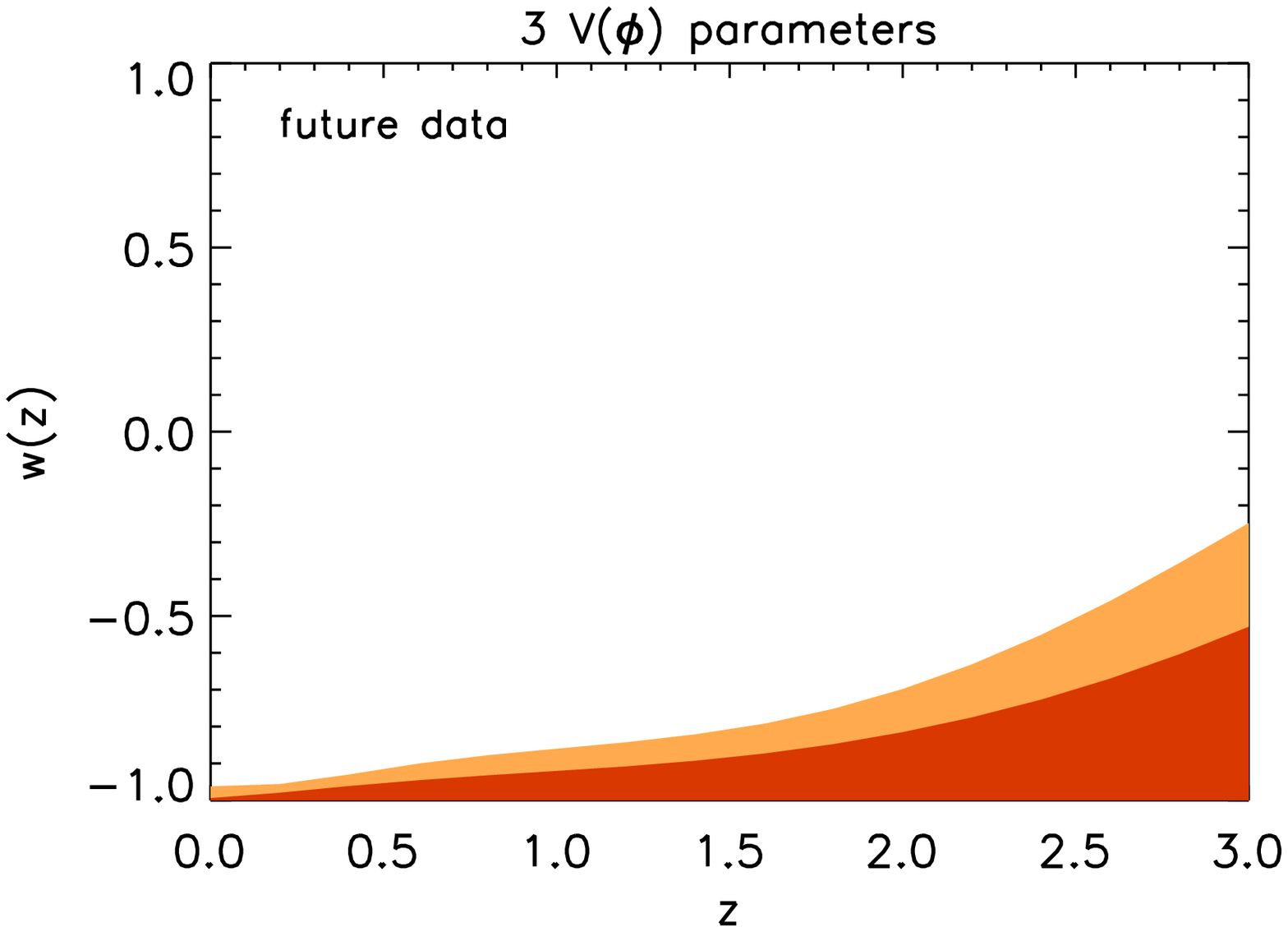,width=3.5in}
\caption{Constraints obtained assuming $V(\phi)$ is a polynomial of 3rd
order. The top left panel shows the constraints on $\epsnow$, $\etanow$,
$\odenow$, $\wnow$ and $\xinow$ (i.e., at $z=0$), while the top right panel
shows the same parameters evaluated at $\zstart=3$. The middle row shows the
constraints in the $(w_0, w_a)$ plane (left panel) and the $(w_{\rm pivot},
w_a)$ plane (right panel), while the bottom row show the reconstructed $w(z)$
for the current data (left panel) and future data (right panel).  }
\label{fig:3param_constraints}
\end{figure*}

In section \ref{sec:scalar_field_eqs} we have briefly mentioned that the low
redshift constraints are largely independent of the starting redshift of
integration $\zstart$.  We further illustrate this statement in
Fig.~\ref{fig:wz_rec_sensitivity} where the shaded regions show the
reconstruction of the equation of state $w(z)$ for the current data as in the
left panel of Fig.~\ref{fig:wz_reconstr}. However now the dashed lines show the
reconstruction assuming the starting redshift $\zstart=5$ rather than
$\zstart=3$ (the epochs $3\leq z\leq 5$ are not shown for clarity). Clearly,
the constraints at $z\lesssim 0.7$ are essentially insensitive to the starting
redshift.

Solid lines in Fig.~\ref{fig:wz_rec_sensitivity} show the reconstruction of $w(z)$ also with
$\zstart=5$, but now forcing the scalar field to start at rest (so that $\wstart=-1$).
Clearly, the field rolls slowly in the beginning as it gains speed, and the equation of state
is close to $-1$. Nevertheless, we find again that memory of this transient behavior is lost
at low redshift  ($z\lesssim 0.7$) where the constraints are unchanged relative to the fiducial
case with unconstrained $\wstart$. 

These two exercises suggest that the low redshift constraints are sensitive
only to the details of the potential and not initial conditions. However, we
have decided to leave a full study of this issue, including imposing physically
motivated constraints on the initial conditions that take into account the
field evolution at $z>\zstart$, for future work. Here we have adopted a
completely empirical approach and allowed maximally general initial conditions
of the scalar field.

\section{Cosmological Implications}\label{sec:implications}

We now present some implications of our constraints regarding the figures of
merit, the classification of the dynamics of our models, and the preference of
\LCDM over the more complicated scalar field models.  To begin with, it is
clear from all results presented so far that the \LCDM model is a very good fit
to the current data.  This result is in full accord with previous analyses
\cite{Spergel_2003,Spergel_2006, Tegmark_SDSS,Seljak_SDSS,Tegmark_LRG}.

\subsection{Current vs.\ future data}\label{sec:current_vs_future}

One of the questions we set out to answer with our approach was whether it is
worth pursuing future experiments, given that the current constraint are
increasingly converging on the \LCDM model. A view of
Fig.~\ref{fig:PC_constraints} shows that the answer is clearly affirmative. For
example, the inverse area of the 2-dimensional 68\% CL contour in the
$w_0$-$w_a$ plane, sometimes taken as the figure of merit for the power of
cosmological constraints \cite{Huterer_Turner,DETF}, is about 8 times better
for the future constraints than the current ones; it is 12 times better if we
compare the 95\% CL contour areas.  For the first two principal components, the FoM is a
similar factor of $\sim 10$ better for the future dataset. Clearly, if dark
energy is observationally distinguishable from the \LCDM model, upcoming
surveys will have a significantly greater power to reveal this. Therefore, we
eagerly expect the next generation of experiments, led by Planck, Joint Dark
Energy Mission (JDEM) and Large Survey Telescope (LST).

\subsection{Thaw or freeze?}\label{sec:thawfreeze}

There has recently been a surge of interest in classification of dark energy
models.  In particular, it has been emphasized that well known dark energy
models --- scalar field models with potentials that are power laws or
exponential functions of the field --- naturally fall in classes of ``thawing''
or ``freezing'' \cite{Caldwell_Linder,Linder_paths,Scherrer,Chiba} depending on
whether they are asymptotically receding from or approaching the state of zero
kinetic energy where the equation of state is $-1$\footnote{Note that
Ref.~\cite{Caldwell_Linder} excludes consideration of potentials with a flat
direction, $V'(\phi)=0$, on the grounds that these possess a hidden
cosmological constant; more general quintessence potentials allowing for flat
directions have been considered in the literature, e.g.\
\cite{Gonzalez-Diaz}.}.  Note that freezing dark energy models have been
studied at least as far back as \cite{Ratra_Peebles}, while thawing models date
back at least to \cite{Frieman_PNGB}.

With the benefit of our framework we can quantitatively answer questions about
the division of models.  Our parameterization is more general than that in
most previous studies as we consider all quadratic/cubic polynomials for $V(\phi)$
with maximally uninformative priors for the initial speed and energy density of
the field. On the other hand, we concentrate mainly on the low redshift
universe and do not attempt to model the effective potential during the ``dark
ages'' of the universe and earlier.  Therefore it is interesting to
compare our findings with those from previous studies.

We define a given model as ``thawing'' if $dw/d\ln a>0$ uniformly, and
``freezing'' if $dw/d\ln a<0$ uniformly, and as neither thawing nor freezing if
$dw/d\ln a$ changes sign, in the interval $\zstart=3$ to $z=0$.  From our prior
alone, with no data cut applied (as in left panels Fig.~\ref{fig:wz_prior}) we
find that about 40\% models are freezing, 0.4\% are thawing, and about 60\% are
neither. If we consider the models in our chains, i.e.\ in the posterior
distribution (as in Fig.~\ref{fig:wz_posterior}), we find that about 74\% are
freezing, a negligible fraction (0.05\%) are thawing, and 26\% fall into
neither category.  Hence the thawing fraction, small to begin with, is nearly
completely eliminated once the data have been applied.

One might consider that, since thawing models have $\wstart \simeq -1$ and
$\odestart \ll 1$ and we have uniform priors in those quantities, that the
prior is biased against generating these models. However, it is not necessarily
true that the {\it posterior} will therefore be biased against them. If the
data clearly favored thawing models over freezing models, the MCMC would
``learn'' to concentrate very close to the initial conditions that favor these
models. Instead we see that the fraction of freezing models in the posterior
increases over that of the prior, and the already-small fraction of thawing
models becomes yet smaller after the application of cosmological data, which
favor models that have $w(z)\simeq -1$ at low redshifts.

This situation is analogous to the case of Monte Carlo reconstruction applied
to inflation. In that prior, about 92\% of models are driven to the $r=0$
late-time ``attractor'' solution (using a uniform prior in $\epsilon$)
\cite{Kinney_2003}.  However, in the posterior, the models with tensor/scalar
ratio $r$ up to the upper limit allowed by the data, which is significantly
larger than zero, are explored very well \cite{Peiris_Easther, PE2}. In fact, a
class of models which has $r\simeq0$ and a very blue spectral index, which
forms a fraction of about 90\% of models generated by the prior, is completely
ruled out by the data. In this sense, as long as the data have constraining
power, the Metropolis-Hastings sampler does not necessarily reproduce biases
due to the measure on the prior which are seen in a naive Monte-Carlo process.

\begin{figure}[]
\psfig{file=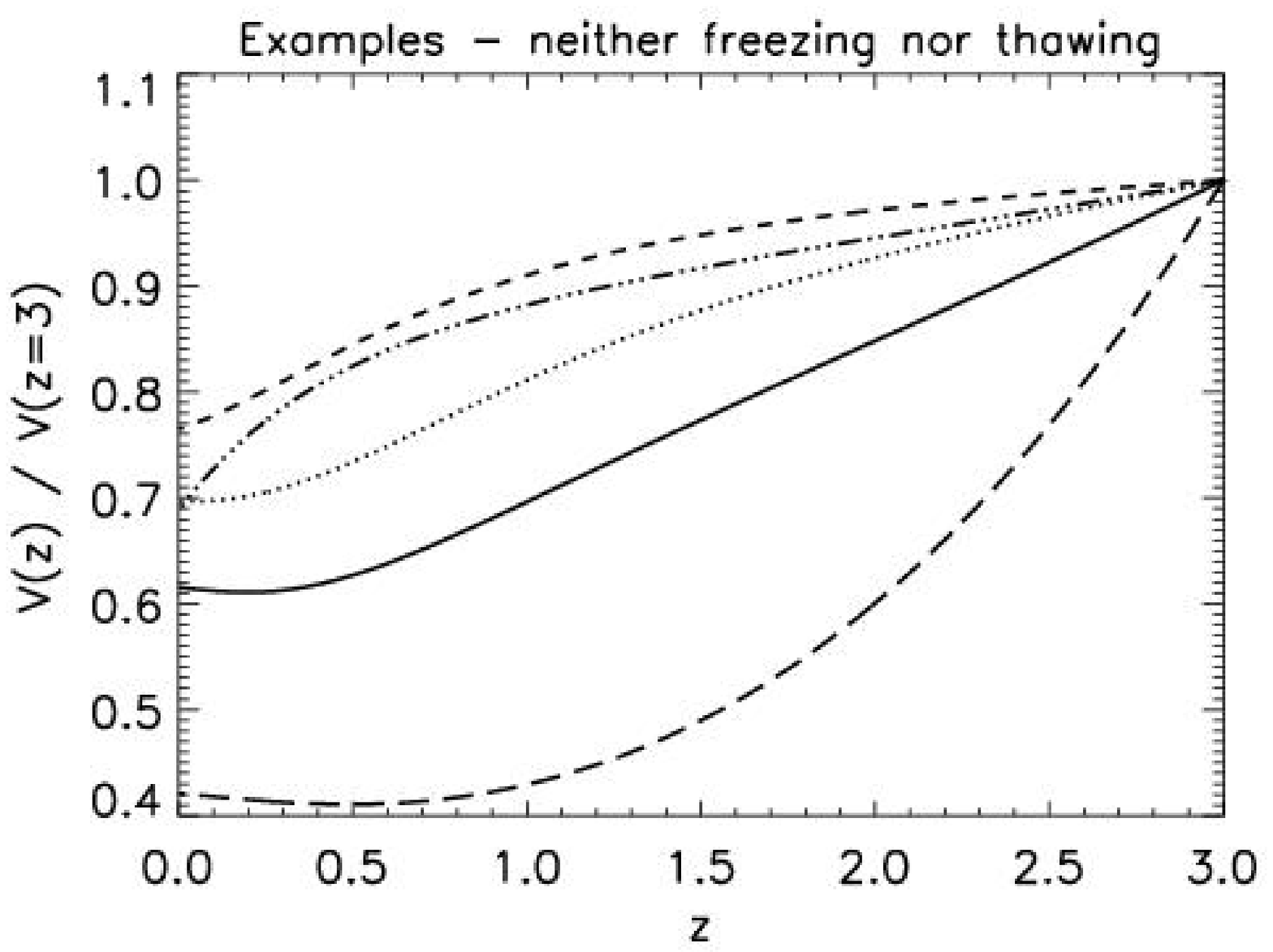,width=3.5in}\\
\psfig{file=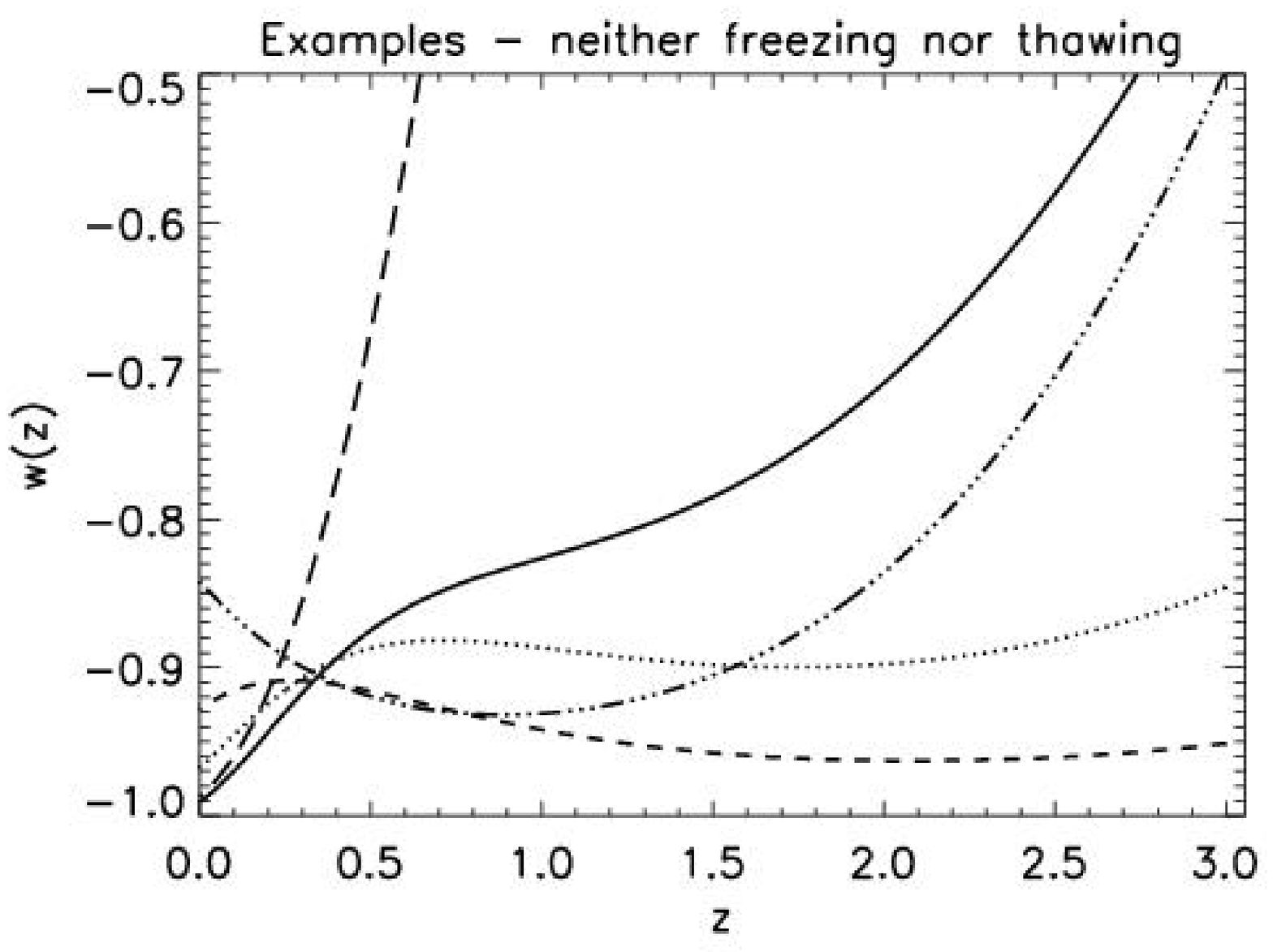,width=3.5in}\\
\psfig{file=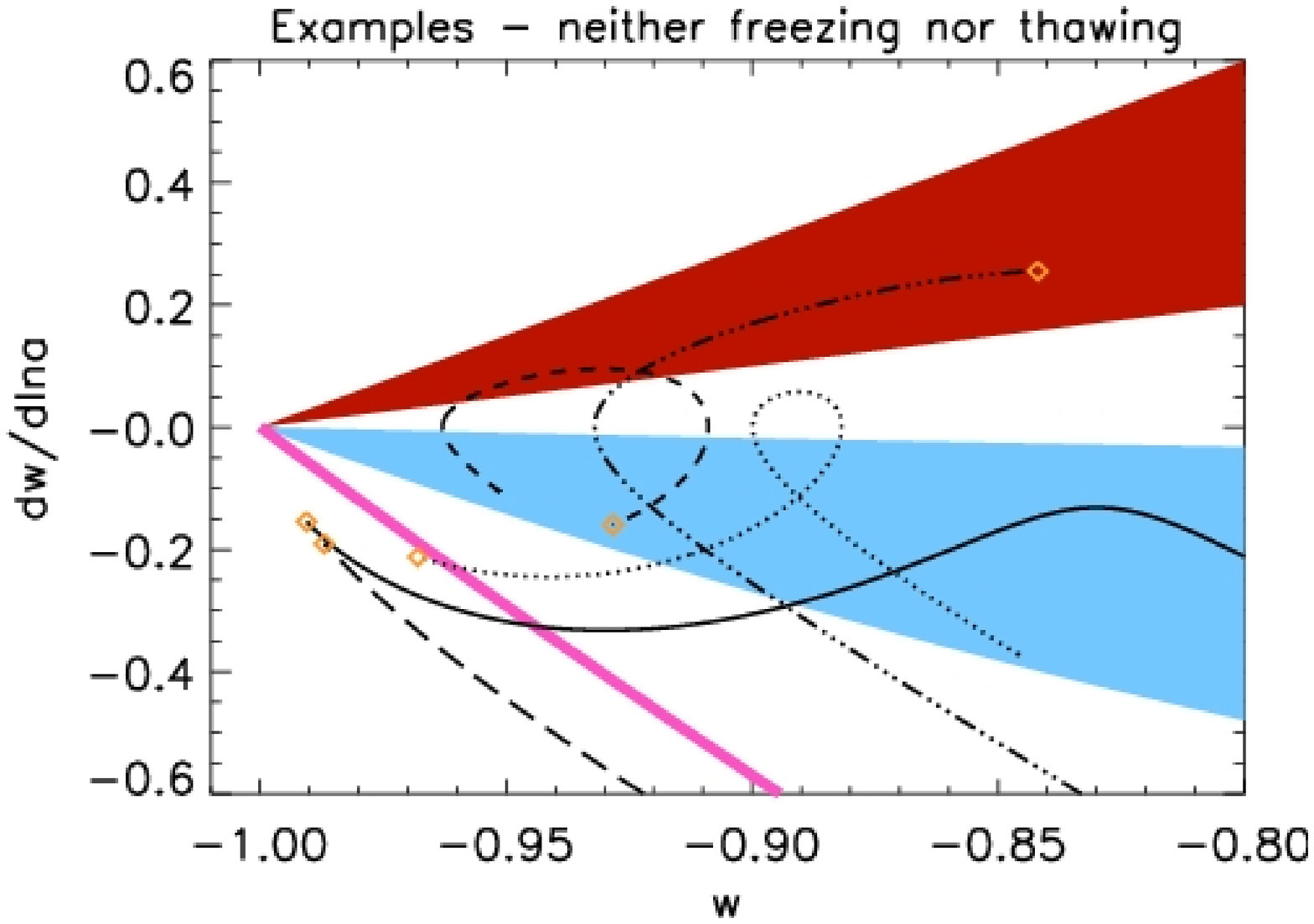,width=3.5in}
\caption{Several dark energy histories from our chains that are relevant to the
thawing/freezing discussion. Top panel: the potentials, $V(z)/V_0$.  Middle
panel: $w(z)$ histories of these models.  Bottom panel: trajectories in the
$w-dw/d\ln a$ plane, together with shaded regions that should be occupied by
thawing and freezing models according to Ref.~\cite{Caldwell_Linder}, the
orange diamonds again denoting $z=0$.  The bottom panel plot shows that our
models often do not lie solely in either region, nor do they necessarily retain
their purely thawing or freezing behavior.
Reasons for the disagreement are discussed in the text.   
The thick (pink) line in this panel shows a more general lower bound on
monotonic quintessence potentials from \cite{Scherrer} which is obeyed by all
our models satisfying this assumption.  }
\label{fig:thaw_freeze}
\end{figure}

Notice however that our thawing/freezing classification was initiated at the
time when we set initial conditions for the scalar field dynamics, $z=\zstart$,
and did not allow the field to settle past the transient and into the late-time
behavior.  If, instead, we require $dw/d\ln a$ to have a given sign only at
$z<z_*$ where $z_*<\zstart$, then the effect of the initial conditions will not
be as important. With $z_*=1$, and looking
at the posterior distribution of models, we find that the thawing fraction of
models rises to 2\% - still small, but much bigger than when $z_*=\zstart$.

\begin{figure*}[t]
\psfig{file=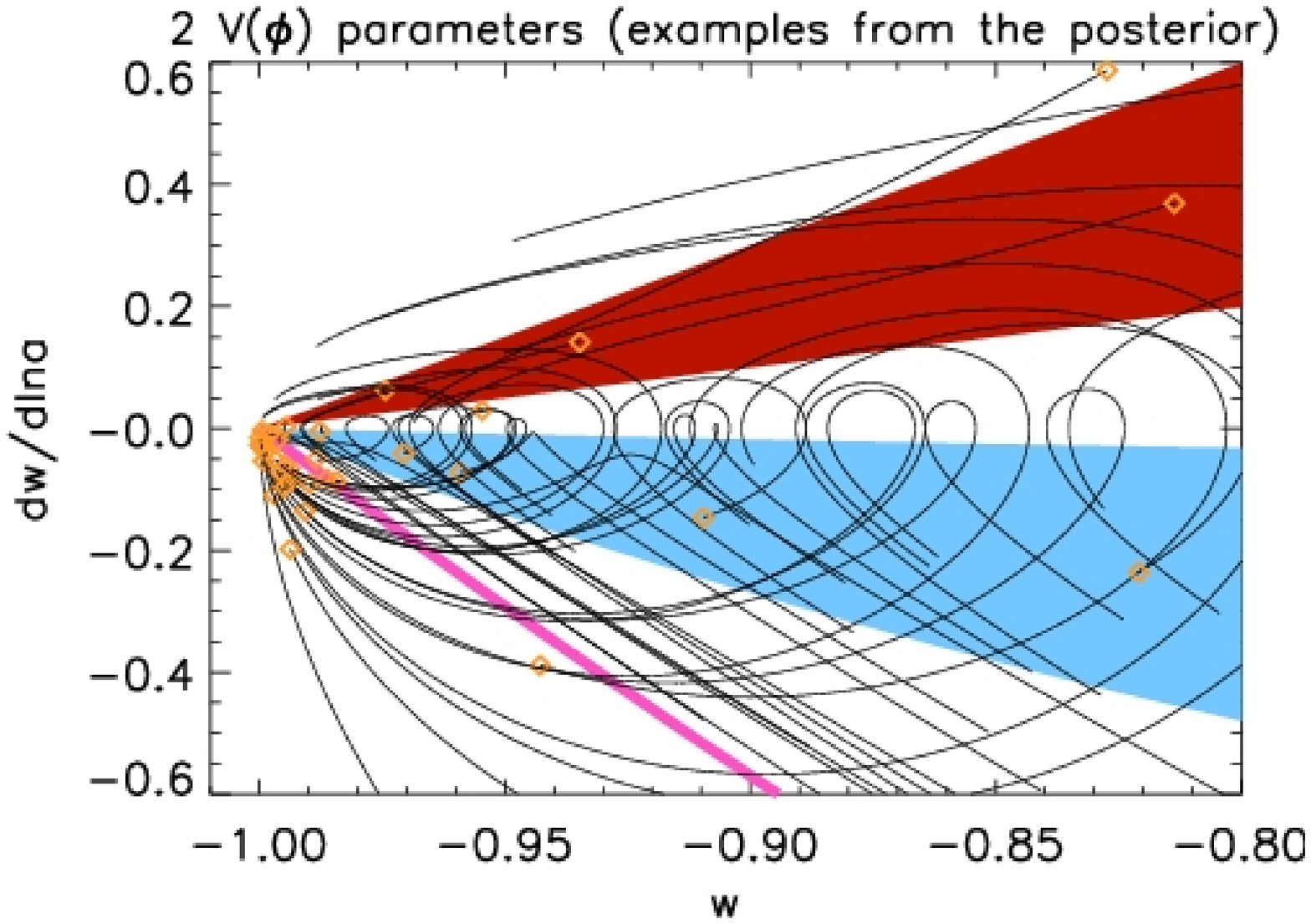 ,width=3.5in}\hfill
\psfig{file=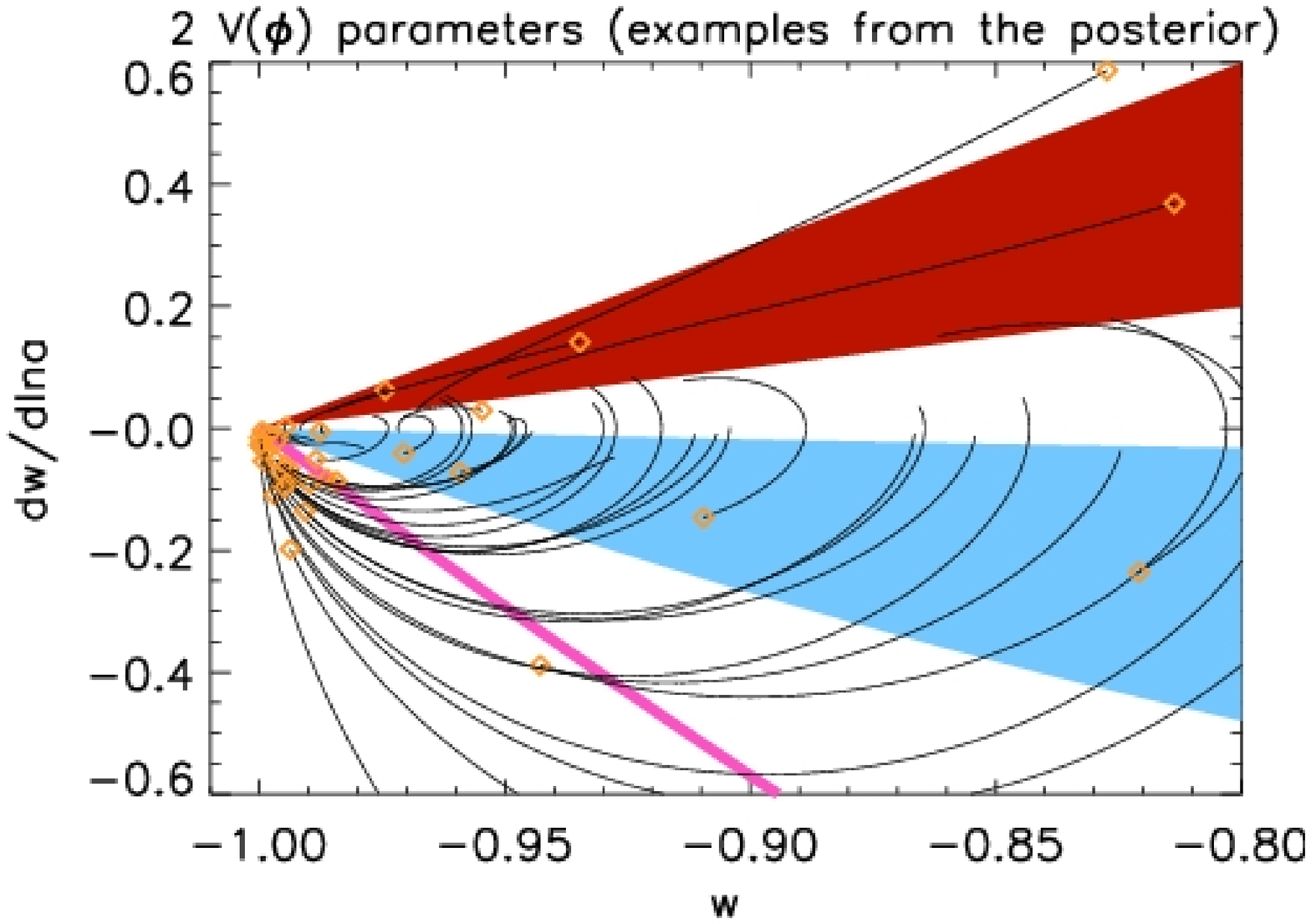,width=3.5in}\\
\caption{Typical trajectories in the $w-dw/d\ln a$ plane for a number of models
in the posterior, together with shaded regions that should be occupied by
thawing and freezing models according to Ref.~\cite{Caldwell_Linder}, the
orange diamonds again denoting $z=0$. The left panel shows the trajectories
starting at $\zstart=3$ and the right panel shows just the region between $z=1$
and $z=0$ for the same models. The thick (pink)
line in both panels shows a more general lower bound on monotonic quintessence
potentials from \cite{Scherrer} which is obeyed by all our models satisfying
this assumption. }
\label{fig:tf2}
\end{figure*}

Several examples of dark energy models from our chains that are relevant to the
thawing/freezing discussion are shown in Fig.~\ref{fig:thaw_freeze}. The top
panel shows the potential $V(z)/V_0$, the middle panel shows the $w(z)$
histories, and the bottom panel shows the trajectories of these models in the
$w-dw/d\ln a$ plane, together with shaded regions that should be occupied by
thawing and freezing models according to Ref.~\cite{Caldwell_Linder}. Clearly,
the latter plot shows that our models often do not lie solely in either region,
nor do they necessarily retain their purely thawing or freezing behavior. This
remains true even if we only restrict to very late-time evolution ($z<z_*=1$,
say). Fig.~\ref{fig:thaw_freeze} also shows a thick (pink) line, a more general
lower bound \cite{Scherrer} which is obeyed by all monotonic quintessence
potentials.  This latter bound, unlike the one from \cite{Caldwell_Linder}, is
obeyed by all our models except for those where the field samples a section of
the potential that is not monotonic.  Fig.~\ref{fig:tf2} shows the evolution on
the $w$ -- $dw/d\ln a$ plane for a number of models, with the left panel
showing the entire evolution from $\zstart=3$ to $z=0$, and the right panel
shows only the evolution for $z<z_*=1$ for the same models.  As the various
examples in these figures illustrate, the disagreement between our models and
the phenomenologically expected bounds from \cite{Caldwell_Linder} is due to
the interplay between the initial conditions of the scalar field, the effects
of Hubble friction, and the shape of the effective potential (specifically,
non-monotonicity in the second derivative) that our more general class of
models possesses. For example, the short dashed example in
Fig.~\ref{fig:thaw_freeze} shows freezing behavior early on due to the Hubble
friction; then the field velocity increases and a thawing period ensues when
$H(z)$ has fallen sufficiently; but at the last, most recent stage the
potential happens to flatten off and the field enters the freezing regime
again. Note that the more general bound from \cite{Scherrer} is again obeyed as
long as the sampled potential is monotonic.

Clearly, the thawing/freezing/neither fractions depend on the particular priors
-- the assumed form of $V(\phi)$ and the initial conditions.  Therefore it
seems that, at this stage when we do not have a good theoretical understanding
of dark energy, it is not useful to specialize into studying models that are
thawing or freezing, because their presence or lack thereof is almost certainly
not robust to fundamental assumptions. Conversely, if we ever obtain a precise
measurement with a decidedly positive or negative value of $dw/d\ln a$ (or
$w_a$, or $\alpha_2$), the thawing/freezing picture can help obtain a fuller
theoretical understanding --- a point also made by others advocating this approach
\cite{Caldwell_Linder,Linder_paths,Scherrer,Chiba}.

\subsection{How many dark energy parameters?}\label{sec:howmany}

More parameters describing the dark energy sector will lead to an improved fit
to the data, but are they {\it required} by the data? This question can be
answered using the Bayesian Information Criterion (BIC) statistic,

\begin{equation}
{\rm BIC} = -2\ln \mathcal{L}_{\rm max} + D\ln N_{\rm data},
\label{eq:BIC}
\end{equation}

\noindent where $\mathcal{L}_{\rm max}$ is the maximum likelihood in a given
class of models, $D$ is the number of parameters in a model, and $N_{\rm data}$
the number of data points used. A BIC difference of 2 or more indicates good
evidence, and 6 and more strong evidence in favor of a model with the smaller
BIC.

Here we compare the \LCDM models which have $D=3$ parameters ($\odestart$,
$\theta_A$ and $\Omega_b h^2$) and the scalar field models with two or
more additional parameters (so $D=6$ or more) describing the shape of the
effective potential ($\wstart$, $\epsstart$ and $\etastart$, plus any higher derivatives
of the potential). The \LCDM model has a slightly higher value of the best fit
$\chi^2=-2\ln \mathcal{L}_{\rm max}$, by about unity (about 108, compared to
about 107 for the two-extra-parameter scalar field models, with 119 degrees of
freedom). However the penalty from the second term in Eq.~(\ref{eq:BIC}) for
the scalar field models is large, $3\ln 119\approx 14.3$, and completely
overwhelms the gain from the improved likelihood fit. Therefore, the overall
BIC evidence (of $+13.3$) is clear: using the current data, there is no compelling
evidence whatsoever for models more complicated than \LCDM. A more comprehensive 
recent analysis using Bayesian model selection, albeit using the empirical 
variables ($w_0$, $w_a$) to describe the evolution of the dark energy, can be 
found in Ref.~\cite{Liddle2006}.

While \LCDM is clearly an excellent fit to the current data, there are two
important caveats to keep in mind. First, it is possible that models different
from ones we considered have a lower BIC than the $\sim 6$-parameter scalar
field models we considered (although it is still highly unlikely they will be
favored by the BIC criterion). Second, we have shown that the future data will
provide a much sharper test of dark energy histories, and may possibly provide
the traction to see a distinct deviation from \LCDM. Such a development would
surely provide a much needed hint about the physics of dark energy.

\subsection{Physical observables}\label{sec:distances}

Since we have obtained constraints using a variety of cosmological probes, it
is interesting to ask which physical quantities (or alternatively, which
dynamical aspects of our scalar field models) are well constrained by the
data. This is a fairly complex question that we largely leave for future work
--- however, before concluding we make an interesting observation. We found
that distances out to intermediate redshifts, $z=1$-$3$, are determined to
2-3\% for the current data. For the future data, the distances will be
determined to about 0.5\%. Therefore, any scalar field model that does not preserve the
distance to redshift $z\sim 2$ to these accuracies will be ruled out. As with
all other parameter constraints we presented, we can expect that these
accuracies to be somewhat degraded once we allow a completely general dark energy
history, not just the class of scalar field models with a smooth
potential. Nevertheless, the aforementioned numbers suggest, for example, that
a cosmological probe that has ambition to significantly improve upon the current
constraints should determine the mean distance to a (single) redshift $z\gtrsim 1$
to $\sim 1$\% or better.

\section{Conclusions}\label{sec:conclusions}

The main motivation behind this work was to study {\it all} possible dark
energy histories within a broad class of models --- chosen here to be the
scalar field models with an effective potential described by a polynomial
series. Adapting the Monte Carlo reconstruction formalism (previously applied
to inflation) to scan a wide range of dark energy models, we have generated
millions of dark energy models and constrained them with the current
compilation of data from the Supernova Legacy Survey \cite{Astier}, baryon
oscillation results from the Sloan Digital Sky Survey \cite{Eisenstein}, cosmic
microwave background constraints from the WMAP experiment \cite{Spergel_2006},
along with the measurement of the Hubble constant from the Hubble Key Project
\cite{HKP}. We have also simulated expected future data from the same
cosmological probes. Instead of attempting to study comparative merits of
specific cosmological probes or optimal survey designs, we have concentrated on
addressing a more general set of issues regarding the future quest for dark
energy.

In particular, we have addressed how the theoretical prior --- working within a
specific class of scalar field models parameterized with a polynomial series in
$V(\phi)$ --- combines with the cosmological data to produce the constraints on
cosmological parameters. We found that the theoretical prior is significant, as
only specific smooth dark energy histories are generated, and that their
qualitative nature does not vary significantly with the order of the polynomial
$V(\phi)$; see Figs.~\ref{fig:wz_prior} and \ref{fig:wz_posterior}.  Moreover,
the same theoretical prior excludes certain regions of parameter space
(especially in the equation of state parameters), and makes the posterior
distribution highly non-Gaussian.  As a consequence, the commonly used Fisher
matrix formalism applied to the equation of state parameters, which assumes
Gaussianity of the likelihood, could be a terrible approximation to the exact
likelihood in a specific class of dark energy models, and would significantly
bias the constraints. A more general approach such as the one we have adopted,
Markov Chain Monte Carlo combined with an exact integration of the dark energy
evolution equations, is needed.  The constraints we have obtained on the
fundamental parameters are shown in Figs.~\ref{fig:like_4x4_2params} and
\ref{fig:like_4x4_2params_future}.

We have computed the principal components of the equation of state for our data
compilation.  As noted in Sec.~\ref{sec:PC_define}, and for reasons listed
there, these principal components have been pre-computed for the same survey
but assuming a general DE history; therefore they are not
uncorrelated. Nevertheless, we find them very useful --- for example, Figure
\ref{fig:PC_constraints} shows that the current data roughly imply
$\alpha_1\lesssim -0.85$, thereby imposing an effective upper limit to the
weighted average of the equation of state.  Moreover, in
Sec.~\ref{sec:w0wa_define} we propose a shortcut to compute the parameters
$w_0$ and $w_a$ that approximately describe the equation of state history
$w(z)$ at low redshift where the constraining power of the data is the
greatest. Instead of fitting $(w_0, w_a)$, we obtain them directly from the
first two principal components. The left panel of Fig.~\ref{fig:wz_reconstr}
illustrates that the approximate $(w_0, w_a)$ description of the models fits
the dark energy histories almost perfectly at $z\lesssim 1$; we found that
even distances to $z=3$ are recovered to within 2-3\%. These results suggest
that converting from the principal components to an arbitrary parametrization
can be performed with very small biases, and essentially at no additional
computational cost.

The constraints in the $(w_0, w_a)$ or $(w_{\rm pivot}, w_a)$ planes indicate
that the constraints with future data will have the (inverse area) figure of
merit improvement of about an order of magnitude over the current data. Given
that our models approach the currently favored \LCDM cosmology arbitrarily
closely, this means that efforts to pursue upcoming experimental efforts, such
as Planck, JDEM, LST, the planned probes of baryon acoustic oscillations, and
the improved Hubble constant measurements are extremely valuable and likely to
lead to significant improvement in sweeping away a large fraction of currently
viable models.

We have also considered the classification of models into ``thawing'' and ``freezing''
depending on whether they are asymptotically receding from or approaching the
state of zero kinetic energy where the equation of state is $-1$. We found that
the thawing and freezing limits on monomial-potential scalar field models,
presented in Ref.~\cite{Caldwell_Linder}, are largely not obeyed by our models.
As illustrated in Fig.~\ref{fig:thaw_freeze}, this is due to the interplay
between the initial conditions of the scalar field, the effects of Hubble
friction, and the shape of the effective potential (non-monotonicity in the
second derivative) that empirically generated models possess. As we do not have
a good understanding behind the physical mechanism that powers dark energy, we
think that the division of models into thawing and freezing, or any similar
nomenclature, is useful only in the context of future experimental constraints:
if the constrained phase space ends up favoring one of these subclasses, this division
can help obtain a fuller theoretical understanding

Finally note that the approach that we outlined can be trivially generalized
to consider, for example, the density perturbations of dark energy. By solving the 
perturbation equations for each model along with the background evolution equations 
as currently done, we can generate trajectories using the MCMC
approach. Therefore, a exhaustive ``scan'' through the possible perturbation
scenarios can be obtained, paired with the background evolution histories, and
then constrained with cosmological data. This approach would nicely complement the
previous analyses that largely relied on standard phenomenological descriptions
of the dark energy sector \cite{Bean_Dore,Weller_Lewis,Hannestad_sound,Battye_Moss}.

This is an excellent time to perform numerically rigorous, comprehensive
analyses of dark energy models because the data is finally allowing precision
tests of dark energy.  Not only are the statistical errors respectable (see
Table \ref{tab:constraints}), but most of the experiments are now reporting
careful analyses of systematic errors. As we have shown in this paper, data
expected in the next $\sim 10$ years will lead to precision measurements of the
parameters describing the effects of dark energy on the expansion history of
the universe.  One can sincerely hope, perhaps even expect, that these efforts
will lead to important hints as to the nature and the origin of dark energy.

\section*{Acknowledgments}
We thank Wayne Hu for useful conversations and questions that inspired a number
of investigations in the paper. We also thank Eric Linder for discussions on
freezing/thawing models and a thorough reading of the manuscript, Antony Lewis
for information about the expected Planck errors, and the Dark Energy Task
Force for promptly making their findings public.  Finally, we thank Charles
Bennett, Simon DeDeo, Josh Frieman, Andrew Liddle, Pia Mukherjee, David
Parkinson, Martin Sahl\'{e}n, Bob Scherrer, Licia Verde and Ned Wright for useful conversations. A
modified version of the GetDist parameter estimation package was used for some
of the plots. We acknowledge the use of the Legacy Archive for Microwave
Background Data Analysis (LAMBDA). HVP acknowledges the hospitality of the
Institute of Astronomy, Cambridge where part of this work was carried out. DH
is supported by the NSF Astronomy and Astrophysics Postdoctoral Fellowship
under Grant No.\ 0401066. HVP is supported by NASA through Hubble Fellowship
grant
\#HF-01177.01-A awarded by the Space Telescope Science Institute, which is
operated by the Association of Universities for Research in Astronomy, Inc.,
for NASA, under contract NAS 5-26555.

\appendix
\section{Current  Cosmological data and the likelihood function}\label{app:current}

Here we report on the current cosmological observations that we use in order to
constrain the cosmological models. We use a combination of SNe Ia, baryon
oscillation and Hubble constant measurements, together with WMAP constraints on
the angular size of the sound horizon, and the physical matter and baryon
densities.

The SNa Ia data we use are Supernova Legacy Survey data from Ref.~\cite{Astier}
with $N=115$ SNa measurements, which include the low-redshift Calan-Tololo
sample.  The observed apparent magnitude is given by
\begin{equation}
m_i = 5 \log H_0 d_L(z_i) + \mathcal{M} + \alpha(s_i-1)+\beta c_i+\epsilon_i \,,
\end{equation}
where $i=1\ldots 115$ runs through the observed SNe. The measured values of the
apparent magnitude of supernova $m_i$, its redshift $z_i$, stretch $s_i$, color
$c_i$ and total error $\epsilon_i$ are given in the SNLS paper
\cite{Astier}. The nuisance parameter $\mathcal{M}$ is a combination of Hubble
parameter and absolute magnitude of SNe \cite{Perlmutter_1999}.  The 
likelihood of measuring the cosmological parameter set ${\boldtheta}$ is given by

\begin{eqnarray}
\mathcal{L}_{\rm SNe}({\boldtheta})  &=& \int \exp(-\chi^2/2)\, 
\mathcal{P}(\alpha) \mathcal{P}(\beta)\, d\mathcal{M}\,d\alpha\, d\beta
\\[0.1cm]
\chi^2({\boldtheta})       &=& \sum_{i=1}^N{(m_i-m({\boldtheta}, z_i))^2\over \epsilon_i^2},
\label{eq:SNe_like}
\end{eqnarray}

\noindent where $\mathcal{P}(\alpha)$ and $\mathcal{P}(\beta)$ are priors given to
the stretch and color coefficients.  Since $\mathcal{M}$ is given a flat prior,
the $\mathcal{M}$ integral can be done analytically \cite{Nesseris} to obtain

\begin{eqnarray}
\mathcal{L}_{\rm SNe}({\boldtheta}) &=& \int\exp(-(A-B^2/C)/2)\,
\mathcal{P}(\alpha) \mathcal{P}(\beta)\,d\alpha\, d\beta\nonumber \\[0.1cm]
A             &\equiv & \sum_{i=1}^N {(m_i-m({\boldtheta}, z_i))^2\over\epsilon_i^2}\\[0.1cm]
B             &\equiv & \sum_{i=1}^N { m_i-m({\boldtheta}, z_i)   \over\epsilon_i^2}\\[0.1cm]
C             &\equiv & \sum_{i=1}^N {1                          \over\epsilon_i^2}
\end{eqnarray}

Finally, we follow the SNLS analysis and adopt the \LCDM values for the stretch
and color parameters, $\alpha=1.52\pm 0.14$ and $\beta=1.57\pm 0.15$, and
marginalize with flat priors over the $\pm 1$-$\sigma$ values in $\alpha$ and
$\beta$.  After experimenting with several alternative prior widths for the
stretch and color coefficients, we conclude that their details introduce
unobservable changes in our results.

SNe Ia provide the best constraints at low redshifts ($z\lesssim 1.7$).
However, Figure \ref{fig:wz_prior} shows that there exist classes of models
that look roughly like the cosmological constant at those redshifts but
dominate the energy density at $z\gtrsim 2$. Those models are not ruled out by
the SN data, but nevertheless are clearly inconsistent with the observed
universe as they prevent sufficient structure formation.  In order to protect
against such models, we use the information provided by baryon oscillations and
CMB measurements of the angular size of the acoustic horizon and the physical
matter density.  More precisely, we use the quantity probed by baryon
oscillations

\begin{equation}
D_V(z=0.35)\equiv \left [ r(z)^2 {cz\over H(z)}\right ]^{1/3}=(1370\pm 64)\,\, {\rm Mpc}
\end{equation}

\noindent where $r(z)$ is the comoving angular diameter distance to redshift $z$ and the
measurement comes from the Sloan Digital Sky Survey \cite{Eisenstein}. 

To represent the CMB measurements given by the WMAP experiment
\cite{Spergel_2003,Spergel_2006}, we do not use the full likelihood function as
its evaluation would be very time consuming for the millions of models we
consider. Instead, an excellent approximation is to use the single derived
quantity that is sensitive to dark energy, the angular size of the sound
horizon \cite{Hu_Fukugita,Frieman}, which is equal to the ratio of the physical
size of the sound horizon and the distance to recombination

\begin{equation}
\theta_A\equiv {s_H\over r(z=1089)}=(0.595\pm 0.002)\,\,{\rm deg},
\end{equation}

\noindent where the sound horizon $s_H$ is given by

\begin{equation}
s_H={c\over \sqrt{3}H_0}\int_0^{a_{\rm dec}} 
\left [\left (1+{3\Omega_b\over 4\Omega_{\gamma}}a\right )
\left (\Omega_m\, a + \Omega_R\right )\right ]^{-1/2}\,da,
\end{equation}

\noindent where $\Omega_\gamma$ and $\Omega_R$ are the energy densities in
photons and radiation (photons plus neutrinos) respectively. The quantities
$\Omega_\gamma$ and $\Omega_R$ are given in terms of the CMB temperature and
are exceedingly accurately measured, so we hold them fixed.  

Finally, since $\theta_A$, $\Omega_m h^2$ and $\Omega_b h^2$ are our original
parameters, and we compute the equation of state history $w(z)$ for
each model, the Hubble constant $H_0$ is necessarily a derived parameter. To
compute the distance to decoupling $r(z=1089)$ we use not the exact
$w(z)$, but rather the constant effective equation of state $w_{\rm eff}$,
defined as

\begin{equation}
w_{\rm eff}\equiv {\ds \int_{1/(1+1089)}^1 \ode(a)\, w(a)\, da\over
\ds \int_{1/(1+1089)}^1 \ode(a)\, da}.
\label{eq:weff}
\end{equation}

\noindent The reason we use $w_{\rm eff}$ is that the original WMAP analysis 
\cite{Spergel_2006} uses a constant equation of state to arrive at their constraints. 

We adopt the the physical baryon density $\Omega_b h^2$ as a free parameter and
give it the prior consistent with WMAP, $\Omega_b h^2=0.0223\pm 0.00074$. The
physical matter density, $\Omega_m h^2$, is also a free parameter accurately
constrained by the CMB.  Since our models allow more complex behavior in the dark
energy sector, we conservatively double the reported \LCDM error 
in this parameter, adopting $\Omega_m h^2=0.127\pm 0.02$.

Finally we adopt the Hubble Key project measurement of the Hubble constant,
$H_0=0.72\pm 0.08$ \cite{HKP}. The final likelihood is

\begin{equation}
\mathcal{L}({\boldtheta})=
\mathcal{L}_{\rm SNe}({\boldtheta})\times  
\mathcal{L}_{\rm BAO}({\boldtheta})\times  
\mathcal{L}_{\rm CMB}({\boldtheta})\times  
\mathcal{L}_{\rm H_0}({\boldtheta})
\label{eq:like}
\end{equation}

\noindent where the SNe likelihood is specified in Eq.~(\ref{eq:SNe_like}), the
BAO and $H_0$ are Gaussian with means and standard deviations specified above,
and the CMB likelihood consists of $\theta_A$, $\Omega_b h^2$ and $\Omega_m
h^2$ constraints, all of which are Gaussian with means and standard deviations
specified above.

\section{Future data}\label{app:future}

To simulate the future cosmological data, we consider the same cosmological
probes as for the current data: type Ia supernovae, baryon acoustic
oscillations, CMB measurements of the sound horizon, $\Omega_m h^2$ and
$\Omega_b h^2$, and the measurements of the Hubble constant. Since we obviously
do not know which cosmological model will be favored by future data, we center
all observables on the \LCDM model with $\Omega_m=1-\Omega_{\rm DE}(z=0)=0.25$,
$w(z)=-1$ and $H_0=0.72$. As described below, we add best-guess systematic
errors for all future measurements.

For the SNa data we assume a SNAP-type experiment with 2800 SNe
distributed in redshift out to $z=1.7$ as given by 
the middle curve of Fig. 9 in Ref.~\cite{SNAP}, and combined
with 300 local supernovae uniformly distributed in the $z=0.03-0.08$ range. We
add systematic errors in quadrature with intrinsic random Gaussian errors 
of 0.15 mag per SNe.  The systematic errors create an effective error floor 
of $0.02\,(1+z_i)/2.7$ mag per bin of $\Delta z=0.1$ centered at redshift
$z_i$. 

Specifications of proposed future baryon oscillation surveys and their expected
statistical and systematic errors are described in the Dark Energy Task Force
report \cite{DETF} (this is partly based on other studies, such as
\cite{Blake}). To a very good approximation, baryon oscillations provide
measurements of the angular diameter distance $d_A(z)$ {\it and} the expansion
rate $H(z)$ out to the redshift(s) where the source population of galaxies is
located.  We adopt the DETF findings and assume a JDEM-type survey covering the
redshift range $z\in[0.5, 2.0]$ of 10,000 square degrees. For simplicity we
assume only three redshift bins centered at $z=0.75$, $1.25$ and $1.75$ with
widths $\Delta z=0.5$ each.  With these parameters, formulae in the baryon
oscillation section of the DETF report indicate that the total expected error
(statistical plus systematic) for a spectroscopic survey is about 1.5\% for the
measurements of the angular diameter distance in each bin, and 2\% for
measurements of the expansion rate $H(z)$. Therefore we assume a total of six
BAO observables -- three distances and three expansion rates -- with the
aforementioned errors.

Note that, in the DETF error budget, a conservative 1\% systematic error is
assumed in each bin for each BAO observable. Controlling the systematics to a
level lower than this, as well as increasing the observed sky area, would
enable a future JDEM BAO mission to achieve significantly higher sensitivities
than assumed here (C. Bennett, private communication). Similarly, if the JDEM
SNa systematics can be understood to better than the systematics floor we
assumed, the sensitivity of that experiment could be higher.

For the future CMB experiment we assume Planck \cite{Planck} which is scheduled
for launch in 2008. Planck's fiducial error in $\Omega_m h^2$, for a flat \LCDM
universe, is about 1\%. Since we are considering models with a complicated
dark energy sector, we conservatively double this uncertainty, just as in the
case with current data. With the assumed cosmology, this amounts to assuming
$\Omega_m h^2=0.1296\pm 0.00270$. The angular size of the acoustic horizon for
the fiducial cosmology and its expected Planck error are $\theta_A=(0.595\pm
0.00017)$ deg. Similarly, the baryon fraction is assumed to be $\Omega_b
h^2=0.023\pm 0.00015$

Finally, we conservatively assume independent future measurements of the Hubble
constant will reach the 5\% level (more accurate measurements are feasible in
principle, but the systematics remain the primary obstacles).  Therefore
we take $H_0=0.72\pm 0.036$.

The full likelihood function is given again by Eq.~(\ref{eq:like}).

\bibliography{flowroll}

\begin{thebibliography}{102}
\expandafter\ifx\csname natexlab\endcsname\relax\def\natexlab#1{#1}\fi
\expandafter\ifx\csname bibnamefont\endcsname\relax
  \def\bibnamefont#1{#1}\fi
\expandafter\ifx\csname bibfnamefont\endcsname\relax
  \def\bibfnamefont#1{#1}\fi
\expandafter\ifx\csname citenamefont\endcsname\relax
  \def\citenamefont#1{#1}\fi
\expandafter\ifx\csname url\endcsname\relax
  \def\url#1{\texttt{#1}}\fi
\expandafter\ifx\csname urlprefix\endcsname\relax\def\urlprefix{URL }\fi
\providecommand{\bibinfo}[2]{#2}
\providecommand{\eprint}[2][]{\url{#2}}

\bibitem[{\citenamefont{Perlmutter et~al.}(1999)}]{Perlmutter_1999}
\bibinfo{author}{\bibfnamefont{S.}~\bibnamefont{Perlmutter}}
  \bibnamefont{et~al.} (\bibinfo{collaboration}{Supernova Cosmology Project}),
  \bibinfo{journal}{Astrophys. J.} \textbf{\bibinfo{volume}{517}},
  \bibinfo{pages}{565} (\bibinfo{year}{1999}), \eprint{astro-ph/9812133}.

\bibitem[{\citenamefont{Riess et~al.}(1998)}]{Riess_1998}
\bibinfo{author}{\bibfnamefont{A.~G.} \bibnamefont{Riess}} \bibnamefont{et~al.}
  (\bibinfo{collaboration}{Supernova Search Team}), \bibinfo{journal}{Astron.
  J.} \textbf{\bibinfo{volume}{116}}, \bibinfo{pages}{1009}
  (\bibinfo{year}{1998}), \eprint{astro-ph/9805201}.

\bibitem[{\citenamefont{Knop et~al.}(2003)}]{Knop}
\bibinfo{author}{\bibfnamefont{R.~A.} \bibnamefont{Knop}} \bibnamefont{et~al.}
  (\bibinfo{collaboration}{Supernova Cosmology Project}),
  \bibinfo{journal}{Astrophys. J.} \textbf{\bibinfo{volume}{598}},
  \bibinfo{pages}{102} (\bibinfo{year}{2003}), \eprint{astro-ph/0309368}.

\bibitem[{\citenamefont{Riess et~al.}(2004)}]{Riess_2004}
\bibinfo{author}{\bibfnamefont{A.~G.} \bibnamefont{Riess}} \bibnamefont{et~al.}
  (\bibinfo{collaboration}{Supernova Search Team}),
  \bibinfo{journal}{Astrophys. J.} \textbf{\bibinfo{volume}{607}},
  \bibinfo{pages}{665} (\bibinfo{year}{2004}), \eprint{astro-ph/0402512}.

\bibitem[{\citenamefont{Astier et~al.}(2006)}]{Astier}
\bibinfo{author}{\bibfnamefont{P.}~\bibnamefont{Astier}} \bibnamefont{et~al.},
  \bibinfo{journal}{Astron. Astrophys.} \textbf{\bibinfo{volume}{447}},
  \bibinfo{pages}{31} (\bibinfo{year}{2006}), \eprint{astro-ph/0510447}.

\bibitem[{\citenamefont{Cooray and Huterer}(1999)}]{Cooray_Huterer}
\bibinfo{author}{\bibfnamefont{A.~R.} \bibnamefont{Cooray}} \bibnamefont{and}
  \bibinfo{author}{\bibfnamefont{D.}~\bibnamefont{Huterer}},
  \bibinfo{journal}{Astrophys. J.} \textbf{\bibinfo{volume}{513}},
  \bibinfo{pages}{L95} (\bibinfo{year}{1999}), \eprint{astro-ph/9901097}.

\bibitem[{\citenamefont{Linder}(2003)}]{Linder_wa}
\bibinfo{author}{\bibfnamefont{E.~V.} \bibnamefont{Linder}},
  \bibinfo{journal}{Phys. Rev. Lett.} \textbf{\bibinfo{volume}{90}},
  \bibinfo{pages}{091301} (\bibinfo{year}{2003}), \eprint{astro-ph/0208512}.

\bibitem[{\citenamefont{Corasaniti and Copeland}(2003)}]{Corasaniti_Copeland}
\bibinfo{author}{\bibfnamefont{P.~S.} \bibnamefont{Corasaniti}}
  \bibnamefont{and} \bibinfo{author}{\bibfnamefont{E.~J.}
  \bibnamefont{Copeland}}, \bibinfo{journal}{Phys. Rev.}
  \textbf{\bibinfo{volume}{D67}}, \bibinfo{pages}{063521}
  (\bibinfo{year}{2003}), \eprint{astro-ph/0205544}.

\bibitem[{\citenamefont{Hannestad and Mortsell}(2004)}]{Hannestad_Mortsell}
\bibinfo{author}{\bibfnamefont{S.}~\bibnamefont{Hannestad}} \bibnamefont{and}
  \bibinfo{author}{\bibfnamefont{E.}~\bibnamefont{Mortsell}},
  \bibinfo{journal}{JCAP} \textbf{\bibinfo{volume}{0409}}, \bibinfo{pages}{001}
  (\bibinfo{year}{2004}), \eprint{astro-ph/0407259}.

\bibitem[{\citenamefont{Huterer and Turner}(1999)}]{reconstr}
\bibinfo{author}{\bibfnamefont{D.}~\bibnamefont{Huterer}} \bibnamefont{and}
  \bibinfo{author}{\bibfnamefont{M.~S.} \bibnamefont{Turner}},
  \bibinfo{journal}{Phys. Rev.} \textbf{\bibinfo{volume}{D60}},
  \bibinfo{pages}{081301} (\bibinfo{year}{1999}), \eprint{astro-ph/9808133}.

\bibitem[{\citenamefont{Nakamura and Chiba}(1999)}]{Nakamura_Chiba}
\bibinfo{author}{\bibfnamefont{T.}~\bibnamefont{Nakamura}} \bibnamefont{and}
  \bibinfo{author}{\bibfnamefont{T.}~\bibnamefont{Chiba}},
  \bibinfo{journal}{Mon. Not. Roy. Astron. Soc.}
  \textbf{\bibinfo{volume}{306}}, \bibinfo{pages}{696} (\bibinfo{year}{1999}),
  \eprint{astro-ph/9810447}.

\bibitem[{\citenamefont{Starobinsky}(1998)}]{Starobinsky}
\bibinfo{author}{\bibfnamefont{A.~A.} \bibnamefont{Starobinsky}},
  \bibinfo{journal}{JETP Lett.} \textbf{\bibinfo{volume}{68}},
  \bibinfo{pages}{757} (\bibinfo{year}{1998}), \eprint{astro-ph/9810431}.

\bibitem[{\citenamefont{Eisenstein et~al.}(1999)\citenamefont{Eisenstein, Hu,
  and Tegmark}}]{EHT}
\bibinfo{author}{\bibfnamefont{D.~J.} \bibnamefont{Eisenstein}},
  \bibinfo{author}{\bibfnamefont{W.}~\bibnamefont{Hu}}, \bibnamefont{and}
  \bibinfo{author}{\bibfnamefont{M.}~\bibnamefont{Tegmark}},
  \bibinfo{journal}{Astrophys. J.} \textbf{\bibinfo{volume}{518}},
  \bibinfo{pages}{2} (\bibinfo{year}{1999}), \eprint{astro-ph/9807130}.

\bibitem[{\citenamefont{Huterer and Turner}(2001)}]{Huterer_Turner}
\bibinfo{author}{\bibfnamefont{D.}~\bibnamefont{Huterer}} \bibnamefont{and}
  \bibinfo{author}{\bibfnamefont{M.~S.} \bibnamefont{Turner}},
  \bibinfo{journal}{Phys. Rev.} \textbf{\bibinfo{volume}{D64}},
  \bibinfo{pages}{123527} (\bibinfo{year}{2001}), \eprint{astro-ph/0012510}.

\bibitem[{\citenamefont{Weller and Albrecht}(2002)}]{Weller_Albrecht}
\bibinfo{author}{\bibfnamefont{J.}~\bibnamefont{Weller}} \bibnamefont{and}
  \bibinfo{author}{\bibfnamefont{A.}~\bibnamefont{Albrecht}},
  \bibinfo{journal}{Phys. Rev.} \textbf{\bibinfo{volume}{D65}},
  \bibinfo{pages}{103512} (\bibinfo{year}{2002}), \eprint{astro-ph/0106079}.

\bibitem[{\citenamefont{Weller and Albrecht}(2001)}]{Weller:2000pf}
\bibinfo{author}{\bibfnamefont{J.}~\bibnamefont{Weller}} \bibnamefont{and}
  \bibinfo{author}{\bibfnamefont{A.}~\bibnamefont{Albrecht}},
  \bibinfo{journal}{Phys. Rev. Lett.} \textbf{\bibinfo{volume}{86}},
  \bibinfo{pages}{1939} (\bibinfo{year}{2001}), \eprint{astro-ph/0008314}.

\bibitem[{\citenamefont{Maor et~al.}(2002)\citenamefont{Maor, Brustein,
  McMahon, and Steinhardt}}]{Maor}
\bibinfo{author}{\bibfnamefont{I.}~\bibnamefont{Maor}},
  \bibinfo{author}{\bibfnamefont{R.}~\bibnamefont{Brustein}},
  \bibinfo{author}{\bibfnamefont{J.}~\bibnamefont{McMahon}}, \bibnamefont{and}
  \bibinfo{author}{\bibfnamefont{P.~J.} \bibnamefont{Steinhardt}},
  \bibinfo{journal}{Phys. Rev.} \textbf{\bibinfo{volume}{D65}},
  \bibinfo{pages}{123003} (\bibinfo{year}{2002}), \eprint{astro-ph/0112526}.

\bibitem[{\citenamefont{Kujat et~al.}(2002)\citenamefont{Kujat, Linn, Scherrer,
  and Weinberg}}]{Kujat}
\bibinfo{author}{\bibfnamefont{J.}~\bibnamefont{Kujat}},
  \bibinfo{author}{\bibfnamefont{A.~M.} \bibnamefont{Linn}},
  \bibinfo{author}{\bibfnamefont{R.~J.} \bibnamefont{Scherrer}},
  \bibnamefont{and} \bibinfo{author}{\bibfnamefont{D.~H.}
  \bibnamefont{Weinberg}}, \bibinfo{journal}{Astrophys. J.}
  \textbf{\bibinfo{volume}{572}}, \bibinfo{pages}{1} (\bibinfo{year}{2002}),
  \eprint{astro-ph/0112221}.

\bibitem[{\citenamefont{Bassett et~al.}(2004)\citenamefont{Bassett, Corasaniti,
  and Kunz}}]{Bassett_compression}
\bibinfo{author}{\bibfnamefont{B.~A.} \bibnamefont{Bassett}},
  \bibinfo{author}{\bibfnamefont{P.~S.} \bibnamefont{Corasaniti}},
  \bibnamefont{and} \bibinfo{author}{\bibfnamefont{M.}~\bibnamefont{Kunz}},
  \bibinfo{journal}{Astrophys. J.} \textbf{\bibinfo{volume}{617}},
  \bibinfo{pages}{L1} (\bibinfo{year}{2004}), \eprint{astro-ph/0407364}.

\bibitem[{\citenamefont{Bassett}(2005)}]{Bassett_optimize}
\bibinfo{author}{\bibfnamefont{B.~A.} \bibnamefont{Bassett}},
  \bibinfo{journal}{Phys. Rev.} \textbf{\bibinfo{volume}{D71}},
  \bibinfo{pages}{083517} (\bibinfo{year}{2005}), \eprint{astro-ph/0407201}.

\bibitem[{\citenamefont{Knox et~al.}(2006)\citenamefont{Knox, Song, and
  Zhan}}]{Knox_Song_Zhan}
\bibinfo{author}{\bibfnamefont{L.}~\bibnamefont{Knox}},
  \bibinfo{author}{\bibfnamefont{Y.-S.} \bibnamefont{Song}}, \bibnamefont{and}
  \bibinfo{author}{\bibfnamefont{H.}~\bibnamefont{Zhan}}
  (\bibinfo{year}{2006}), \eprint{astro-ph/0605536}.

\bibitem[{\citenamefont{Virey and Ealet}(2006)}]{Virey}
\bibinfo{author}{\bibfnamefont{J.-M.} \bibnamefont{Virey}} \bibnamefont{and}
  \bibinfo{author}{\bibfnamefont{A.}~\bibnamefont{Ealet}}
  (\bibinfo{year}{2006}), \eprint{astro-ph/0607589}.

\bibitem[{\citenamefont{Melchiorri et~al.}(2003)\citenamefont{Melchiorri,
  Mersini-Houghton, Odman, and Trodden}}]{Melchiorri_state}
\bibinfo{author}{\bibfnamefont{A.}~\bibnamefont{Melchiorri}},
  \bibinfo{author}{\bibfnamefont{L.}~\bibnamefont{Mersini-Houghton}},
  \bibinfo{author}{\bibfnamefont{C.~J.} \bibnamefont{Odman}}, \bibnamefont{and}
  \bibinfo{author}{\bibfnamefont{M.}~\bibnamefont{Trodden}},
  \bibinfo{journal}{Phys. Rev.} \textbf{\bibinfo{volume}{D68}},
  \bibinfo{pages}{043509} (\bibinfo{year}{2003}), \eprint{astro-ph/0211522}.

\bibitem[{\citenamefont{Corasaniti et~al.}(2004)\citenamefont{Corasaniti, Kunz,
  Parkinson, Copeland, and Bassett}}]{Corasaniti_foundations}
\bibinfo{author}{\bibfnamefont{P.~S.} \bibnamefont{Corasaniti}},
  \bibinfo{author}{\bibfnamefont{M.}~\bibnamefont{Kunz}},
  \bibinfo{author}{\bibfnamefont{D.}~\bibnamefont{Parkinson}},
  \bibinfo{author}{\bibfnamefont{E.~J.} \bibnamefont{Copeland}},
  \bibnamefont{and} \bibinfo{author}{\bibfnamefont{B.~A.}
  \bibnamefont{Bassett}}, \bibinfo{journal}{Phys. Rev.}
  \textbf{\bibinfo{volume}{D70}}, \bibinfo{pages}{083006}
  (\bibinfo{year}{2004}), \eprint{astro-ph/0406608}.

\bibitem[{\citenamefont{Spergel et~al.}(2003)}]{Spergel_2003}
\bibinfo{author}{\bibfnamefont{D.~N.} \bibnamefont{Spergel}}
  \bibnamefont{et~al.} (\bibinfo{collaboration}{WMAP}),
  \bibinfo{journal}{Astrophys. J. Suppl.} \textbf{\bibinfo{volume}{148}},
  \bibinfo{pages}{175} (\bibinfo{year}{2003}), \eprint{astro-ph/0302209}.

\bibitem[{\citenamefont{Spergel et~al.}(2006)}]{Spergel_2006}
\bibinfo{author}{\bibfnamefont{D.~N.} \bibnamefont{Spergel}}
  \bibnamefont{et~al.} (\bibinfo{year}{2006}), \eprint{astro-ph/0603449}.

\bibitem[{\citenamefont{Tegmark et~al.}(2004)}]{Tegmark_SDSS}
\bibinfo{author}{\bibfnamefont{M.}~\bibnamefont{Tegmark}} \bibnamefont{et~al.}
  (\bibinfo{collaboration}{SDSS}), \bibinfo{journal}{Phys. Rev.}
  \textbf{\bibinfo{volume}{D69}}, \bibinfo{pages}{103501}
  (\bibinfo{year}{2004}), \eprint{astro-ph/0310723}.

\bibitem[{\citenamefont{Seljak et~al.}(2005)}]{Seljak_SDSS}
\bibinfo{author}{\bibfnamefont{U.}~\bibnamefont{Seljak}} \bibnamefont{et~al.}
  (\bibinfo{collaboration}{SDSS}), \bibinfo{journal}{Phys. Rev.}
  \textbf{\bibinfo{volume}{D71}}, \bibinfo{pages}{103515}
  (\bibinfo{year}{2005}), \eprint{astro-ph/0407372}.

\bibitem[{\citenamefont{Upadhye et~al.}(2005)\citenamefont{Upadhye, Ishak, and
  Steinhardt}}]{Upadhye}
\bibinfo{author}{\bibfnamefont{A.}~\bibnamefont{Upadhye}},
  \bibinfo{author}{\bibfnamefont{M.}~\bibnamefont{Ishak}}, \bibnamefont{and}
  \bibinfo{author}{\bibfnamefont{P.~J.} \bibnamefont{Steinhardt}},
  \bibinfo{journal}{Phys. Rev.} \textbf{\bibinfo{volume}{D72}},
  \bibinfo{pages}{063501} (\bibinfo{year}{2005}), \eprint{astro-ph/0411803}.

\bibitem[{\citenamefont{Xia et~al.}(2006)\citenamefont{Xia, Zhao, Feng, Li, and
  Zhang}}]{Xia}
\bibinfo{author}{\bibfnamefont{J.-Q.} \bibnamefont{Xia}},
  \bibinfo{author}{\bibfnamefont{G.-B.} \bibnamefont{Zhao}},
  \bibinfo{author}{\bibfnamefont{B.}~\bibnamefont{Feng}},
  \bibinfo{author}{\bibfnamefont{H.}~\bibnamefont{Li}}, \bibnamefont{and}
  \bibinfo{author}{\bibfnamefont{X.}~\bibnamefont{Zhang}},
  \bibinfo{journal}{Phys. Rev.} \textbf{\bibinfo{volume}{D73}},
  \bibinfo{pages}{063521} (\bibinfo{year}{2006}), \eprint{astro-ph/0511625}.

\bibitem[{\citenamefont{Zhao et~al.}(2006)\citenamefont{Zhao, Xia, Feng, and
  Zhang}}]{Zhao}
\bibinfo{author}{\bibfnamefont{G.-B.} \bibnamefont{Zhao}},
  \bibinfo{author}{\bibfnamefont{J.-Q.} \bibnamefont{Xia}},
  \bibinfo{author}{\bibfnamefont{B.}~\bibnamefont{Feng}}, \bibnamefont{and}
  \bibinfo{author}{\bibfnamefont{X.}~\bibnamefont{Zhang}}
  (\bibinfo{year}{2006}), \eprint{astro-ph/0603621}.

\bibitem[{\citenamefont{Nesseris and Perivolaropoulos}(2005)}]{Nesseris}
\bibinfo{author}{\bibfnamefont{S.}~\bibnamefont{Nesseris}} \bibnamefont{and}
  \bibinfo{author}{\bibfnamefont{L.}~\bibnamefont{Perivolaropoulos}},
  \bibinfo{journal}{Phys. Rev.} \textbf{\bibinfo{volume}{D72}},
  \bibinfo{pages}{123519} (\bibinfo{year}{2005}), \eprint{astro-ph/0511040}.

\bibitem[{\citenamefont{Jarvis et~al.}(2006)\citenamefont{Jarvis, Jain,
  Bernstein, and Dolney}}]{Jarvis}
\bibinfo{author}{\bibfnamefont{M.}~\bibnamefont{Jarvis}},
  \bibinfo{author}{\bibfnamefont{B.}~\bibnamefont{Jain}},
  \bibinfo{author}{\bibfnamefont{G.}~\bibnamefont{Bernstein}},
  \bibnamefont{and} \bibinfo{author}{\bibfnamefont{D.}~\bibnamefont{Dolney}},
  \bibinfo{journal}{Astrophys. J.} \textbf{\bibinfo{volume}{644}},
  \bibinfo{pages}{71} (\bibinfo{year}{2006}), \eprint{astro-ph/0502243}.

\bibitem[{\citenamefont{Doran et~al.}(2006)\citenamefont{Doran, Robbers, and
  Wetterich}}]{Doran}
\bibinfo{author}{\bibfnamefont{M.}~\bibnamefont{Doran}},
  \bibinfo{author}{\bibfnamefont{G.}~\bibnamefont{Robbers}}, \bibnamefont{and}
  \bibinfo{author}{\bibfnamefont{C.}~\bibnamefont{Wetterich}}
  (\bibinfo{year}{2006}), \eprint{astro-ph/0609814}.

\bibitem[{\citenamefont{Saini et~al.}(2000)\citenamefont{Saini, Raychaudhury,
  Sahni, and Starobinsky}}]{Saini_reconstr}
\bibinfo{author}{\bibfnamefont{T.~D.} \bibnamefont{Saini}},
  \bibinfo{author}{\bibfnamefont{S.}~\bibnamefont{Raychaudhury}},
  \bibinfo{author}{\bibfnamefont{V.}~\bibnamefont{Sahni}}, \bibnamefont{and}
  \bibinfo{author}{\bibfnamefont{A.~A.} \bibnamefont{Starobinsky}},
  \bibinfo{journal}{Phys. Rev. Lett.} \textbf{\bibinfo{volume}{85}},
  \bibinfo{pages}{1162} (\bibinfo{year}{2000}), \eprint{astro-ph/9910231}.

\bibitem[{\citenamefont{Huterer}(2002)}]{Huterer_thesis}
\bibinfo{author}{\bibfnamefont{D.}~\bibnamefont{Huterer}},
  \bibinfo{journal}{Phys. Rev.} \textbf{\bibinfo{volume}{D65}},
  \bibinfo{pages}{063001} (\bibinfo{year}{2002}), \eprint{astro-ph/0106399}.

\bibitem[{\citenamefont{Takada and Jain}(2004)}]{Takada_Jain}
\bibinfo{author}{\bibfnamefont{M.}~\bibnamefont{Takada}} \bibnamefont{and}
  \bibinfo{author}{\bibfnamefont{B.}~\bibnamefont{Jain}},
  \bibinfo{journal}{Mon. Not. Roy. Astron. Soc.}
  \textbf{\bibinfo{volume}{348}}, \bibinfo{pages}{897} (\bibinfo{year}{2004}),
  \eprint{astro-ph/0310125}.

\bibitem[{\citenamefont{Seo and Eisenstein}(2003)}]{Seo_Eisenstein}
\bibinfo{author}{\bibfnamefont{H.-J.} \bibnamefont{Seo}} \bibnamefont{and}
  \bibinfo{author}{\bibfnamefont{D.~J.} \bibnamefont{Eisenstein}},
  \bibinfo{journal}{Astrophys. J.} \textbf{\bibinfo{volume}{598}},
  \bibinfo{pages}{720} (\bibinfo{year}{2003}), \eprint{astro-ph/0307460}.

\bibitem[{\citenamefont{Alam et~al.}(2004)\citenamefont{Alam, Sahni, and
  Starobinsky}}]{Alam}
\bibinfo{author}{\bibfnamefont{U.}~\bibnamefont{Alam}},
  \bibinfo{author}{\bibfnamefont{V.}~\bibnamefont{Sahni}}, \bibnamefont{and}
  \bibinfo{author}{\bibfnamefont{A.~A.} \bibnamefont{Starobinsky}},
  \bibinfo{journal}{JCAP} \textbf{\bibinfo{volume}{0406}}, \bibinfo{pages}{008}
  (\bibinfo{year}{2004}), \eprint{astro-ph/0403687}.

\bibitem[{\citenamefont{Jonsson et~al.}(2004)\citenamefont{Jonsson, Goobar,
  Amanullah, and Bergstrom}}]{Jonsson}
\bibinfo{author}{\bibfnamefont{J.}~\bibnamefont{Jonsson}},
  \bibinfo{author}{\bibfnamefont{A.}~\bibnamefont{Goobar}},
  \bibinfo{author}{\bibfnamefont{R.}~\bibnamefont{Amanullah}},
  \bibnamefont{and} \bibinfo{author}{\bibfnamefont{L.}~\bibnamefont{Bergstrom}}
  (\bibinfo{year}{2004}), \eprint{astro-ph/0404468}.

\bibitem[{\citenamefont{Feng et~al.}(2005)\citenamefont{Feng, Wang, and
  Zhang}}]{Feng}
\bibinfo{author}{\bibfnamefont{B.}~\bibnamefont{Feng}},
  \bibinfo{author}{\bibfnamefont{X.-L.} \bibnamefont{Wang}}, \bibnamefont{and}
  \bibinfo{author}{\bibfnamefont{X.-M.} \bibnamefont{Zhang}},
  \bibinfo{journal}{Phys. Lett.} \textbf{\bibinfo{volume}{B607}},
  \bibinfo{pages}{35} (\bibinfo{year}{2005}), \eprint{astro-ph/0404224}.

\bibitem[{\citenamefont{Jassal et~al.}(2005)\citenamefont{Jassal, Bagla, and
  Padmanabhan}}]{Jassal}
\bibinfo{author}{\bibfnamefont{H.~K.} \bibnamefont{Jassal}},
  \bibinfo{author}{\bibfnamefont{J.~S.} \bibnamefont{Bagla}}, \bibnamefont{and}
  \bibinfo{author}{\bibfnamefont{T.}~\bibnamefont{Padmanabhan}},
  \bibinfo{journal}{Mon. Not. Roy. Astron. Soc.}
  \textbf{\bibinfo{volume}{356}}, \bibinfo{pages}{L11} (\bibinfo{year}{2005}),
  \eprint{astro-ph/0404378}.

\bibitem[{\citenamefont{Wang and Tegmark}(2005)}]{Wang_Tegmark_2005}
\bibinfo{author}{\bibfnamefont{Y.}~\bibnamefont{Wang}} \bibnamefont{and}
  \bibinfo{author}{\bibfnamefont{M.}~\bibnamefont{Tegmark}},
  \bibinfo{journal}{Phys. Rev.} \textbf{\bibinfo{volume}{D71}},
  \bibinfo{pages}{103513} (\bibinfo{year}{2005}), \eprint{astro-ph/0501351}.

\bibitem[{\citenamefont{Wang and Mukherjee}(2006)}]{Wang_Mukherjee}
\bibinfo{author}{\bibfnamefont{Y.}~\bibnamefont{Wang}} \bibnamefont{and}
  \bibinfo{author}{\bibfnamefont{P.}~\bibnamefont{Mukherjee}}
  (\bibinfo{year}{2006}), \eprint{astro-ph/0604051}.

\bibitem[{\citenamefont{Shafieloo et~al.}(2006)\citenamefont{Shafieloo, Alam,
  Sahni, and Starobinsky}}]{Shafieloo}
\bibinfo{author}{\bibfnamefont{A.}~\bibnamefont{Shafieloo}},
  \bibinfo{author}{\bibfnamefont{U.}~\bibnamefont{Alam}},
  \bibinfo{author}{\bibfnamefont{V.}~\bibnamefont{Sahni}}, \bibnamefont{and}
  \bibinfo{author}{\bibfnamefont{A.~A.} \bibnamefont{Starobinsky}},
  \bibinfo{journal}{Mon. Not. Roy. Astron. Soc.}
  \textbf{\bibinfo{volume}{366}}, \bibinfo{pages}{1081} (\bibinfo{year}{2006}),
  \eprint{astro-ph/0505329}.

\bibitem[{\citenamefont{Huterer and Cooray}(2005)}]{Huterer_Cooray}
\bibinfo{author}{\bibfnamefont{D.}~\bibnamefont{Huterer}} \bibnamefont{and}
  \bibinfo{author}{\bibfnamefont{A.}~\bibnamefont{Cooray}},
  \bibinfo{journal}{Phys. Rev.} \textbf{\bibinfo{volume}{D71}},
  \bibinfo{pages}{023506} (\bibinfo{year}{2005}), \eprint{astro-ph/0404062}.

\bibitem[{\citenamefont{Sahlen et~al.}(2005)\citenamefont{Sahlen, Liddle, and
  Parkinson}}]{Sahlen}
\bibinfo{author}{\bibfnamefont{M.}~\bibnamefont{Sahlen}},
  \bibinfo{author}{\bibfnamefont{A.~R.} \bibnamefont{Liddle}},
  \bibnamefont{and}
  \bibinfo{author}{\bibfnamefont{D.}~\bibnamefont{Parkinson}},
  \bibinfo{journal}{Phys. Rev.} \textbf{\bibinfo{volume}{D72}},
  \bibinfo{pages}{083511} (\bibinfo{year}{2005}), \eprint{astro-ph/0506696}.

\bibitem[{\citenamefont{Ratra and Peebles}(1988)}]{Ratra_Peebles}
\bibinfo{author}{\bibfnamefont{B.}~\bibnamefont{Ratra}} \bibnamefont{and}
  \bibinfo{author}{\bibfnamefont{P.~J.~E.} \bibnamefont{Peebles}},
  \bibinfo{journal}{Phys. Rev.} \textbf{\bibinfo{volume}{D37}},
  \bibinfo{pages}{3406} (\bibinfo{year}{1988}).

\bibitem[{\citenamefont{Wetterich}(1988)}]{Wetterich}
\bibinfo{author}{\bibfnamefont{C.}~\bibnamefont{Wetterich}},
  \bibinfo{journal}{Nucl. Phys.} \textbf{\bibinfo{volume}{B302}},
  \bibinfo{pages}{668} (\bibinfo{year}{1988}).

\bibitem[{\citenamefont{Frieman et~al.}(1995)\citenamefont{Frieman, Hill,
  Stebbins, and Waga}}]{Frieman_PNGB}
\bibinfo{author}{\bibfnamefont{J.~A.} \bibnamefont{Frieman}},
  \bibinfo{author}{\bibfnamefont{C.~T.} \bibnamefont{Hill}},
  \bibinfo{author}{\bibfnamefont{A.}~\bibnamefont{Stebbins}}, \bibnamefont{and}
  \bibinfo{author}{\bibfnamefont{I.}~\bibnamefont{Waga}},
  \bibinfo{journal}{Phys. Rev. Lett.} \textbf{\bibinfo{volume}{75}},
  \bibinfo{pages}{2077} (\bibinfo{year}{1995}), \eprint{astro-ph/9505060}.

\bibitem[{\citenamefont{Coble et~al.}(1997)\citenamefont{Coble, Dodelson, and
  Frieman}}]{Coble}
\bibinfo{author}{\bibfnamefont{K.}~\bibnamefont{Coble}},
  \bibinfo{author}{\bibfnamefont{S.}~\bibnamefont{Dodelson}}, \bibnamefont{and}
  \bibinfo{author}{\bibfnamefont{J.~A.} \bibnamefont{Frieman}},
  \bibinfo{journal}{Phys. Rev.} \textbf{\bibinfo{volume}{D55}},
  \bibinfo{pages}{1851} (\bibinfo{year}{1997}), \eprint{astro-ph/9608122}.

\bibitem[{\citenamefont{Ferreira and Joyce}(1998)}]{Ferreira_Joyce}
\bibinfo{author}{\bibfnamefont{P.~G.} \bibnamefont{Ferreira}} \bibnamefont{and}
  \bibinfo{author}{\bibfnamefont{M.}~\bibnamefont{Joyce}},
  \bibinfo{journal}{Phys. Rev.} \textbf{\bibinfo{volume}{D58}},
  \bibinfo{pages}{023503} (\bibinfo{year}{1998}), \eprint{astro-ph/9711102}.

\bibitem[{\citenamefont{Zlatev et~al.}(1999)\citenamefont{Zlatev, Wang, and
  Steinhardt}}]{Zlatev}
\bibinfo{author}{\bibfnamefont{I.}~\bibnamefont{Zlatev}},
  \bibinfo{author}{\bibfnamefont{L.-M.} \bibnamefont{Wang}}, \bibnamefont{and}
  \bibinfo{author}{\bibfnamefont{P.~J.} \bibnamefont{Steinhardt}},
  \bibinfo{journal}{Phys. Rev. Lett.} \textbf{\bibinfo{volume}{82}},
  \bibinfo{pages}{896} (\bibinfo{year}{1999}), \eprint{astro-ph/9807002}.

\bibitem[{\citenamefont{Liddle and Scherrer}(1998)}]{Liddle_Scherrer}
\bibinfo{author}{\bibfnamefont{A.~R.} \bibnamefont{Liddle}} \bibnamefont{and}
  \bibinfo{author}{\bibfnamefont{R.~J.} \bibnamefont{Scherrer}},
  \bibinfo{journal}{Phys. Rev.} \textbf{\bibinfo{volume}{D59}},
  \bibinfo{pages}{023509} (\bibinfo{year}{1998}), \eprint{astro-ph/9809272}.

\bibitem[{\citenamefont{Liddle et~al.}(1994)\citenamefont{Liddle, Parsons, and
  Barrow}}]{LPB}
\bibinfo{author}{\bibfnamefont{A.~R.} \bibnamefont{Liddle}},
  \bibinfo{author}{\bibfnamefont{P.}~\bibnamefont{Parsons}}, \bibnamefont{and}
  \bibinfo{author}{\bibfnamefont{J.~D.} \bibnamefont{Barrow}},
  \bibinfo{journal}{Phys. Rev.} \textbf{\bibinfo{volume}{D50}},
  \bibinfo{pages}{7222} (\bibinfo{year}{1994}), \eprint{astro-ph/9408015}.

\bibitem[{\citenamefont{Kinney}(2002)}]{Kinney_2003}
\bibinfo{author}{\bibfnamefont{W.~H.} \bibnamefont{Kinney}},
  \bibinfo{journal}{Phys. Rev.} \textbf{\bibinfo{volume}{D66}},
  \bibinfo{pages}{083508} (\bibinfo{year}{2002}), \eprint{astro-ph/0206032}.

\bibitem[{\citenamefont{Grivell and Liddle}(2000)}]{Grivell_Liddle}
\bibinfo{author}{\bibfnamefont{I.~J.} \bibnamefont{Grivell}} \bibnamefont{and}
  \bibinfo{author}{\bibfnamefont{A.~R.} \bibnamefont{Liddle}},
  \bibinfo{journal}{Phys. Rev.} \textbf{\bibinfo{volume}{D61}},
  \bibinfo{pages}{081301} (\bibinfo{year}{2000}), \eprint{astro-ph/9906327}.

\bibitem[{\citenamefont{Hoffman and Turner}(2001)}]{Hoffman_Turner}
\bibinfo{author}{\bibfnamefont{M.~B.} \bibnamefont{Hoffman}} \bibnamefont{and}
  \bibinfo{author}{\bibfnamefont{M.~S.} \bibnamefont{Turner}},
  \bibinfo{journal}{Phys. Rev.} \textbf{\bibinfo{volume}{D64}},
  \bibinfo{pages}{023506} (\bibinfo{year}{2001}), \eprint{astro-ph/0006321}.

\bibitem[{\citenamefont{Easther and Kinney}(2003)}]{Easther_Kinney}
\bibinfo{author}{\bibfnamefont{R.}~\bibnamefont{Easther}} \bibnamefont{and}
  \bibinfo{author}{\bibfnamefont{W.~H.} \bibnamefont{Kinney}},
  \bibinfo{journal}{Phys. Rev.} \textbf{\bibinfo{volume}{D67}},
  \bibinfo{pages}{043511} (\bibinfo{year}{2003}), \eprint{astro-ph/0210345}.

\bibitem[{\citenamefont{Peiris et~al.}(2003)}]{Peiris_2003}
\bibinfo{author}{\bibfnamefont{H.~V.} \bibnamefont{Peiris}}
  \bibnamefont{et~al.}, \bibinfo{journal}{Astrophys. J. Suppl.}
  \textbf{\bibinfo{volume}{148}}, \bibinfo{pages}{213} (\bibinfo{year}{2003}),
  \eprint{astro-ph/0302225}.

\bibitem[{\citenamefont{Peiris and
  Easther}(2006{\natexlab{a}})}]{Peiris_Easther}
\bibinfo{author}{\bibfnamefont{H.}~\bibnamefont{Peiris}} \bibnamefont{and}
  \bibinfo{author}{\bibfnamefont{R.}~\bibnamefont{Easther}},
  \bibinfo{journal}{JCAP} \textbf{\bibinfo{volume}{0607}}, \bibinfo{pages}{002}
  (\bibinfo{year}{2006}{\natexlab{a}}), \eprint{astro-ph/0603587}.

\bibitem[{\citenamefont{Peiris and Easther}(2006{\natexlab{b}})}]{PE2}
\bibinfo{author}{\bibfnamefont{H.}~\bibnamefont{Peiris}} \bibnamefont{and}
  \bibinfo{author}{\bibfnamefont{R.}~\bibnamefont{Easther}}
  (\bibinfo{year}{2006}{\natexlab{b}}), \eprint{astro-ph/0609003}.

\bibitem[{\citenamefont{Liddle}(2003)}]{Liddle_flow}
\bibinfo{author}{\bibfnamefont{A.~R.} \bibnamefont{Liddle}},
  \bibinfo{journal}{Phys. Rev.} \textbf{\bibinfo{volume}{D68}},
  \bibinfo{pages}{103504} (\bibinfo{year}{2003}), \eprint{astro-ph/0307286}.

\bibitem[{\citenamefont{Christensen and Meyer}(2000)}]{Christensen:2000ji}
\bibinfo{author}{\bibfnamefont{N.}~\bibnamefont{Christensen}} \bibnamefont{and}
  \bibinfo{author}{\bibfnamefont{R.}~\bibnamefont{Meyer}}
  (\bibinfo{year}{2000}), \eprint{astro-ph/0006401}.

\bibitem[{\citenamefont{Christensen et~al.}(2001)\citenamefont{Christensen,
  Meyer, Knox, and Luey}}]{Christensen:2001gj}
\bibinfo{author}{\bibfnamefont{N.}~\bibnamefont{Christensen}},
  \bibinfo{author}{\bibfnamefont{R.}~\bibnamefont{Meyer}},
  \bibinfo{author}{\bibfnamefont{L.}~\bibnamefont{Knox}}, \bibnamefont{and}
  \bibinfo{author}{\bibfnamefont{B.}~\bibnamefont{Luey}},
  \bibinfo{journal}{Class. Quant. Grav.} \textbf{\bibinfo{volume}{18}},
  \bibinfo{pages}{2677} (\bibinfo{year}{2001}), \eprint{astro-ph/0103134}.

\bibitem[{\citenamefont{Knox et~al.}(2001)\citenamefont{Knox, Christensen, and
  Skordis}}]{Knox:2001fz}
\bibinfo{author}{\bibfnamefont{L.}~\bibnamefont{Knox}},
  \bibinfo{author}{\bibfnamefont{N.}~\bibnamefont{Christensen}},
  \bibnamefont{and} \bibinfo{author}{\bibfnamefont{C.}~\bibnamefont{Skordis}},
  \bibinfo{journal}{Astrophys. J.} \textbf{\bibinfo{volume}{563}},
  \bibinfo{pages}{L95} (\bibinfo{year}{2001}), \eprint{astro-ph/0109232}.

\bibitem[{\citenamefont{Lewis and Bridle}(2002)}]{Lewis:2002ah}
\bibinfo{author}{\bibfnamefont{A.}~\bibnamefont{Lewis}} \bibnamefont{and}
  \bibinfo{author}{\bibfnamefont{S.}~\bibnamefont{Bridle}},
  \bibinfo{journal}{Phys. Rev.} \textbf{\bibinfo{volume}{D66}},
  \bibinfo{pages}{103511} (\bibinfo{year}{2002}), \eprint{astro-ph/0205436}.

\bibitem[{\citenamefont{Kosowsky et~al.}(2002)\citenamefont{Kosowsky,
  Milosavljevic, and Jimenez}}]{Kosowsky:2002zt}
\bibinfo{author}{\bibfnamefont{A.}~\bibnamefont{Kosowsky}},
  \bibinfo{author}{\bibfnamefont{M.}~\bibnamefont{Milosavljevic}},
  \bibnamefont{and} \bibinfo{author}{\bibfnamefont{R.}~\bibnamefont{Jimenez}},
  \bibinfo{journal}{Phys. Rev.} \textbf{\bibinfo{volume}{D66}},
  \bibinfo{pages}{063007} (\bibinfo{year}{2002}), \eprint{astro-ph/0206014}.

\bibitem[{\citenamefont{Verde et~al.}(2003)}]{Verde:2003ey}
\bibinfo{author}{\bibfnamefont{L.}~\bibnamefont{Verde}} \bibnamefont{et~al.},
  \bibinfo{journal}{Astrophys. J. Suppl.} \textbf{\bibinfo{volume}{148}},
  \bibinfo{pages}{195} (\bibinfo{year}{2003}), \eprint{astro-ph/0302218}.

\bibitem[{\citenamefont{Gelman and Rubin}(1992)}]{gelman/rubin}
\bibinfo{author}{\bibfnamefont{A.}~\bibnamefont{Gelman}} \bibnamefont{and}
  \bibinfo{author}{\bibfnamefont{D.}~\bibnamefont{Rubin}},
  \bibinfo{journal}{Statistical Science} \textbf{\bibinfo{volume}{7}},
  \bibinfo{pages}{452} (\bibinfo{year}{1992}).

\bibitem[{\citenamefont{Eisenstein et~al.}(2005)}]{Eisenstein}
\bibinfo{author}{\bibfnamefont{D.~J.} \bibnamefont{Eisenstein}}
  \bibnamefont{et~al.}, \bibinfo{journal}{Astrophys. J.}
  \textbf{\bibinfo{volume}{633}}, \bibinfo{pages}{560} (\bibinfo{year}{2005}),
  \eprint{astro-ph/0501171}.

\bibitem[{\citenamefont{Freedman et~al.}(2001)}]{HKP}
\bibinfo{author}{\bibfnamefont{W.~L.} \bibnamefont{Freedman}}
  \bibnamefont{et~al.}, \bibinfo{journal}{Astrophys. J.}
  \textbf{\bibinfo{volume}{553}}, \bibinfo{pages}{47} (\bibinfo{year}{2001}),
  \eprint{astro-ph/0012376}.

\bibitem[{\citenamefont{Caldwell and Linder}(2005)}]{Caldwell_Linder}
\bibinfo{author}{\bibfnamefont{R.~R.} \bibnamefont{Caldwell}} \bibnamefont{and}
  \bibinfo{author}{\bibfnamefont{E.~V.} \bibnamefont{Linder}},
  \bibinfo{journal}{Phys. Rev. Lett.} \textbf{\bibinfo{volume}{95}},
  \bibinfo{pages}{141301} (\bibinfo{year}{2005}), \eprint{astro-ph/0505494}.

\bibitem[{\citenamefont{Chevallier and Polarski}(2001)}]{Chevallier_Polarski}
\bibinfo{author}{\bibfnamefont{M.}~\bibnamefont{Chevallier}} \bibnamefont{and}
  \bibinfo{author}{\bibfnamefont{D.}~\bibnamefont{Polarski}},
  \bibinfo{journal}{Int. J. Mod. Phys.} \textbf{\bibinfo{volume}{D10}},
  \bibinfo{pages}{213} (\bibinfo{year}{2001}), \eprint{gr-qc/0009008}.

\bibitem[{\citenamefont{Huterer and Starkman}(2003)}]{Huterer_Starkman}
\bibinfo{author}{\bibfnamefont{D.}~\bibnamefont{Huterer}} \bibnamefont{and}
  \bibinfo{author}{\bibfnamefont{G.}~\bibnamefont{Starkman}},
  \bibinfo{journal}{Phys. Rev. Lett.} \textbf{\bibinfo{volume}{90}},
  \bibinfo{pages}{031301} (\bibinfo{year}{2003}), \eprint{astro-ph/0207517}.

\bibitem[{\citenamefont{Crittenden and Pogosian}(2005)}]{Crittenden_Pogosian}
\bibinfo{author}{\bibfnamefont{R.~G.} \bibnamefont{Crittenden}}
  \bibnamefont{and} \bibinfo{author}{\bibfnamefont{L.}~\bibnamefont{Pogosian}}
  (\bibinfo{year}{2005}), \eprint{astro-ph/0510293}.

\bibitem[{\citenamefont{Linder and Huterer}(2005)}]{Linder_Huterer_howmany}
\bibinfo{author}{\bibfnamefont{E.~V.} \bibnamefont{Linder}} \bibnamefont{and}
  \bibinfo{author}{\bibfnamefont{D.}~\bibnamefont{Huterer}},
  \bibinfo{journal}{Phys. Rev.} \textbf{\bibinfo{volume}{D72}},
  \bibinfo{pages}{043509} (\bibinfo{year}{2005}), \eprint{astro-ph/0505330}.

\bibitem[{\citenamefont{Simpson and Bridle}(2006)}]{Simpson_Bridle_PC}
\bibinfo{author}{\bibfnamefont{F.}~\bibnamefont{Simpson}} \bibnamefont{and}
  \bibinfo{author}{\bibfnamefont{S.}~\bibnamefont{Bridle}},
  \bibinfo{journal}{Phys. Rev.} \textbf{\bibinfo{volume}{D73}},
  \bibinfo{pages}{083001} (\bibinfo{year}{2006}), \eprint{astro-ph/0602213}.

\bibitem[{\citenamefont{Hu}(2002)}]{Hu_PC}
\bibinfo{author}{\bibfnamefont{W.}~\bibnamefont{Hu}}, \bibinfo{journal}{Phys.
  Rev.} \textbf{\bibinfo{volume}{D66}}, \bibinfo{pages}{083515}
  (\bibinfo{year}{2002}), \eprint{astro-ph/0208093}.

\bibitem[{\citenamefont{Stephan-Otto}(2006)}]{Stephan-Otto}
\bibinfo{author}{\bibfnamefont{C.}~\bibnamefont{Stephan-Otto}},
  \bibinfo{journal}{Phys. Rev.} \textbf{\bibinfo{volume}{D74}},
  \bibinfo{pages}{023507} (\bibinfo{year}{2006}), \eprint{astro-ph/0605403}.

\bibitem[{\citenamefont{Dick et~al.}(2006)\citenamefont{Dick, Knox, and
  Chu}}]{Dick}
\bibinfo{author}{\bibfnamefont{J.}~\bibnamefont{Dick}},
  \bibinfo{author}{\bibfnamefont{L.}~\bibnamefont{Knox}}, \bibnamefont{and}
  \bibinfo{author}{\bibfnamefont{M.}~\bibnamefont{Chu}},
  \bibinfo{journal}{JCAP} \textbf{\bibinfo{volume}{0607}}, \bibinfo{pages}{001}
  (\bibinfo{year}{2006}), \eprint{astro-ph/0603247}.

\bibitem[{\citenamefont{Shapiro and Turner}(2005)}]{Shapiro_Turner}
\bibinfo{author}{\bibfnamefont{C.}~\bibnamefont{Shapiro}} \bibnamefont{and}
  \bibinfo{author}{\bibfnamefont{M.~S.} \bibnamefont{Turner}}
  (\bibinfo{year}{2005}), \eprint{astro-ph/0512586}.

\bibitem[{\citenamefont{Gerke and Efstathiou}(2002)}]{Gerke_Efstathiou}
\bibinfo{author}{\bibfnamefont{B.~F.} \bibnamefont{Gerke}} \bibnamefont{and}
  \bibinfo{author}{\bibfnamefont{G.}~\bibnamefont{Efstathiou}},
  \bibinfo{journal}{Mon. Not. Roy. Astron. Soc.}
  \textbf{\bibinfo{volume}{335}}, \bibinfo{pages}{33} (\bibinfo{year}{2002}),
  \eprint{astro-ph/0201336}.

\bibitem[{\citenamefont{Daly and Djorgovski}(2004)}]{Daly_Djorgovski}
\bibinfo{author}{\bibfnamefont{R.~A.} \bibnamefont{Daly}} \bibnamefont{and}
  \bibinfo{author}{\bibfnamefont{S.~G.} \bibnamefont{Djorgovski}},
  \bibinfo{journal}{Astrophys. J.} \textbf{\bibinfo{volume}{612}},
  \bibinfo{pages}{652} (\bibinfo{year}{2004}), \eprint{astro-ph/0403664}.

\bibitem[{\citenamefont{Simon et~al.}(2005)\citenamefont{Simon, Verde, and
  Jimenez}}]{Simon}
\bibinfo{author}{\bibfnamefont{J.}~\bibnamefont{Simon}},
  \bibinfo{author}{\bibfnamefont{L.}~\bibnamefont{Verde}}, \bibnamefont{and}
  \bibinfo{author}{\bibfnamefont{R.}~\bibnamefont{Jimenez}},
  \bibinfo{journal}{Phys. Rev.} \textbf{\bibinfo{volume}{D71}},
  \bibinfo{pages}{123001} (\bibinfo{year}{2005}), \eprint{astro-ph/0412269}.

\bibitem[{\citenamefont{Sahni and Starobinsky}(2006)}]{Sahni_review}
\bibinfo{author}{\bibfnamefont{V.}~\bibnamefont{Sahni}} \bibnamefont{and}
  \bibinfo{author}{\bibfnamefont{A.}~\bibnamefont{Starobinsky}}
  (\bibinfo{year}{2006}), \eprint{astro-ph/0610026}.

\bibitem[{\citenamefont{Tegmark et~al.}(2006)}]{Tegmark_LRG}
\bibinfo{author}{\bibfnamefont{M.}~\bibnamefont{Tegmark}} \bibnamefont{et~al.}
  (\bibinfo{year}{2006}), \eprint{astro-ph/0608632}.

\bibitem[{\citenamefont{Albrecht et~al.}(2006)}]{DETF}
\bibinfo{author}{\bibfnamefont{A.}~\bibnamefont{Albrecht}} \bibnamefont{et~al.}
  (\bibinfo{year}{2006}), \eprint{astro-ph/0609591}.

\bibitem[{\citenamefont{Linder}(2006)}]{Linder_paths}
\bibinfo{author}{\bibfnamefont{E.~V.} \bibnamefont{Linder}},
  \bibinfo{journal}{Phys. Rev.} \textbf{\bibinfo{volume}{D73}},
  \bibinfo{pages}{063010} (\bibinfo{year}{2006}), \eprint{astro-ph/0601052}.

\bibitem[{\citenamefont{Scherrer}(2006)}]{Scherrer}
\bibinfo{author}{\bibfnamefont{R.~J.} \bibnamefont{Scherrer}},
  \bibinfo{journal}{Phys. Rev.} \textbf{\bibinfo{volume}{D73}},
  \bibinfo{pages}{043502} (\bibinfo{year}{2006}), \eprint{astro-ph/0509890}.

\bibitem[{\citenamefont{Chiba}(2006)}]{Chiba}
\bibinfo{author}{\bibfnamefont{T.}~\bibnamefont{Chiba}},
  \bibinfo{journal}{Phys. Rev.} \textbf{\bibinfo{volume}{D73}},
  \bibinfo{pages}{063501} (\bibinfo{year}{2006}), \eprint{astro-ph/0510598}.

\bibitem[{\citenamefont{Gonzalez-Diaz}(2000)}]{Gonzalez-Diaz}
\bibinfo{author}{\bibfnamefont{P.~F.} \bibnamefont{Gonzalez-Diaz}},
  \bibinfo{journal}{Phys. Rev.} \textbf{\bibinfo{volume}{D62}},
  \bibinfo{pages}{023513} (\bibinfo{year}{2000}), \eprint{astro-ph/0004125}.

\bibitem[{\citenamefont{Liddle et~al.}(2006)\citenamefont{Liddle, Mukherjee,
  Parkinson, and Wang}}]{Liddle2006}
\bibinfo{author}{\bibfnamefont{A.~R.} \bibnamefont{Liddle}},
  \bibinfo{author}{\bibfnamefont{P.}~\bibnamefont{Mukherjee}},
  \bibinfo{author}{\bibfnamefont{D.}~\bibnamefont{Parkinson}},
  \bibnamefont{and} \bibinfo{author}{\bibfnamefont{Y.}~\bibnamefont{Wang}}
  (\bibinfo{year}{2006}), \eprint{astro-ph/0610126}.

\bibitem[{\citenamefont{Bean and Dore}(2004)}]{Bean_Dore}
\bibinfo{author}{\bibfnamefont{R.}~\bibnamefont{Bean}} \bibnamefont{and}
  \bibinfo{author}{\bibfnamefont{O.}~\bibnamefont{Dore}},
  \bibinfo{journal}{Phys. Rev.} \textbf{\bibinfo{volume}{D69}},
  \bibinfo{pages}{083503} (\bibinfo{year}{2004}), \eprint{astro-ph/0307100}.

\bibitem[{\citenamefont{Weller and Lewis}(2003)}]{Weller_Lewis}
\bibinfo{author}{\bibfnamefont{J.}~\bibnamefont{Weller}} \bibnamefont{and}
  \bibinfo{author}{\bibfnamefont{A.~M.} \bibnamefont{Lewis}},
  \bibinfo{journal}{Mon. Not. Roy. Astron. Soc.}
  \textbf{\bibinfo{volume}{346}}, \bibinfo{pages}{987} (\bibinfo{year}{2003}),
  \eprint{astro-ph/0307104}.

\bibitem[{\citenamefont{Hannestad}(2005)}]{Hannestad_sound}
\bibinfo{author}{\bibfnamefont{S.}~\bibnamefont{Hannestad}},
  \bibinfo{journal}{Phys. Rev.} \textbf{\bibinfo{volume}{D71}},
  \bibinfo{pages}{103519} (\bibinfo{year}{2005}), \eprint{astro-ph/0504017}.

\bibitem[{\citenamefont{Battye and Moss}(2006)}]{Battye_Moss}
\bibinfo{author}{\bibfnamefont{R.~A.} \bibnamefont{Battye}} \bibnamefont{and}
  \bibinfo{author}{\bibfnamefont{A.}~\bibnamefont{Moss}},
  \bibinfo{journal}{Phys. Rev.} \textbf{\bibinfo{volume}{D74}},
  \bibinfo{pages}{041301} (\bibinfo{year}{2006}), \eprint{astro-ph/0602377}.

\bibitem[{\citenamefont{Hu et~al.}(2001)\citenamefont{Hu, Fukugita,
  Zaldarriaga, and Tegmark}}]{Hu_Fukugita}
\bibinfo{author}{\bibfnamefont{W.}~\bibnamefont{Hu}},
  \bibinfo{author}{\bibfnamefont{M.}~\bibnamefont{Fukugita}},
  \bibinfo{author}{\bibfnamefont{M.}~\bibnamefont{Zaldarriaga}},
  \bibnamefont{and} \bibinfo{author}{\bibfnamefont{M.}~\bibnamefont{Tegmark}},
  \bibinfo{journal}{Astrophys. J.} \textbf{\bibinfo{volume}{549}},
  \bibinfo{pages}{669} (\bibinfo{year}{2001}), \eprint{astro-ph/0006436}.

\bibitem[{\citenamefont{Frieman et~al.}(2003)\citenamefont{Frieman, Huterer,
  Linder, and Turner}}]{Frieman}
\bibinfo{author}{\bibfnamefont{J.~A.} \bibnamefont{Frieman}},
  \bibinfo{author}{\bibfnamefont{D.}~\bibnamefont{Huterer}},
  \bibinfo{author}{\bibfnamefont{E.~V.} \bibnamefont{Linder}},
  \bibnamefont{and} \bibinfo{author}{\bibfnamefont{M.~S.}
  \bibnamefont{Turner}}, \bibinfo{journal}{Phys. Rev.}
  \textbf{\bibinfo{volume}{D67}}, \bibinfo{pages}{083505}
  (\bibinfo{year}{2003}), \eprint{astro-ph/0208100}.

\bibitem[{\citenamefont{Aldering et~al.}(2004)}]{SNAP}
\bibinfo{author}{\bibfnamefont{G.}~\bibnamefont{Aldering}} \bibnamefont{et~al.}
  (\bibinfo{collaboration}{SNAP}) (\bibinfo{year}{2004}),
  \eprint{astro-ph/0405232}.

\bibitem[{\citenamefont{Blake et~al.}(2006)}]{Blake}
\bibinfo{author}{\bibfnamefont{C.}~\bibnamefont{Blake}} \bibnamefont{et~al.},
  \bibinfo{journal}{Mon. Not. Roy. Astron. Soc.}
  \textbf{\bibinfo{volume}{365}}, \bibinfo{pages}{255} (\bibinfo{year}{2006}),
  \eprint{astro-ph/0510239}.

\bibitem[{\citenamefont{{The Planck Collaboration}}()}]{Planck}
\bibinfo{author}{\bibnamefont{{The Planck Collaboration}}},
  \urlprefix\url{http://www.rssd.esa.int/index.php?project=PLANCK}.

\end{thebibliography}

\end{document}